\documentclass[aps,prd,twocolumn,groupedaddress,showpacs,eqsecnum,floatfix,nofootinbib]{revtex4-1}

\usepackage{amsmath}
\usepackage{amssymb}
\usepackage{graphicx}

\usepackage{color}

\begin{document}

\title{Solitons and hairy black holes in Einstein-non-Abelian-Proca theory\\ in anti-de Sitter space-time}

\author{Supakchai Ponglertsakul}
\email{smp12sp@sheffield.ac.uk}

\affiliation{Consortium for Fundamental Physics, School of Mathematics and Statistics,
The University of Sheffield, Hicks Building, Hounsfield Road, Sheffield S3 7RH, United Kingdom}

\author{Elizabeth Winstanley}
\email{e.winstanley@sheffield.ac.uk}

\affiliation{Consortium for Fundamental Physics, School of Mathematics and Statistics,
The University of Sheffield, Hicks Building, Hounsfield Road, Sheffield S3 7RH, United Kingdom}

\affiliation{Department of Physics and Astronomy, University of Canterbury, Private Bag 4800, Christchurch 8140, New Zealand}

\date{\today}

\begin{abstract}
We present new soliton and hairy black hole solutions of Einstein-non-Abelian-Proca theory in asymptotically anti-de Sitter space-time with gauge group ${\mathfrak {su}}(2)$.
For static, spherically symmetric configurations, we show that the gauge field must be purely magnetic, and solve the resulting field equations numerically.
The equilibrium gauge field is described by a single function $\omega (r)$, which must have at least one zero.
The solitons and hairy black holes share many properties with the corresponding solutions in asymptotically flat space-time.
In particular, all the solutions we study are unstable under linear, spherically symmetric, perturbations of the metric and gauge field.
\end{abstract}


\maketitle

\section{Introduction}
\label{sec:intro}

The study, over the past twenty-five years, of black holes and solitons in classical non-Abelian gauge theories coupled to gravity has revealed
many surprises.
The first surprise was the discovery \cite{Bartnik:1988am} of nontrivial regular soliton solutions of ${\mathfrak {su}}(2)$ Einstein-Yang-Mills (EYM) theory in asymptotically flat space-time.
This was a surprise because there are no regular soliton solutions of Einstein-Maxwell theory in four space-time dimensions, other than the trivial solution, pure Minkowski space-time.
Soon after, corresponding nontrivial hairy black hole solutions of asymptotically flat ${\mathfrak {su}}(2)$ EYM theory were also discovered \cite{Bizon:1990sr}.
Although the asymptotically flat soliton and black hole solutions are unstable under linear, spherically symmetric perturbations of the metric and non-Abelian gauge field \cite{Straumann:1989tf,Lavrelashvili:1994rp}, their discovery sparked what is now an extensive literature on classical soliton and black hole solutions of the EYM equations, see for example \cite{Volkov:1998cc} for reviews.
For the ${\mathfrak {su}}(2)$ gauge group and asymptotically flat space-times, it can be proven that the gauge field must be purely magnetic for nontrivial configurations \cite{Ershov:1991nv,Bizon:1992pi} and is described by a single function $\omega (r)$ of the radial coordinate $r$.
Furthermore, the function $\omega (r)$ must have at least one zero \cite{Bartnik:1988am,Bizon:1990sr,Breitenlohner:1993es}.
The solutions can be parameterized by $r_{h}$, the radius of the event horizon (for the black hole case, setting $r_{h}=0$ gives the soliton case) and $n$, the number of zeros of the gauge field function $\omega (r)$.

More complicated matter models involving Yang-Mills gauge fields have also been extensively studied.
For example, in Ref.~\cite{Greene:1992fw} asymptotically flat soliton and hairy black hole solutions are found in two models where the Yang-Mills gauge symmetry is broken.
The first model, Einstein-non-Abelian-Proca (ENAP) theory, is an effective theory in which the Yang-Mills gauge field is given a nonzero mass term in the Lagrangian.  In the second model, Einstein-Yang-Mills-Higgs (EYMH) theory, the gauge field acquires a mass dynamically via its coupling to a Higgs field in the fundamental representation \footnote{Throughout this paper ``EYMH'' refers to Einstein-Yang-Mills-Higgs theory with a doublet-Higgs field in the fundamental representation.  There are also solutions of Einstein-Yang-Mills-Higgs theory with a triplet-Higgs field in the adjoint representation in both asymptotically flat and asymptotically AdS space-times, but their properties are rather different \cite{Aichelburg:1992st}.}.
The black holes and solitons in the two models presented in \cite{Greene:1992fw} are very similar and share many properties with the pure EYM solutions, in particular the ${\mathfrak {su}}(2)$ gauge field is purely magnetic and described by a single function $\omega (r)$ which must have at least one zero.
However, the phase space of solutions is more complicated, with two branches of solutions existing, so that there are two solutions for each value of $r_{h}$ (with $r_{h}=0$ for solitons) and $n$.
Solutions on the first branch, denoted the $n=i$ branch (for $i=1,2,\ldots $), are very similar to the corresponding $n=i$ EYM solitons or black holes, as applicable, and do not vary much as either the Proca field mass (for ENAP) or Higgs coupling (for EYMH) vary.  In particular, as either the Proca field mass or Higgs coupling tends to zero, the solutions approach the relevant $n=i$ solution of pure EYM theory in asymptotically flat space-time.
The second branch is denoted the quasi-$n=i-1$ branch (for $i=1,2,\ldots $).  For solutions on this branch, the value of $r$ at the outermost zero of $\omega (r)$ tends to infinity as either the Proca field mass or Higgs coupling tends to zero and the solutions approach the corresponding $n=i-1$ solution of pure EYM theory.
The stability of the asymptotically flat EYMH solitons and black holes has been extensively studied \cite{Maeda:1993ap,Mavromatos:1995kc} - all the solutions presented in \cite{Greene:1992fw} are unstable under linear, spherically symmetric, perturbations of the metric and matter fields.

Further surprises emerged from the study of solitons and black holes in EYM theory in asymptotically anti-de Sitter (AdS) space-time.
Considering purely magnetic configurations with an ${\mathfrak {su}}(2)$ gauge group, solutions exist for which the gauge field function $\omega (r)$ has no zeros, at least some of which are stable under linear, spherically symmetric perturbations \cite{Winstanley:1998sn,Bjoraker:1999yd} (and also under general linear perturbations \cite{Sarbach:2001mc}).  The phase space of solutions also has a different structure compared to the asymptotically flat case.
For the ${\mathfrak {su}}(2)$ gauge group, the phase space can be taken to be $r_{h}$ (with $r_{h}=0$ for solitons) and a single additional parameter governing the behaviour of the gauge field either near the origin or near the event horizon, as applicable.
In asymptotically flat space-time, nontrivial EYM solutions occur at discrete values of the additional parameter for fixed $r_{h}$, but in asymptotically adS space-time, there are solutions in continuous regions of the phase space.

A natural question is whether these remarkable differences between soliton and black hole solutions of EYM in asymptotically flat and asymptotically AdS space-times persist if an extended model is considered.
It is perhaps at first sight surprising that this is not the case for the EYMH model \cite{VanderBij:2001ah}.
The asymptotically AdS soliton and black hole solutions of this model behave similarly to those in asymptotically flat space-time. In particular, the gauge field function $\omega (r)$ has at least one zero; there are two branches of solutions for which $\omega (r)$ has $n$ zeros; and the solutions are unstable (proven in \cite{VanderBij:2001ah} for solitons and in \cite{Winstanley:2016taz} for black holes).

The authors of Ref.~\cite{VanderBij:2001ah} anticipated that soliton and hairy black hole solutions of the ENAP model in AdS would have very similar properties to those of the EYMH model, but, to the best of our knowledge, this has not been explored in detail in the literature.  In this paper we close this gap by studying soliton and black hole solutions of ENAP theory in AdS, to see whether they behave like the EYMH solutions or more like the EYM solutions in AdS.

The outline of this paper is as follows.
In Sec.~\ref{sec:ENAP} we introduce ENAP theory in AdS, the field equations and our ansatze for the static, spherically symmetric metric and ${\mathfrak {su}}(2)$ gauge field. We show that there are no nontrivial dyon solutions, so that the gauge field must be purely magnetic. In this case there is a single function $\omega (r)$ describing the gauge field, and we show that this function must have at least one zero.
We present numerical solutions of the equilibrium field equations describing solitons and hairy black holes in Sec.~\ref{sec:solutions}.
The stability of these solutions under linear, spherically symmetric, perturbations of the metric and gauge field is studied in Sec.~\ref{sec:stab}.
Our conclusions are in Sec.~\ref{sec:conc}.

\section{Einstein-non-Abelian-Proca theory}
\label{sec:ENAP}

\subsection{Action, ansatz and field equations}
\label{sec:ansatz}

The ${\mathfrak {su}}(2)$ ENAP theory in four-dimensional asymptotically AdS space-time is described by the action
\begin{equation}
S_{{\text {ENAP}}} = \int d^{4}x \sqrt{-g} \left( \frac{\mathcal{R}}{16\pi}-2\Lambda  + \mathcal{L}_{{\text {NAP}}} \right)
\label{ENAPaction},
\end{equation}
where the matter field Lagrangian density is
\begin{equation}
\mathcal{L}_{{\text {NAP}}} = -\frac{1}{16\pi}\left( \text{Tr} F_{ab}F^{ab} + 2\mu^{2} \text{Tr} A_{c}A^{c} \right) .
\label{ENAPLagrangian}
\end{equation}
In (\ref{ENAPaction}, \ref{ENAPLagrangian}),
$\mathcal{R}=g^{ab}R_{ab}$ is the Ricci scalar, $\Lambda$ is the cosmological constant, $\mu $ is the non-Abelian Proca (NAP) field mass and the Lie algebra trace is denoted by ${\text {Tr}}$.
The non-Abelian gauge field strength $F_{ab}$ is given terms of the gauge field potential $A_{a}$ as follows
\begin{equation}
F_{ab} = \partial_{a}A_{b} - \partial_{b}A_{a} + e \left[A_{a},A_{b}\right],
\end{equation}
where $e$ is the gauge coupling constant.
Throughout this paper, the metric has signature $(-,+,+,+)$ and we use units in which $c=G=1$.

By varying the action (\ref{ENAPaction}) with respect to the inverse metric $g^{ab}$ and gauge potential $A_{a}$, we obtain two equations of motion
\begin{subequations}
\label{fieldequations}
\begin{align}
R_{ab}-\frac{1}{2}g_{ab}\mathcal{R} + \Lambda g_{ab} &= 8\pi T_{ab},
\label{EFE} \\
\nabla_{a}{F_b}^{a} +  e \left[A_{a},{F_b}^{a}\right] + \mu^2 A_{b} &= 0,
\label{NAPequation}
\end{align}
\end{subequations}
where the energy-momentum tensor of the NAP field is given by
\begin{align}
\label{ENAPstresstensor}
8\pi T_{ab} = & 2\text{Tr}F_{ac}{F_{b}}^{c}-\frac{1}{2}g_{ab}\text{Tr}F_{cd}F^{cd} \nonumber \\
 & + \mu^2\left(2\text{Tr}A_{a}A_{b}-g_{ab}\text{Tr}A_{c}A^{c}\right).
\end{align}
Taking the divergence of the Proca equation (\ref{NAPequation}) we find that the gauge potential must satisfy the constraint
\begin{align}
\nabla_{a} A^{a} &= 0,
\label{gaugeCond}
\end{align}
which restricts our choice of gauge for the potential $A_{a}$.

We consider a spherically symmetric space-time given by the following metric ansatz
\begin{equation}
ds^{2} = -RS^{2}dt^{2}+R^{-1}dr^{2}+r^{2}d\theta^{2}+r^{2}\sin^{2}\theta~d\phi^{2}.
\label{metric}
\end{equation}
The metric functions $R$ and $S$ are functions of the radial coordinate $r$ and time $t$ only.  We can express the metric function $R(t,r)$ as
\begin{equation}
R(t,r) = 1-\frac{2m(t,r)}{r}-\frac{\Lambda r^{2}}{3},
\label{mdef}
\end{equation}
where $m(t,r)$ can be interpreted as the total mass within the given radius $r$.

The most general spherically symmetric $\mathfrak{su}(2)$ gauge potential is given by \cite{Witten:1976ck}
\begin{align}
eA =  & a\hat{\tau}_{r}dt + b\hat{\tau}_{r}dr+[d\hat{\tau}_{\theta}-(1+c)\hat{\tau}_{\phi}]d\theta  \nonumber \\
&  + [(1+c)\hat{\tau}_{\theta}+d\hat{\tau}_{\phi}]\sin\theta \, d\phi ,
\label{generalgaugeconnection}
\end{align}
where the functions  $a,b,c$ and $d$ depend only on $r$ and $t$.
The $\mathfrak{su}(2)$ basis matrices ${\hat {\tau }}_{a}$ satisfy the relations $[\hat{\tau}_{a},\hat{\tau}_{b}] = \epsilon_{abc}\hat{\tau}_{c}$
and can be expressed in spherical coordinates as follows
\begin{align}
\hat{\tau}_{r} &= -\frac{i}{2}\left[\sigma_{1}\sin\theta\cos\phi + \sigma_{2}\sin\theta\sin\phi + \sigma_{3}\cos\theta
\right], \nonumber \\
\hat{\tau}_{\theta} &= -\frac{i}{2}\left[\sigma_{1}\cos\theta\cos\phi + \sigma_{2}\cos\theta\sin\phi - \sigma_{3}\sin\theta
\right], \nonumber \\
\hat{\tau}_{\phi} &= -\frac{i}{2}\left[-\sigma_{1}\sin\phi + \sigma_{2}\cos\phi \right],
\end{align}
where the $\sigma_{j}$'s (with $j\in\left(1,2,3\right)$) are the usual Pauli matrices.

The gauge potential ansatz (\ref{generalgaugeconnection}) has a residual $\mathfrak{u}(1)$ gauge freedom
\begin{equation}
\label{gaugetransform}
A \rightarrow h A h^{-1} + \frac{1}{e}h \,dh^{-1},
\end{equation}
with transformation matrix $h=\exp[\beta(t,r)\hat{\tau}_r]$, under which the gauge potential functions transform as \cite{Greene:1992fw}
\begin{equation}
\left( \begin{array}{ccc}
a \\
b \\
c \\
d \\
\end{array} \right)\rightarrow
\left( \begin{array}{ccc}
\hat{a} \\
\hat{b} \\
\hat{c} \\
\hat{d} \\
\end{array} \right) =
\left( \begin{array}{ccc}
a-\dot{\beta} \\
b-\beta' \\
c \cos\beta-d \sin\beta \\
d \cos\beta+c \sin\beta \\
\end{array} \right),
\label{gaugetransform1}
\end{equation}
where $\dot{}$ and $'$ denote partial derivative with respect to time $t$ and radial coordinate $r$, respectively.
However this residual gauge freedom is restricted by the constraint (\ref{gaugeCond}).

\subsection{Static configurations}
\label{sec:static}

Now consider static, spherically symmetric configurations so that the metric (\ref{metric}) and gauge potential (\ref{generalgaugeconnection})
depend only on the radial coordinate $r$ and not on time $t$.
In this case the constraint (\ref{gaugeCond}) takes the form
\begin{equation}
Rb' + \left[ \frac {2R}{r} + \frac {\left( RS \right) ' }{S} \right] b - \frac {2}{r^{2}} d =0.
\label{gaugeconstraint1}
\end{equation}
One can choose $\beta(r)$ in the gauge transformation (\ref{gaugetransform1}) such that $\hat{b}\equiv0$.
In this case the constraint (\ref{gaugeconstraint1}) implies that ${\hat {d}}=0$.
In keeping with the conventions in the literature \cite{Greene:1992fw}, we rewrite $c(r)=\omega (r)$.

Therefore the gauge potential ansatz (\ref{generalgaugeconnection}) takes the form
\begin{equation}
eA = a{\hat{\tau }}_{r} dt+\left(1+\omega\right)\left[-\hat{\tau}_{\phi}d\theta+\hat{\tau}_{\theta}\sin\theta \, d\phi\right] .
\label{dyon:gaugeform}
\end{equation}
With this gauge potential ansatz, the static field equations (\ref{fieldequations}) take the form
\begin{subequations}
\label{dyonfieldequations}
\begin{align}
a'' = & -\frac{2a'}{r}+\frac{a'S'}{S}+\frac{2a\omega^2}{r^{2}R}+\frac{a\mu^2}{R},
\label{NAPtdyon} \\
\omega'' =  & -\frac{a^{2}\omega}{R^{2}S^{2}}-\frac{\omega'S'}{S}-\frac{\omega'R'}{R}+\frac{\omega\left(\omega^{2}-1\right)}{r^{2}R}
\nonumber \\
&  +\frac{\left(1+\omega\right)\mu^2}{R},
\label{NAPthetadyon} \\
m' = & \frac{r^{2}a'^{2}}{2e^2S^2}+\frac{a^2\omega^2}{e^2RS^2}+\frac{a^2\mu^2r^2}{2e^2RS^2}+\frac{R\omega'^2}{e^2}
\nonumber \\
&+\frac{\left(\omega^2-1\right)^2}{2e^2r^2}+\frac{\left(1+\omega\right)^2\mu^2}{e^2},
\label{EFEm}\\
\frac{S'}{S} = & \frac{2a^2\omega^2}{e^2rR^2S^2}+\frac{ra^2\mu^2}{e^2R^2S^2}+\frac{2\omega'^2}{e^2r}.
\label{EFEdyon}
\end{align}
\end{subequations}
When the Proca mass $\mu $ is set equal to zero, the equations (\ref{dyonfieldequations}) reduce to the usual EYM equations for a dyonic configuration \cite{Bjoraker:1999yd,Nolan:2012ax}.

The field equations (\ref{dyonfieldequations}) are singular at the origin, the event horizon $r=r_{h}$ (if there is one) and as $r\rightarrow \infty $.
We therefore need to impose boundary conditions on the field variables near these singular points.
For globally regular (soliton) solutions, we assume that all quantities are finite at the origin, and furthermore that all curvature invariants are also finite there.
These requirements mean that the magnetic gauge field function $\omega (r) \rightarrow -1$ as $r\rightarrow 0$ \cite{Greene:1992fw} and that the electric gauge field function $a(r)$ must vanish at the origin.
Furthermore, $\omega '(r)$ must vanish at the origin.
Regular Taylor series expansions of the field variables in a neighbourhood of the origin are then given in terms of three arbitrary constants, $a_{1}$, $\omega _{2}$ and $S_{0}$ as:
\begin{align}
a(r) = & a_1 r + \frac{a_1}{5}\left(-2\omega_2+\frac{2a_{1}^2}{e^2S_{0}^2}+\frac{8\omega_{2}^2}{e^2}+\frac{\Lambda}{3}+\frac{\mu^2}{2}\right)r^{3} \nonumber \\
& + O(r^4), \nonumber \\
m(r) = &\left(\frac{a_{1}^2}{2e^2S_{0}^2}+\frac{2\omega_{2}^2}{e^2}\right)r^3 + O(r^4), \nonumber \\
S(r) = & S_0 + \left(\frac{a_{1}^2}{e^2S_0}+\frac{4S_0\omega_{2}^2}{e^2}\right)r^2 + O(r^3), \nonumber \\
\omega(r) = & -1 + \omega_2 r^{2} + O(r^3) .
\label{powerseriesatorigindyon1}
\end{align}
Setting the Proca field mass $\mu $ to zero, the expansions (\ref{powerseriesatorigindyon1}) reduce to those in pure EYM theory in AdS \cite{Bjoraker:1999yd,Nolan:2012ax}.

For black hole solutions, we assume that there is a regular nonextremal event horizon at $r=r_{h}$, where $R(r_{h})=0$ and $R'(r_{h})>0$.
These conditions fix the value of $m(r_{h})$ and it must be the case that $a(r_{h})=0$ to avoid a singularity in the field variables.
Regular Taylor series expansions of the field variables in a neighbourhood of the event horizon then take the following form:
\begin{align}
a(r) = & a'_{h}(r-r_h) + O(r-r_h)^2, \nonumber  \\
m(r) = & \left(\frac{r_h}{2} - \frac{\Lambda r_{h}^{3}}{6}\right) + m'_{h}(r-r_h) + O(r-r_h)^2, \nonumber \\
S(r) = & S_{h}+  S_{h}'(r-r_h) + O(r-r_h)^2,  \nonumber \\
\omega(r) = &\omega _{h} + \omega_{h}'(r-r_h) + O(r-r_h)^2,
\label{powerseriesathorizondyon}
\end{align}
where $a_{h}'$, $S_{h}$ and $\omega _{h}$ are arbitrary constants.
The first derivatives appearing in (\ref{powerseriesathorizondyon}) are given in terms of these three constants:
\begin{align}
m'_h = & \frac{r_{h}^{2}a'^2_{h}}{2e^2S_{h}^2}+\frac{\left(\omega_{h}^2-1\right)^2}{2e^2r_{h}^2}+\frac{\mu^2\left(1+\omega_h\right)^2}{e^2}, \nonumber \\
S'_h = & \frac{2\omega_h^2a'_h}{e^2r_hS_hR'^2_{h}}+\frac{2\omega'^2_{h}S_h}{e^2r_h}+\frac{\mu^2r_ha'^2_{h}}{e^2S_hR'^2_{h}}, \nonumber \\
\omega'_h = & \frac{\omega_h\left(\omega_{h}^2-1\right)}{r_{h}^2R'^2_{h}}+\frac{\mu^{2}\left(1+\omega_h\right)}{R'_h},
\end{align}
where $R_{h}'=R'(r_{h})$ depends on $r_{h}$ and $m'_{h}$.
Again, the expansions (\ref{powerseriesathorizondyon}) reduce to those for dyon solutions of EYM theory \cite{Bjoraker:1999yd,Nolan:2012ax} on setting $\mu =0$.

As $r\rightarrow \infty $, we require that the metric (\ref{metric}) approach that of pure AdS space-time.  This means that $m(r)\rightarrow M$ and $S(r) \rightarrow 1$ as $r\rightarrow \infty $.
For both the ENAP and EYMH equations in asymptotically flat space-time \cite{Greene:1992fw}, the field variables decay exponentially to their asymptotic values as $r\rightarrow \infty $. However, for solutions of EYMH in asymptotically AdS space-time \cite{VanderBij:2001ah}, the field variables have a complicated power-law behaviour as infinity is approached.
For ENAP theory in asymptotically AdS space-time, we find a similar power-law decay, with the field variables having the following behaviour as $r\rightarrow \infty $:
\begin{align}
a(r) = & \frac{\alpha_\infty}{r^\Delta} +...,
\nonumber \\
m(r) = & M+\frac{\left(\Delta^2\Lambda-3\mu^2\right)\left(2\Lambda\omega_\infty^2-3\alpha_\infty^2\right)}{6e^2\Delta\Lambda}\frac{1}{r^{2\Delta-1}} +...,
\nonumber \\
S(r) = & 1-\frac{\left(9\alpha_\infty^2\mu^2+2\Delta^2\Lambda^2\omega_\infty^2\right)}{e^2\Delta \Lambda^2}\frac{1}{r^{2\Delta+2}} +...,
\nonumber \\
\omega(r) = &-1 + \frac{\omega_\infty}{r^\Delta} +....
\label{infinityofdyon}
\end{align}
The expansions (\ref{infinityofdyon}) depend on arbitrary constants $M$, $\alpha _{\infty }$ and $\omega _{\infty }$.
The exponent $\Delta$ is given by
\begin{equation}
\Delta =\Delta_\pm =\frac{1}{2}\pm\frac{1}{2}\sqrt{1-\frac{12\mu^2}{\Lambda}}.
\label{BCinf3}
\end{equation}
We choose the upper root $\Delta =\Delta_{+}$, since $\Delta_{-}<0$.

\subsection{No nontrivial dyon solutions}
\label{sec:nodyons}

We now use an elegant method from Ershov and Galt'sov \cite{Ershov:1991nv} to show that there are no nontrivial dyonic solutions of ENAP theory in asymptotically AdS space-time.
This method assumes that the configurations have finite total energy, so that the boundary conditions (\ref{infinityofdyon}) hold.
In particular, we must have $a(r)\rightarrow 0$ as $r\rightarrow \infty $ otherwise $m'(r)$ (\ref{EFEm}) does not vanish as $r\rightarrow \infty $.
For pure EYM theory in asymptotically flat space-time, Ref.~\cite{Ershov:1991nv} assumes that $a(r)\rightarrow 0$ at infinity, but
this assumption can be relaxed in proving the absence of dyonic solutions \cite{Bizon:1992pi}.

We start by re-writing the field equation (\ref{NAPtdyon}) in the form
\begin{equation}
\left[\frac{r^2a'a}{S}\right]' = \frac{2a^2\omega^2}{RS} + \frac{a^2\mu^2r^2}{RS} + \frac{r^2a'^2}{S}.
\end{equation}
Then we integrate this equation throughout space,
\begin{align}
\frac{r^2a'a}{S}\bigg|_{r_0}^\infty &= \int\limits_{r_0}^{\infty}\frac{r^2}{S}\left(\frac{2a^2\omega^2}{r^2R} + \frac{a^2\mu^2}{R} + a'^2\right)dr, \label{nodyonproof}
\end{align}
where the lower limit of the integrals, $r_{0}$, is zero for regular solitons and $r_{h}$ for black holes.
For soliton solutions, all field variables are regular at $r_{0}=0$ and therefore the contribution to the boundary term on the left-hand-side of (\ref{nodyonproof}) at $r_{0}$ vanishes.
For black hole solutions, all field variables are regular at $r_{0}=r_{h}$ and, from (\ref{powerseriesathorizondyon}), the electric gauge field function $a(r)$ vanishes at the horizon, so again the contribution to the boundary term in (\ref{nodyonproof}) at $r_{0}$ vanishes.
For the contribution to the boundary term coming from $r\rightarrow \infty $, we have $S\rightarrow1$ as $r\rightarrow\infty$ and, using (\ref{infinityofdyon}),
\begin{equation}
r^2a'a \approx -\Delta_{+} \alpha_\infty^2r^{1-2\Delta_{+}} +....
\end{equation}
From (\ref{BCinf3}) we have
\begin{equation}
1-2\Delta_{+}=-\sqrt{1-\frac{12\mu^2}{\Lambda}}<0 ,
\end{equation}
and therefore the contribution to the boundary term in (\ref{nodyonproof}) coming from $r\rightarrow \infty $ also vanishes.

The integrand on  the right-hand-side of (\ref{nodyonproof}) is the sum of positive terms, and thus each term must vanish identically.
In particular, $a'=0$ and hence the electric gauge field function $a(r)$ is a constant.
As a consequence of this, $a(r)$ must be zero everywhere if $\omega \neq 0$ and $\mu\neq 0$.
Therefore there are no nontrivial dyon solutions of ENAP theory in asymptotically AdS space-time.

Our proof extends readily to the asymptotically flat case considered in \cite{Greene:1992fw}, where the boundary conditions as $r\rightarrow \infty $ again ensure the vanishing of the boundary term on the left-hand-side of (\ref{nodyonproof}).
A similar no-dyon theorem has been proven for the EYMH model \cite{VanderBij:2001ah}.
However, our result breaks down for pure EYM theory in asymptotically AdS space-time with $\mu =0$.
In this case the exponent $\Delta =1$ and $a(r)$ does not have to vanish as $r\rightarrow \infty $ for finite energy configurations. Hence the contribution to the boundary term on the left-hand-side of (\ref{nodyonproof}) from $r\rightarrow \infty $ no longer vanishes.
This leaves open the existence of dyonic soliton and black hole solutions of EYM theory in AdS, as expected \cite{Bjoraker:1999yd,Nolan:2012ax}.

\subsection{Purely magnetic configurations}
\label{sec:nodes}

Since we have shown that there are no nontrivial dyon solutions of ENAP in AdS, we now restrict our attention to purely magnetic configurations by setting the electric part of the gauge potential to vanish identically, $a(r)\equiv 0$.
The gauge potential (\ref{dyon:gaugeform}) then takes the form
\begin{equation}
eA = \left[1+\omega(r)\right]\left[-\hat{\tau}_{\phi}d\theta+\hat{\tau}_{\theta}\sin\theta \, d\phi\right] .
\label{reducedgaugeconnection}
\end{equation}
The field equations (\ref{dyonfieldequations}) reduce to
\begin{subequations}
\label{fieldequations1}
\begin{align}
m' = & \frac{R\omega'^2}{e^2}+\frac{(1-\omega^{2})^{2}}{2e^{2}r^{2}} + \frac{\mu^{2}}{e^{2}}(1+\omega)^2, \label{EFE00} \\
\delta ' = & -\frac{2\omega'^{2}}{e^{2}r}, \label{EFE11}\\
0 = & r^{2}R\omega'' \nonumber \\
& + \left[2m-\frac{2r^{3}\Lambda}{3}-\frac{(1-\omega^{2})^{2}}{e^{2}r}-\frac{2\mu^{2}r}{e^{2}}(1+\omega)^{2}\right]\omega' \nonumber \\
& + \left(1-\omega^{2}\right)\omega-\mu^{2}r^{2}\left( 1+ \omega \right) , \label{NAP}
\end{align}
\end{subequations}
where we have introduced a quantity $\delta $ defined by $S\equiv  \exp \left( - \delta \right)$.
Like pure EYM theory, the equation (\ref{EFE11}) for $\delta '$ decouples from the other two equations.
For this reason, in our discussion of numerical solutions of the field equations (\ref{fieldequations1}) in the next section we focus on the metric function $m(r)$ and the gauge field function $\omega (r)$.
In pure EYM theory, the field equations possess a discrete symmetry under $\omega\rightarrow-\omega$, however this symmetry is broken in the ENAP equations (\ref{fieldequations1}) due to the presence of the Proca field mass $\mu $.

The expansions of the field variables near the origin (\ref{powerseriesatorigindyon1}), black hole event horizon (\ref{powerseriesathorizondyon}) and infinity (\ref{infinityofdyon}) also simplify upon setting $a\equiv 0$. Near the origin, the expansions take the form (where we have included some higher-order terms which are useful for our numerical integration of the field equations in Sec.~\ref{sec:solutions})
\begin{align}
m(r) = & \frac{2\omega_{2}^{2}}{e^2} r^3 + \frac{1}{5e^2}\left[-8\omega_{2}^{3} + 3\mu^{2}\omega_{2}^{2} + \frac{8\Lambda\omega_{2}^{2}}{3}\right] r^5  \nonumber \\
&+O(r^6), \nonumber \\
\delta(r) = & \delta _{0} -\frac{4\omega_{2}^{2}}{e^2} r^2 - \frac{4}{5e^2}\left[2\Lambda \omega_{2}^{2} + \mu^2 \omega_{2}^{2} - 3\omega_{2}^{3} + \frac{8\omega_{2}^{4}}{e^{2}}\right] r^4 \nonumber \\
& + O(r^5), \nonumber \\
\omega(r) = & -1 + \omega_{2} r^2 + \frac{1}{10e^2}\left[2e^{2}\Lambda \omega_{2} + e^{2}\mu^{2}\omega_{2} \right. \nonumber \\
&\quad\left. -3e^{2} \omega_{2}^{2}+8 \omega_{2}^{3}\right] r^4 + O(r^5).
\label{BCregular}
\end{align}
In a neighbourhood of the horizon, we write the expansions (\ref{powerseriesathorizondyon}) in terms of $\delta $ and obtain
\begin{align}
m(r) = & \left( \frac{r_h}{2} - \frac{\Lambda r_{h}^{3}}{6} \right) + m'_{h}(r-r_h) + O(r-r_h)^2, \nonumber \\
\delta(r) = &\delta_{h} +  \delta'_{h}(r-r_h) + O(r-r_h)^2, \nonumber \\
\omega(r) = & \omega_{h} + \omega'_{h}(r-r_h) + O(r-r_h)^2,
\label{nearhorizonseries}
\end{align}
with
\begin{align}
m'_{h} = & \frac{(1-\omega_{h}^{2})^{2}}{2e^{2}r_h^{2}}+\frac{\mu^{2}(1+\omega_{h})^{2}}{e^{2}},  \nonumber \\
\delta'_{h} = & -\frac{2\omega'^{2}_{h}}{e^{2}r_{h}}, \nonumber \\
\omega'_{h} = & \frac{\mu^{2}r_{h}^{2}(1+\omega_{h})-(1-\omega_{h}^{2})\omega_{h}}{(r_{h}-\Lambda r_h^{3})-\frac{(1-\omega_{h}^{2})^{2}}{e^{2}r_h}-\frac{2\mu^{2}r_{h}(1+\omega_{h})^{2}}{e^{2}}}.
\label{BCathorizon}
\end{align}
As $r \rightarrow \infty $, the expansions (\ref{infinityofdyon}) again simplify and using the new variable $\delta $ take the form
\begin{align}
m(r) = & M + \frac{\left(\Delta^{2}\Lambda-3\mu^{2}\right)}{3e^{2}\Delta}\frac{\omega _{\infty }^2}{r^{2\Delta-1}}  + ..., \nonumber\\
\delta(r) = & \frac{2\Delta}{e^2}\frac{\omega _{\infty }^2}{r^{2\Delta+2}} + ..., \nonumber \\
\omega(r) = & -1 + \frac{\omega _{\infty }}{r^\Delta} +....
\label{serieexpansionatinfinity}
\end{align}

If we set $\omega (r) \equiv -1$, the functions $m(r)\equiv M$ and $\delta (r) \equiv 0$ are both constants and the Schwarzschild-AdS black hole is a trivial solution of the field equations (\ref{fieldequations1}).  However, unlike EYM theory, the magnetically-charged Reissner-Nordstr\"om black hole is not a solution of the field equations as we cannot set $\omega (r)\equiv 0 $ in the NAP equation (\ref{NAP}).

As discussed in Sec.~\ref{sec:intro}, soliton and black hole solutions of ENAP and EYMH in asymptotically flat space-time are such that the magnetic gauge field function $\omega (r)$ has at least one zero.
In contrast, there exist pure EYM solutions in asymptotically AdS space-time for which the gauge field function $\omega (r)$ is nodeless \cite{Winstanley:1998sn,Bjoraker:1999yd}.
The latter are of particular interest since some of them are stable under linear perturbations of the metric and gauge field functions \cite{Winstanley:1998sn,Bjoraker:1999yd,Sarbach:2001mc}.
Before studying numerical solutions of the  ENAP-AdS field equations (\ref{fieldequations1}) in the next section, we now show that $\omega (r)$ must have at least one zero.

First consider the case of soliton solutions.
From the expansions near the origin (\ref{BCregular}), we see that $\omega (r)\rightarrow -1$ as $r\rightarrow 0$ and that the sign of $\omega '(r)$ sufficiently close to  $r=0$ is the same as the sign of the constant $\omega _{2}$.
Suppose that $\omega _{2}<0$ so that $\omega (r)<-1$ in a neighbourhood of the origin.
From the boundary conditions (\ref{serieexpansionatinfinity}), the gauge field function $\omega (r) \rightarrow -1$ as $r\rightarrow \infty $, and therefore there must be an $r=r_{1}$ at which $\omega (r)$ has a minimum.
Since $\omega '(r_{1})=0$, the NAP equation (\ref{NAP}) gives
\begin{equation}
r_{1}^{2}R(r_{1})\omega ''(r_{1})= \left[ \omega (r_{1})^{2} - 1 \right] \omega (r_{1}) + \left[ 1 + \omega (r_{1}) \right] \mu ^{2}r_{1}^{2}.
\label{maxmin}
\end{equation}
For $\omega (r)$ to have a minimum at $r=r_{1}$, we require $\omega ''(r_{1})>0$, but both terms on the right-hand-side of (\ref{maxmin}) are negative
for $\omega (r_{1})<-1$.
Since the metric function $R(r)$ is positive everywhere, we therefore have a contradiction and it must be the case that $\omega _{2}>0$.

With $\omega _{2}>0$, the gauge field function $\omega (r)>-1$ in a neighbourhood of the origin and  therefore must have a maximum at some $r=r_{1}$
(since $\omega \rightarrow -1$ as $r\rightarrow \infty $).
Suppose that at $r_{1}$ we have $-1<\omega (r_{1})<0$.
Then, both terms on the right-hand-side of (\ref{maxmin}) are positive, and therefore $\omega ''(r_{1})>0$, giving a contradiction with our assumption that
$\omega (r)$ has a maximum at $r=r_{1}$.
Therefore it must be the case that $\omega (r_{1})>0$.
Therefore $\omega (r)$ has at least one zero.
In fact, since $\omega (r)\rightarrow -1$ as both $r\rightarrow 0$ and $r\rightarrow \infty $, we can conclude that $\omega (r)$ has an even number of zeros when we consider soliton solutions.

The argument for black hole solutions proceeds along similar lines.
We start by assuming that $\omega _{h}=\omega (r_{h})<-1$.
The denominator in the expression for $\omega _{h}'$ (\ref{BCathorizon}) is equal to $r_{h}^{2}R'(r_{h})>0$ since we assume that the event horizon is regular and nonextremal.
The numerator in $\omega _{h}'$ (\ref{BCathorizon}) is negative when $\omega _{h}<-1$, so we have $\omega _{h}'<0$.
Therefore $\omega (r)$ must have a minimum at some $r=r_{1}$ where $\omega (r_{1})<-1$ and $\omega '(r_{1})=0$.
Then (\ref{maxmin}) gives $\omega ''(r_{1})<0$ and hence we have a contradiction.
Therefore it must be the case that $\omega _{h}>-1$.

Next suppose that $-1<\omega _{h}<0$. In this case $\omega _{h}'>0$ (\ref{BCathorizon}) and $\omega (r)$ must have a maximum at some $r=r_{1}$.
Then, from (\ref{maxmin}), $\omega ''(r_{1})>0$ if $-1<\omega (r_{1})<0$, yielding a contradiction.
So we conclude that $\omega (r_{1})>0$ and the gauge field function $\omega (r)$ has an even number of zeros.

The remaining possibility is $\omega _{h}>0$. In this case the gauge field function must have an odd number of zeros since $\omega (r)\rightarrow -1$ as $r\rightarrow \infty $.
In summary, we have shown that for both soliton and black hole solutions, the gauge field function $\omega (r)$ must have at least one zero.

\section{Solitons and hairy black holes}
\label{sec:solutions}

We now present numerical solutions of the ENAP-AdS equations (\ref{fieldequations1}) representing solitons and hairy black holes.
For the solutions presented here, the magnetic gauge field function $\omega (r)$ will have either one or two zeros, but we anticipate that solutions in which $\omega (r)$ has more zeros also exist.
In this section we set the gauge coupling constant $e=1$.

\subsection{Solitons}
\label{sec:solitons}

To find numerical soliton solutions, the initial point for integrating the field equations (\ref{fieldequations1}) is taken to be close to the origin
(at typically $r\sim 10^{-3}$).
We use the expansions (\ref{BCregular}) as initial conditions for the field variables.
For fixed Proca field mass $\mu $ and negative cosmological constant $\Lambda $, we use a standard shooting method, scanning for values of $\omega _{2}$ such that $\omega (r)\rightarrow -1$ as $r\rightarrow \infty $.
We find solutions satisfying the boundary conditions at infinity at discrete values of $\omega _{2}$ for fixed $\mu $ and $\Lambda $.

\begin{figure*}
\includegraphics[width=8.5cm]{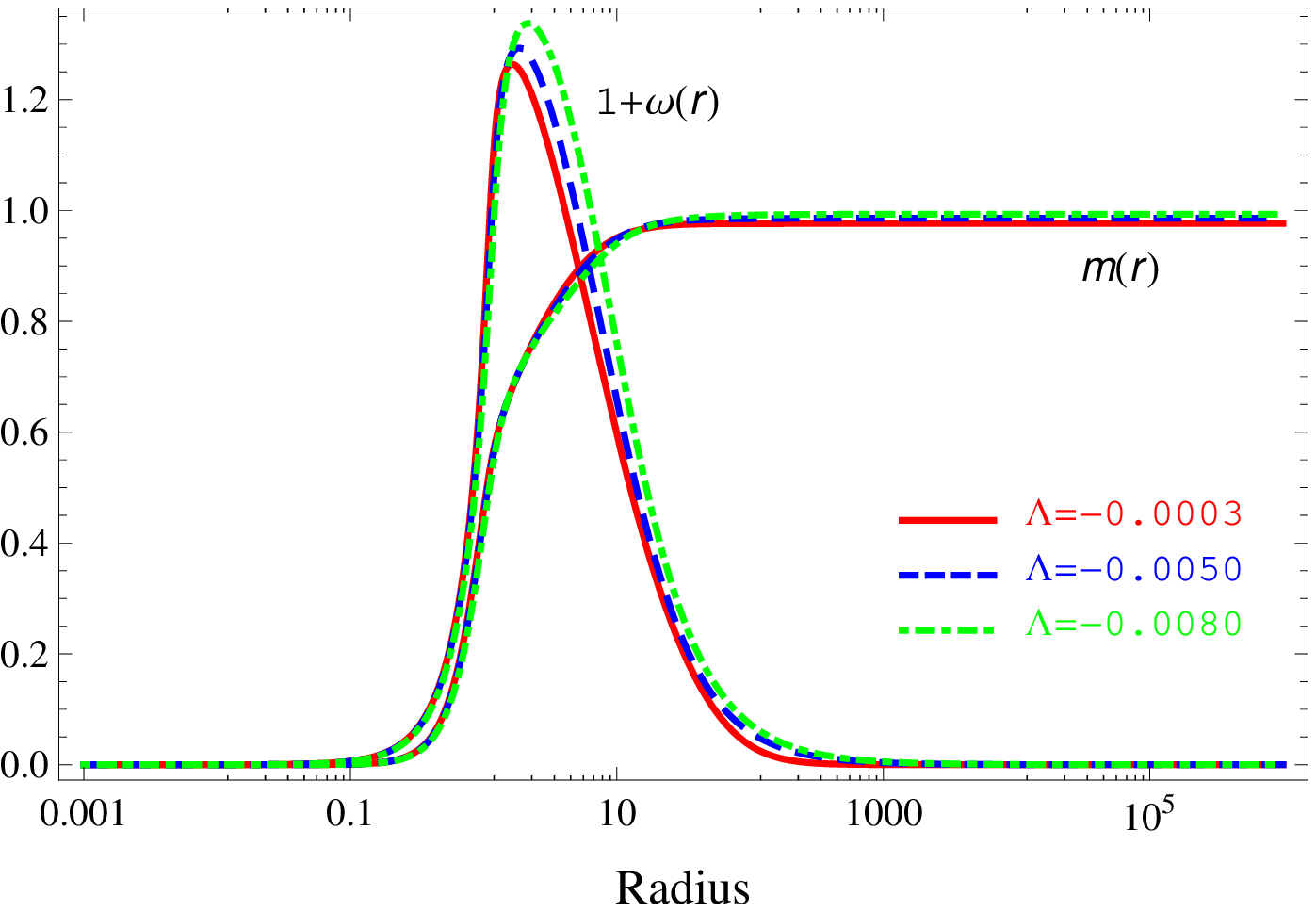}
\includegraphics[width=8.5cm]{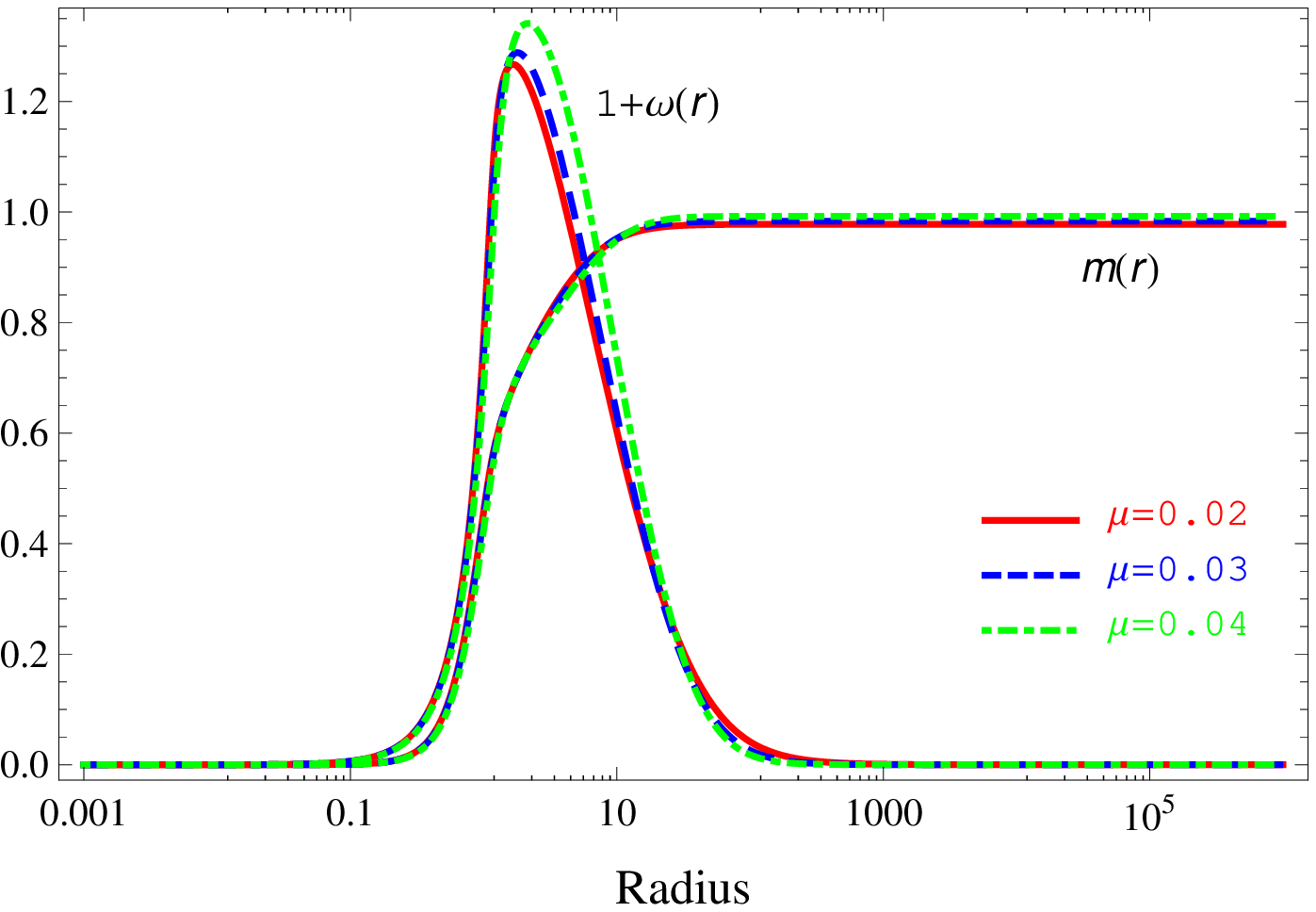}
\includegraphics[width=8.5cm]{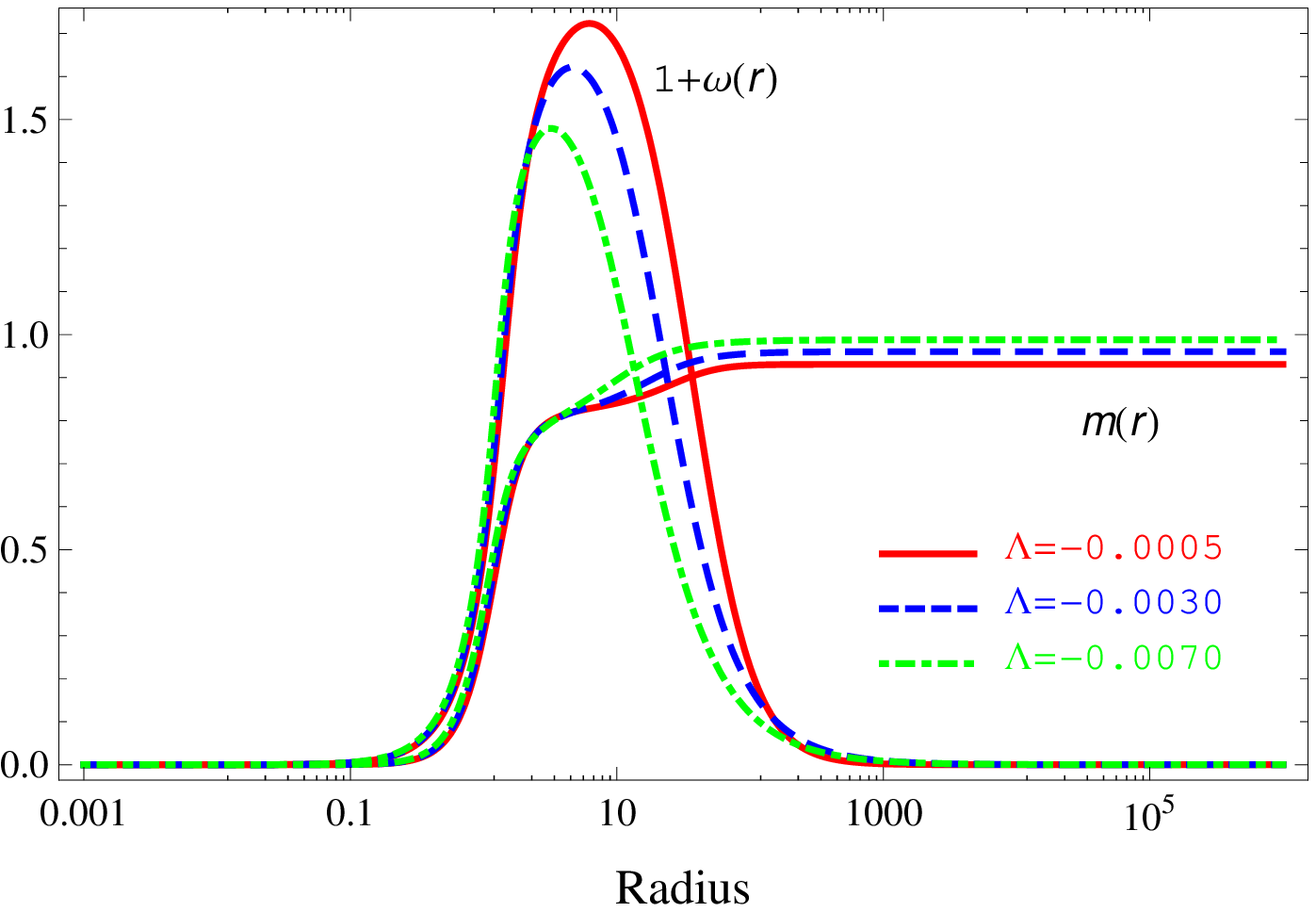}
\includegraphics[width=8.5cm]{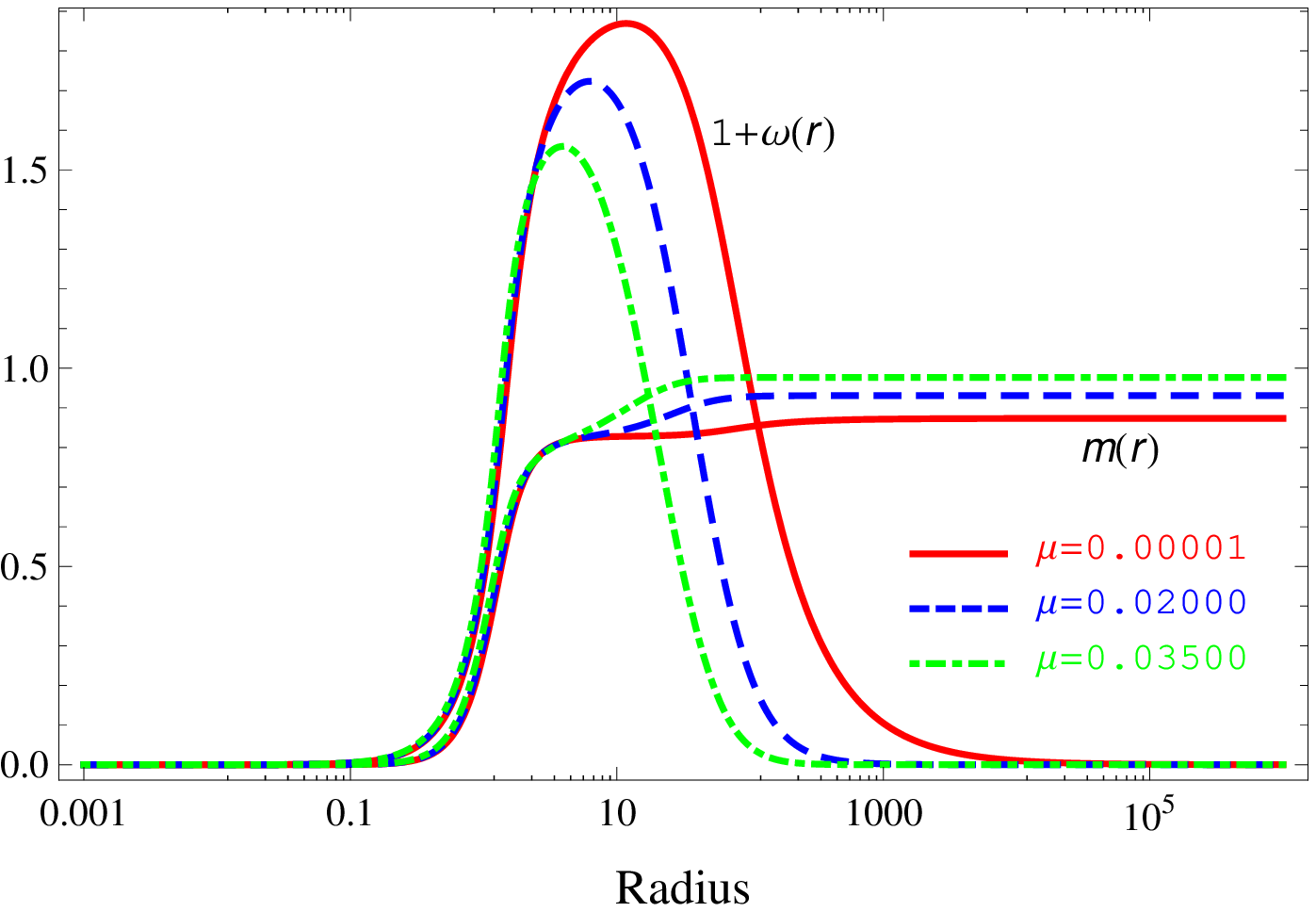}
\caption{Regular solutions of ENAP-AdS theory for which $\omega (r)$ has two zeros. Top row: $n=2$ solutions with (left panel) fixed $\mu =0.02$ and varying $\Lambda $, (right panel) fixed $\Lambda =-0.001$ and varying $\mu $.
Bottom row: quasi-$n=1$ solutions with (left panel) fixed $\mu = 0.02$ and varying $\Lambda $, (right panel) fixed $\Lambda = -0.0005$ and varying $\mu $.}
\label{fig:solitons}
\end{figure*}

As shown in Sec.~\ref{sec:nodes}, for soliton solutions the gauge field function $\omega (r)$ must have an even (nonzero) number of zeros.
Some typical soliton solutions for which $\omega (r)$ has two zeros are shown in Fig.~\ref{fig:solitons}.
We plot the gauge field function $1+\omega (r)$ and metric function $m(r)$ .
We find two branches of soliton solutions, which, following \cite{Greene:1992fw,VanderBij:2001ah}, we term the ``$n=2$'' and ``quasi-$n=1$'' branches (the reasons for this terminology will be explained in more detail below).
Solutions on the $n=2$ branch are shown in the top row in Fig.~\ref{fig:solitons}, while the bottom row shows solutions on the quasi-$n=1$ branch.
In the left-hand plots in Fig.~\ref{fig:solitons} the Proca field mass $\mu $ is fixed and the cosmological constant $\Lambda $ varies; in the right-hand plots the cosmological constant $\Lambda $ is fixed and $\mu $ varies.

Consider first the solutions shown in the top row of plots in Fig.~\ref{fig:solitons}, namely the $n=2$ branch of solutions.
With fixed $\mu $ (left-hand plot), increasing $\left| \Lambda \right| $ increases the maximum value of $\omega $ and the peak moves to slightly larger $r$.  With fixed $\Lambda $ (right-hand plot), increasing $\mu $ also increases the peak value of $\omega $ and the location of the maximum is at larger $r$.
The metric function $m(r)$ is monotonically increasing from the origin to infinity, with slightly larger values as $r\rightarrow \infty $ for larger $\mu $ with fixed $\Lambda $ or larger $\left| \Lambda \right| $ with fixed $\mu $.
The solutions shown in the top row of Fig.~\ref{fig:solitons} do not vary much as either $\mu $ or $\Lambda $ vary.

However, we find different behaviour on the second branch of solutions for which $\omega (r)$ has two nodes, the quasi-$n=1$ branch, shown in the bottom row of plots in Fig.~\ref{fig:solitons}.
For solutions on the quasi-$n=1$ branch, both $m(r)$ and $\omega (r)$ vary more as the parameters $\mu $ and $\Lambda $ vary than for solutions on the $n=2$ branch.
For fixed $\mu $, increasing $\left| \Lambda \right| $ decreases the maximum value of $\omega $ and the peak moves to smaller values of $r$.
Similarly, for fixed $\Lambda $, as $\mu $ increases the maximum value of $\omega $ decreases and the peak in $\omega $ moves closer to the origin.
We find that solutions on the $n=2$ branch have larger values of the shooting parameter $\omega _{2}$ than those on the quasi-$n=1$ branch.
The solitons on the $n=2$ branch also have larger masses than those on the quasi-$n=1$ branch.

This branch structure also occurs for solutions of ENAP and EYMH in asymptotically flat space-time \cite{Greene:1992fw} and EYMH in asymptotically AdS space-time \cite{VanderBij:2001ah}.
For the $n=2$ branch of asymptotically flat ENAP and EYMH solutions, the location of the outermost zero of $\omega (r)$ does not vary much as $\mu \rightarrow 0$, but for the quasi-$n=1$ branch of solutions, the outermost zero of $\omega (r)$ moves towards infinity as $\mu \rightarrow 0$ \cite{Greene:1992fw}.
For both ENAP and EYMH solitons in asymptotically flat space-time, the gauge field function $\omega (r)$ must have an even number of zeros for the boundary conditions at the origin and infinity to be satisfied, but as the outermost zero of $\omega (r)$ moves far from the origin, the function $\omega (r)$ looks very much like that for the first EYM soliton \cite{Bartnik:1988am} (for which $\omega (r)$ has a single zero) for a large interval of values of $r$ before $\omega (r)$ has its second zero.

In Fig.~\ref{fig:solitons}, we find similar behaviour on the quasi-$n=1$ branch of ENAP-AdS solutions; as either $\mu \rightarrow 0$ for fixed $\Lambda $ or $\Lambda \rightarrow 0$ for fixed $\mu $,
the location of the outermost zero of $\omega (r)$ moves to larger values of $r$.
We might have expected that our quasi-$n=1$ solutions of ENAP in AdS behave like those in asymptotically flat space-time, with the second zero of $\omega (r)$ located far from the origin when $\mu $ is sufficiently small.
However, as can be seen in Fig.~\ref{fig:solitons}, this is not the case.
Although the second zero of $\omega (r)$ does move to larger $r$ with decreasing $\mu $ and fixed $\Lambda $, it does not move as far out as in the corresponding asymptotically flat solutions \cite{Greene:1992fw}.
Nonetheless we retain the quasi-$n=1$ terminology for this branch of solutions since they have some features in common with both the asymptotically flat quasi-$n=1$ solutions \cite{Greene:1992fw} and the asymptotically AdS EYMH quasi-$n=1$ solutions \cite{VanderBij:2001ah}.

In \cite{Greene:1992fw}, the branch structure of the asymptotically-flat ENAP solutions is understood as arising from the existence of two length scales in the theory, one set by the gravitational coupling of the NAP field and the other by the NAP field mass. Here we have a third length scale, the AdS radius of curvature $\ell = {\sqrt {-3/\Lambda }}$. The scale of the $n=2$ branch of ENAP-AdS solutions is set by the gravitational coupling and hence these solutions do not change much as either the Proca field mass $\mu $ or cosmological constant $\Lambda $ varies.
In contrast, the scale of the quasi-$n=1$ branch of ENAP-AdS solitons is set by the other two length scales and hence this branch of solutions shows more variation as either $\mu $ or $\Lambda $ vary.

\begin{figure*}
\includegraphics[width=8.5cm]{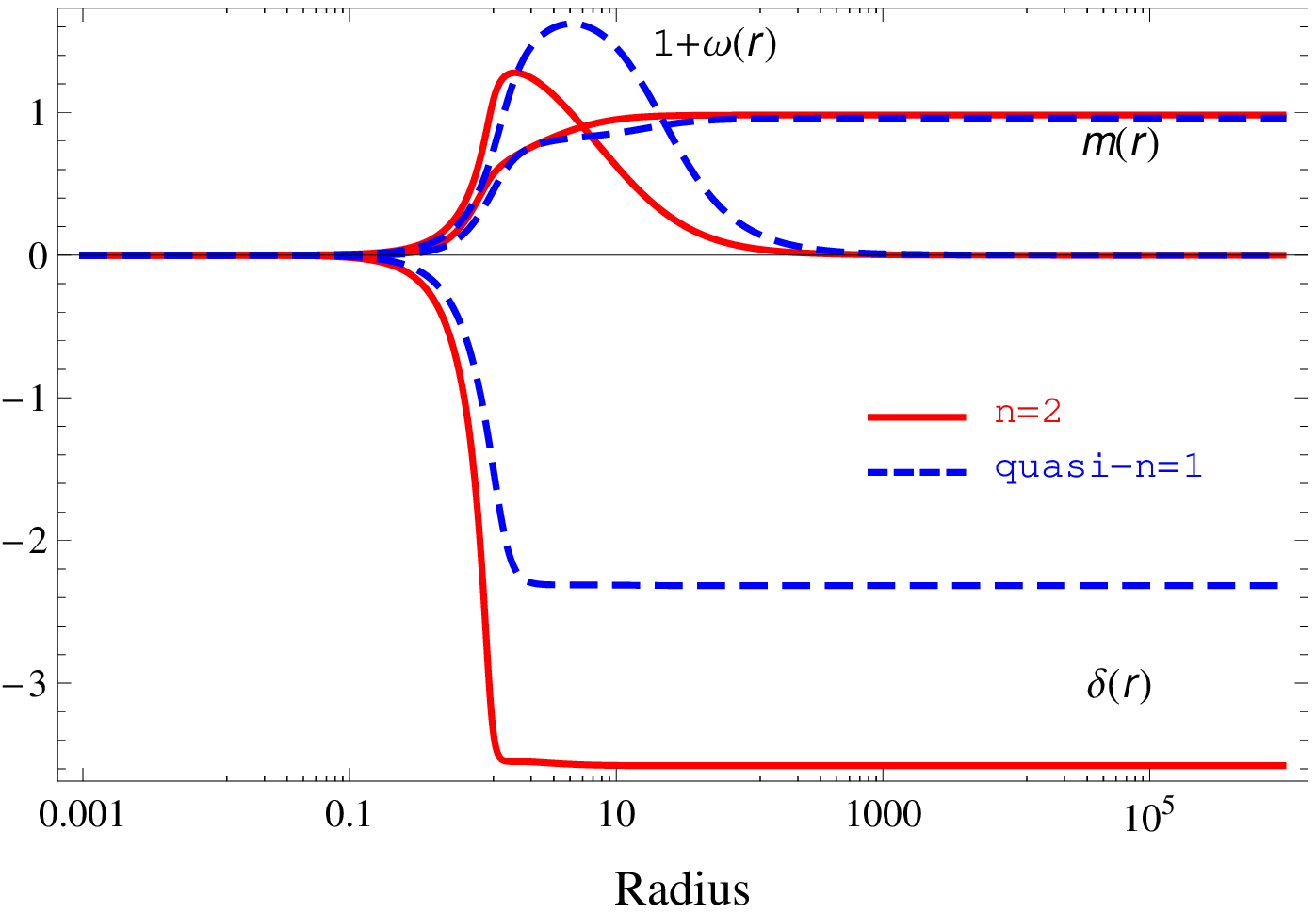}
\includegraphics[width=8.5cm]{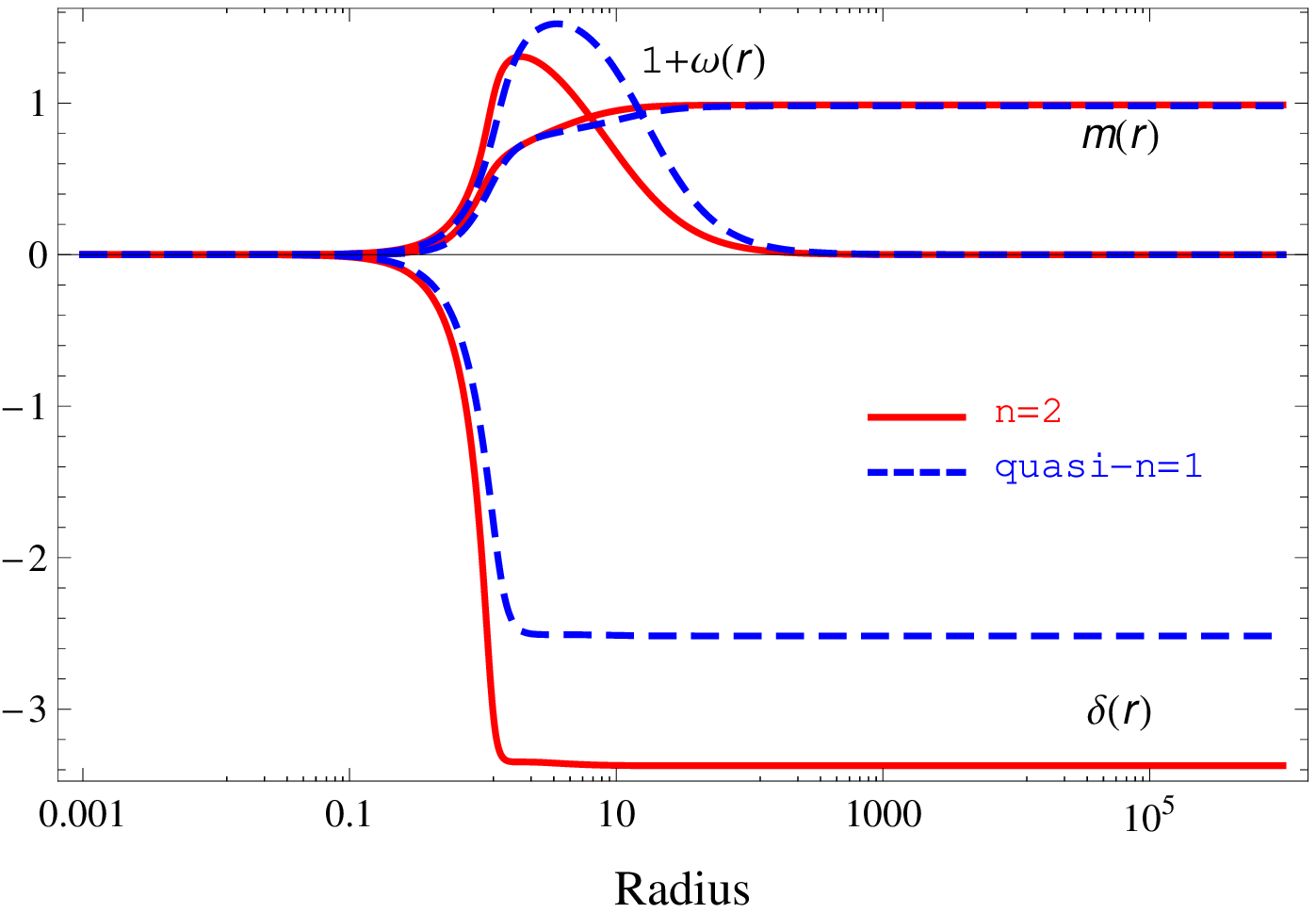}
\caption{Comparison of $n=2$ and quasi-$n=1$ soliton solutions with $\Lambda =-0.003$ and $\mu = 0.02$ (left), $\mu = 0.03$ (right). The functions for the $n=2$ solitons are shown in red (solid curves) and those for the quasi-$n=1$ solitons are shown in blue (dashed curves).}
\label{fig:n2qn1}
\end{figure*}

In Fig.~\ref{fig:n2qn1} we compare soliton solutions on the $n=2$ and quasi-$n=1$ branches with the same values of $\Lambda $ and $\mu $. We plot the gauge field function $1+\omega (r)$ and metric functions $m(r)$ and $\delta (r)$.
To find $\delta (r)$, we have set the parameter $\delta _{0}$ in (\ref{BCregular}) to vanish. This means that $\delta (r)$ does not tend to zero as $r\rightarrow \infty $, as required by the boundary conditions (\ref{serieexpansionatinfinity}).
However, the ENAP equations (\ref{fieldequations1}) depend only on $\delta '$ and not on $\delta $. Therefore we can add a constant to $\delta (r)$ so that $\delta (r) \rightarrow 0$ as $r\rightarrow \infty $.
However, in Fig.~\ref{fig:n2qn1} we have not done this, so that the difference in behaviour of $\delta (r)$ for the two solutions is clearer.
For all our solutions, we find that $\delta (r)$ decreases as $r$ increases, as expected from (\ref{EFE11}).

Fig.~\ref{fig:n2qn1} reveals that the $n=2$ solutions have smaller maximum values of $\omega (r)$ than the corresponding quasi-$n=1$ solutions; the peak in the value of $\omega (r)$ is at a lower value of $r$ for the $n=2$ solutions; the functions $m(r)$ are very similar for the two solutions (but the quasi-$n=1$ solutions have smaller mass than the corresponding $n=2$ solutions); and the difference in the values of $\delta (r)$ as $r\rightarrow \infty $ and at the origin is much larger for the $n=2$ solutions.

\begin{figure}
\includegraphics[width=8.5cm]{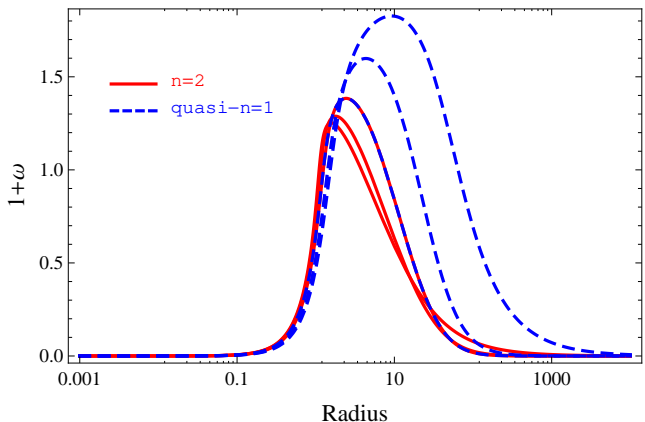}
\caption{Gauge field function $1+\omega (r)$ for a selection of $n=2$ and quasi-$n=1$ solitons with fixed $\Lambda =-0.001$ and varying Proca field mass $\mu = 0.00005$, $0.03$ and $0.042$. Quasi-$n=1$ curves are blue (dashed), while those for $n=2$ are red (solid).
The two branches of solutions merge when $\mu = \mu _{{\text {max}}}=0.042$.}
\label{fig:n2qn1merge}
\end{figure}

Comparing the two plots in Fig.~\ref{fig:n2qn1}, we see that the differences between the solutions on the two branches become less significant as $\mu $ increases for fixed $\Lambda $.
This trend continues as $\mu $ increases further, until the two branches of solutions merge at $\mu = \mu _{\text {max}}$, see Fig.~\ref{fig:n2qn1merge}.
The behaviour depicted in Fig.~\ref{fig:n2qn1merge} is very similar to that seen in \cite{Greene:1992fw} for asymptotically flat ENAP solutions, where the $n=2$ and quasi-$n=1$ branches merge at the maximum value of the Proca field mass $\mu $.
However, the value of $\mu _{\text {max}}$ for fixed $\Lambda <0$ is less than that for $\Lambda =0$.
In the EYMH model in either asymptotically flat \cite{Greene:1992fw} or asymptotically AdS \cite{VanderBij:2001ah} space-time, for fixed $\Lambda $ there is a maximum value of the Higgs vacuum expectation value $\alpha $ (which is essentially the dynamically generated gauge field mass) where the branches of solutions merge.
In general, for fixed $\mu $ we also find a maximum value of $\left| \Lambda \right| $ for which nontrivial ENAP-AdS soliton solutions with $\omega (r)$ having two zeros exist.

\begin{figure}
\includegraphics[width=8.5cm]{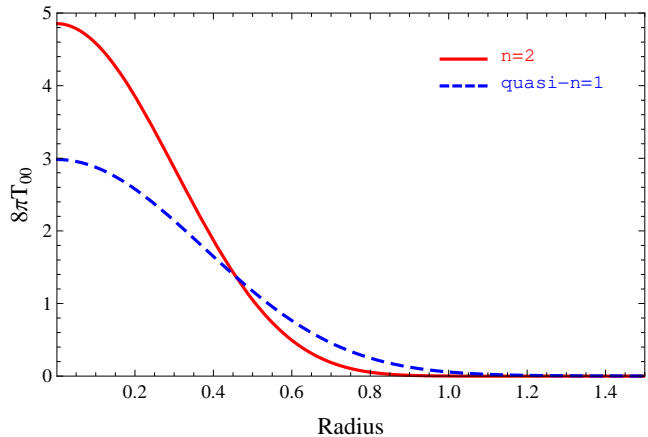}
\caption{$8\pi T_{00}$ (\ref{ENAPstresstensor}) for two regular solitons with $\mu =0.02$ and $\Lambda = -0.003$.}
\label{fig:n2qn1energy}
\end{figure}

We close our comparison of the $n=2$ and quasi-$n=1$ ENAP-AdS solitons by plotting, in Fig.~\ref{fig:n2qn1energy}, the quantity $8\pi T_{00}$ (where the stress-energy tensor is given by (\ref{ENAPstresstensor})) for an $n=2$ and a quasi-$n=1$ soliton with $\mu = 0.02$ and $\Lambda = -0.003$.
Both solutions have a compact region  near the origin where the  energy density is high, outside which the energy density rapidly decreases to zero as $r\rightarrow \infty $.
This is in accordance with our interpretation of these solutions as regular solitons.
The central energy density of the $n=2$ solution is much greater than that of the quasi-$n=1$ solution and the energy density becomes negligible at smaller $r$ for the $n=2$ soliton.
Therefore the $n=2$ solution represents a soliton which is denser and more compact than the corresponding quasi-$n=1$ solution.

\begin{figure*}
\includegraphics[width=8.5cm]{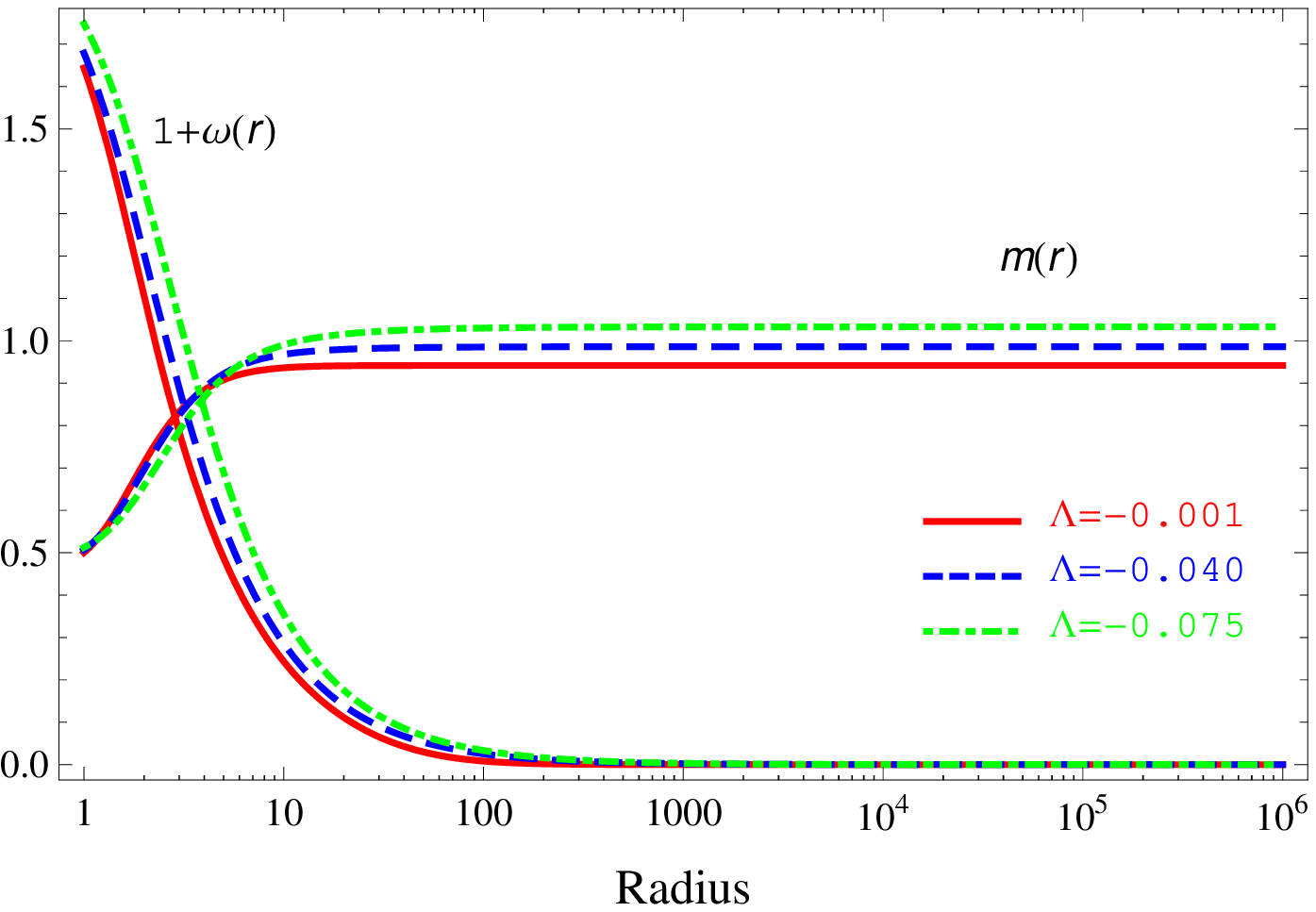}
\includegraphics[width=8.5cm]{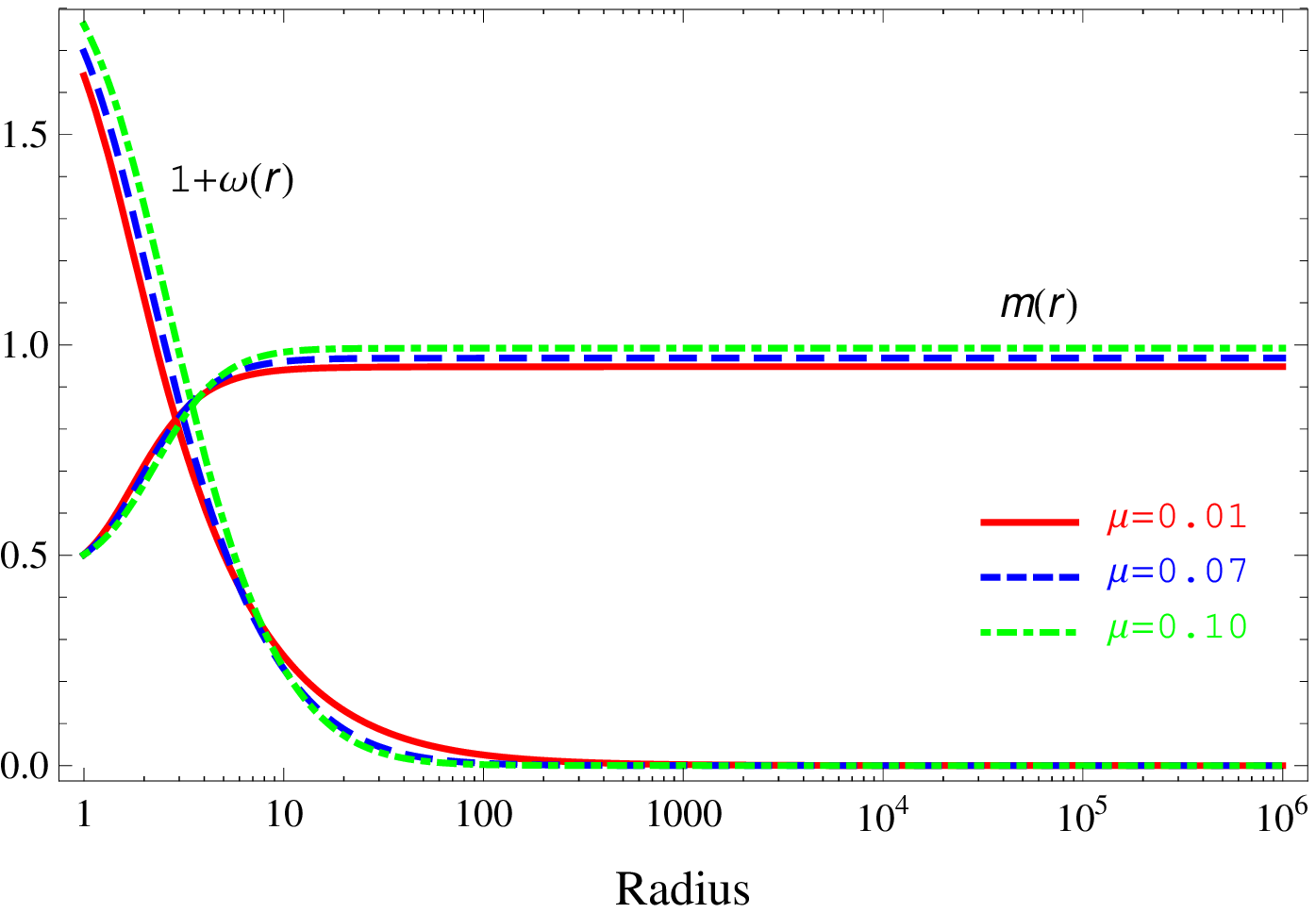}
\includegraphics[width=8.5cm]{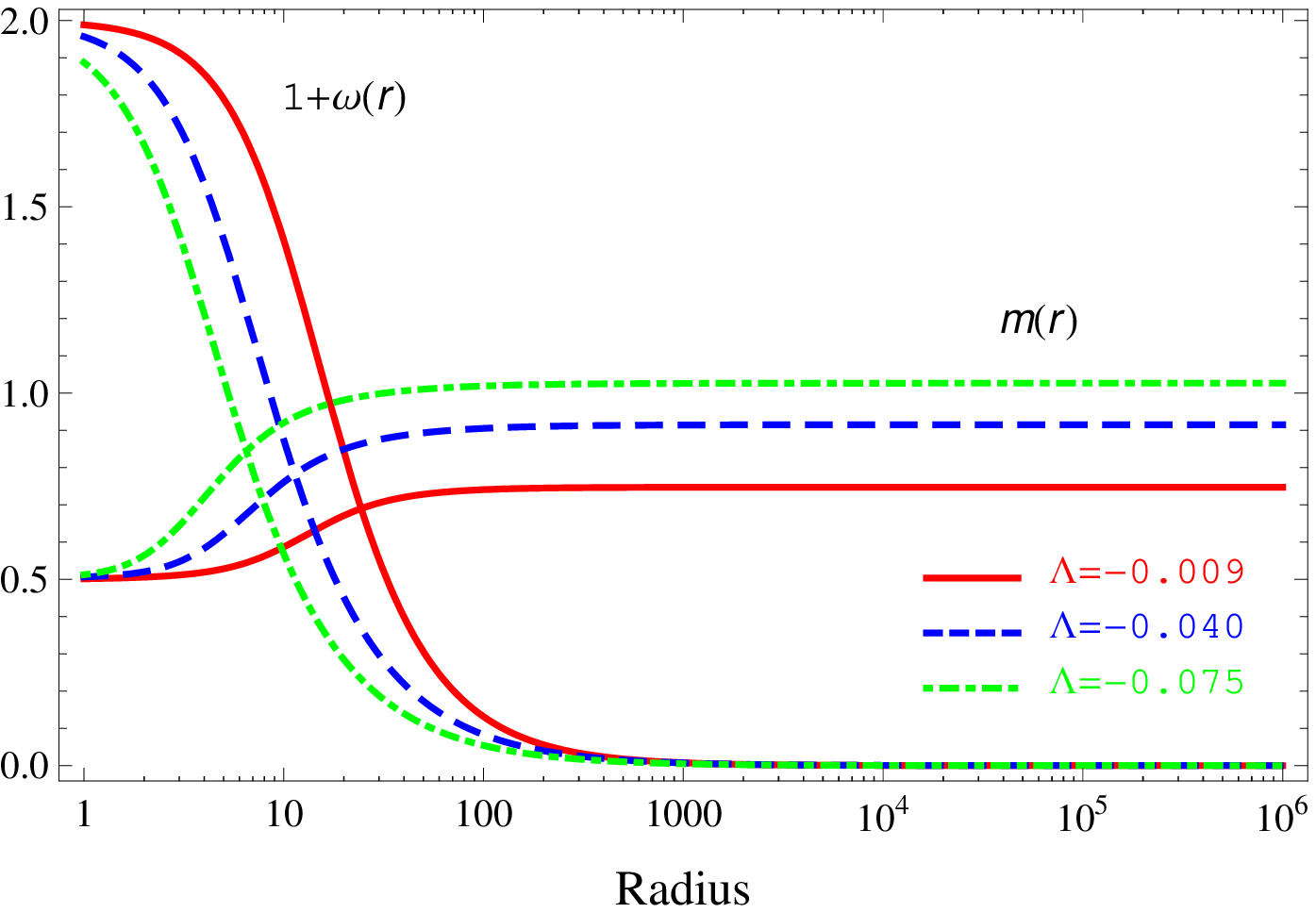}
\includegraphics[width=8.5cm]{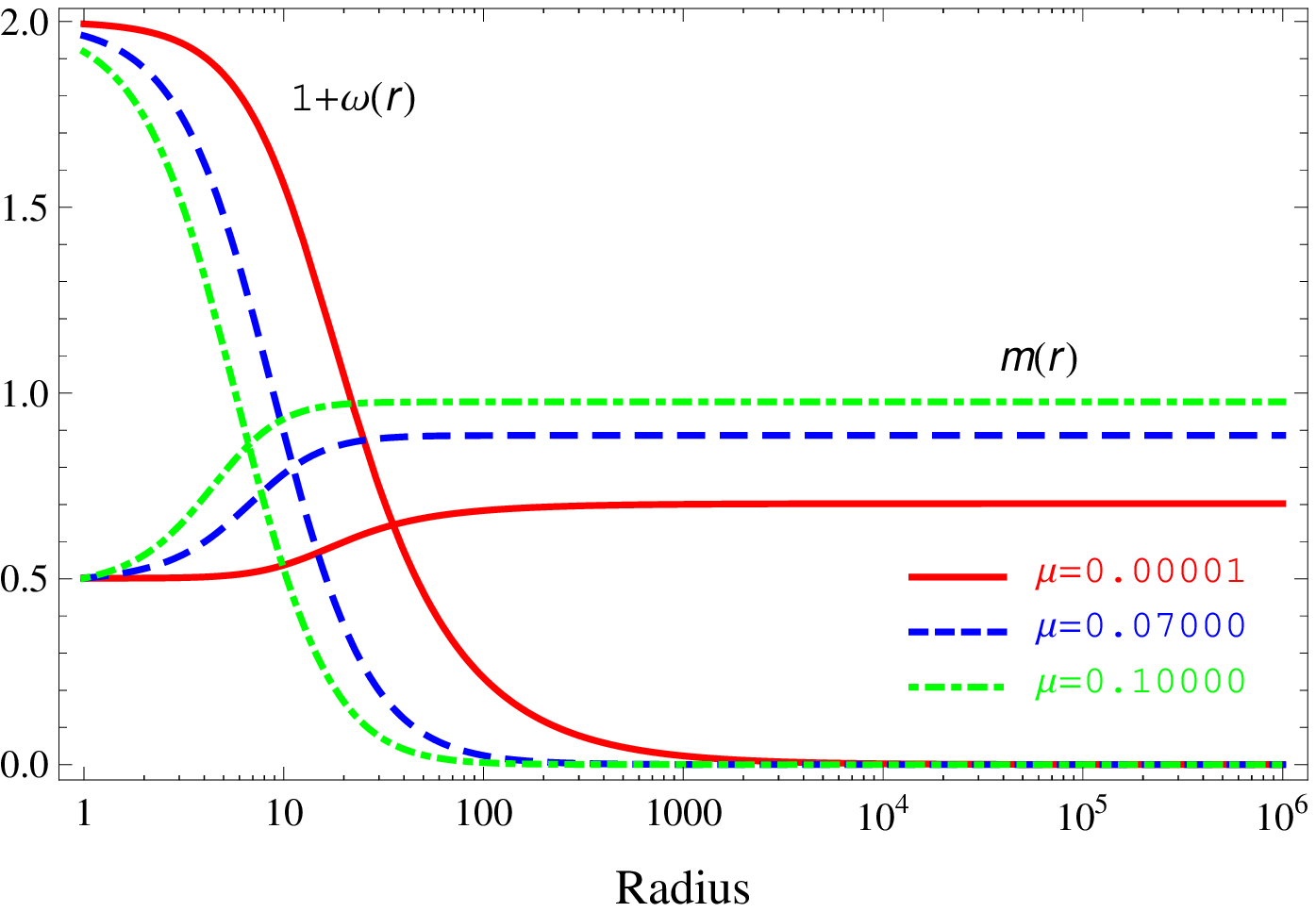}
\caption{Black hole solutions of ENAP-AdS theory for which $\omega (r)$ has one zero. Top row: $n=1$ solutions with (left panel) fixed $\mu =0.03$ and varying $\Lambda $, (right panel) fixed $\Lambda =-0.01$ and varying $\mu $.
Bottom row: quasi-$n=0$ solutions with (left panel) fixed $\mu = 0.03$ and varying $\Lambda $, (right panel) fixed $\Lambda = -0.01$ and varying $\mu $.
The event horizon radius is fixed to be $r_{h}=1$.}
\label{fig:BHn1qn0}
\end{figure*}

In this subsection, we have considered only soliton solutions for which the gauge field function $\omega (r)$ has two zeros. We expect that there are also solitons for which $\omega (r)$ has more than two zeros, but they will be more challenging to find numerically.

\subsection{Black holes}
\label{sec:BH}

\begin{figure*}
\includegraphics[width=8.5cm]{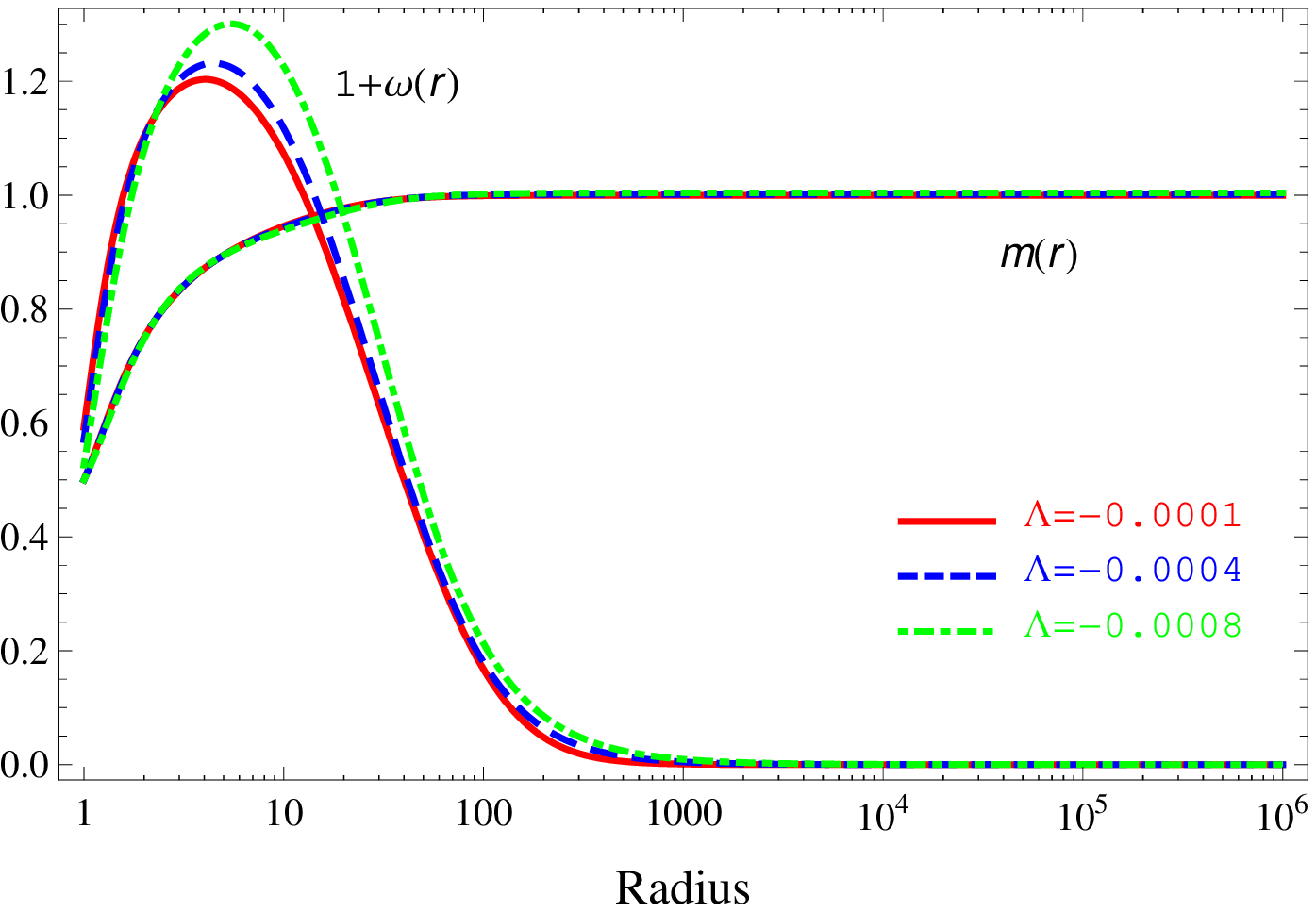}
\includegraphics[width=8.5cm]{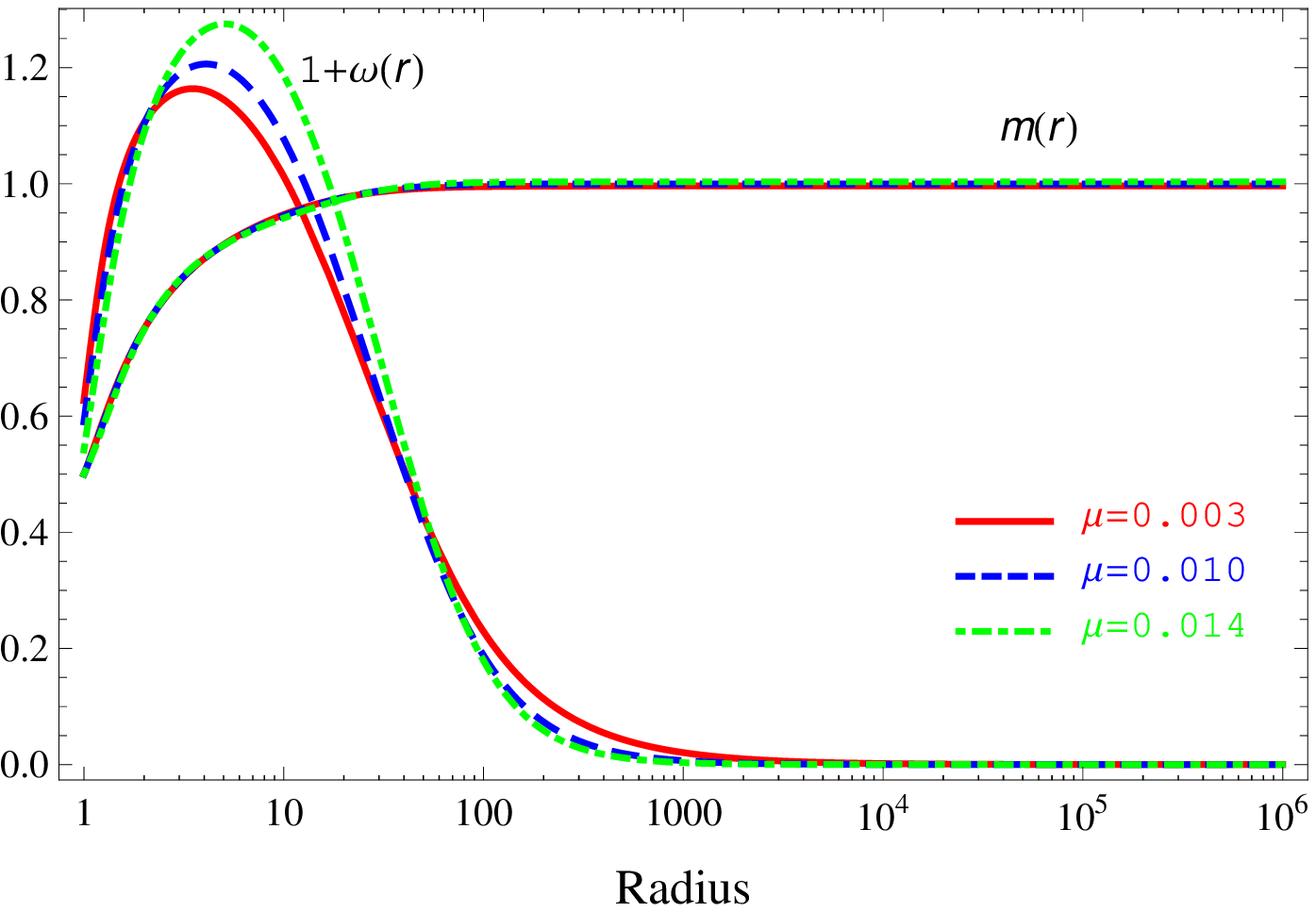}
\includegraphics[width=8.5cm]{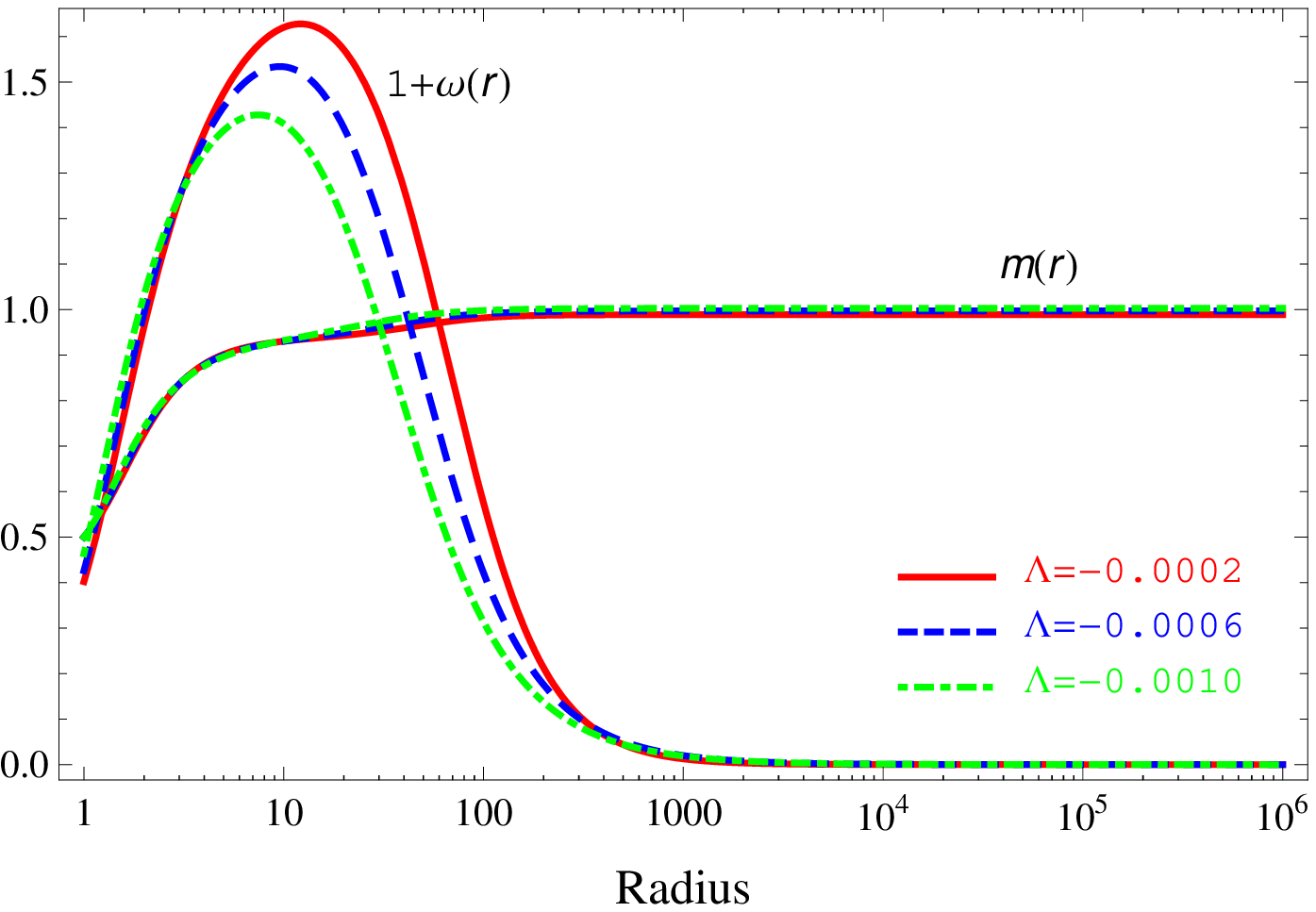}
\includegraphics[width=8.5cm]{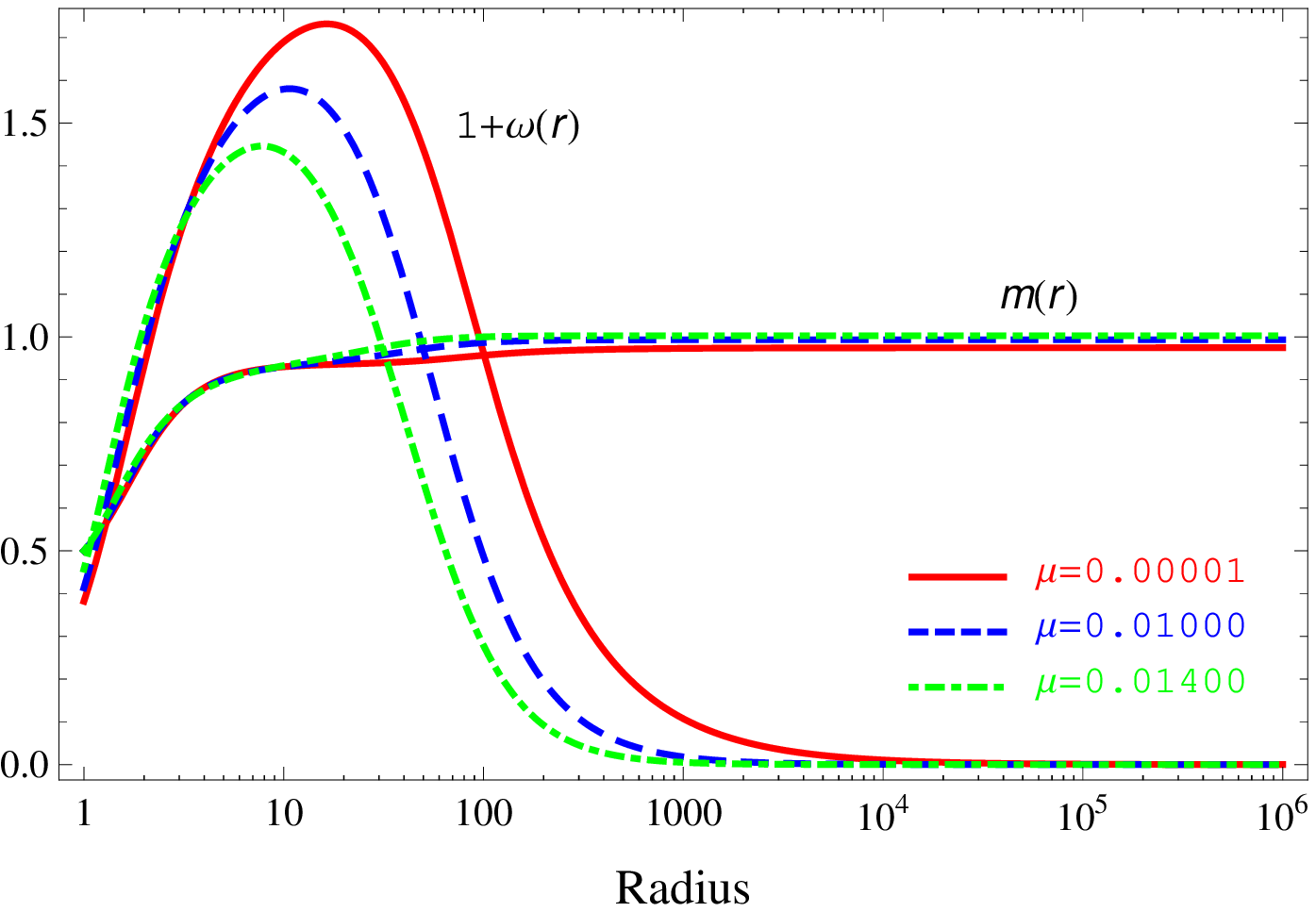}
\caption{Black hole solutions of ENAP-AdS theory for which $\omega (r)$ has two zeros. Top row: $n=2$ solutions with (left panel) fixed $\mu =0.012$ and varying $\Lambda $, (right panel) fixed $\Lambda =-0.0004$ and varying $\mu $.
Bottom row: quasi-$n=1$ solutions with (left panel) fixed $\mu = 0.01$ and varying $\Lambda $, (right panel) fixed $\Lambda = -0.0004$ and varying $\mu $.
The event horizon radius is fixed to be $r_{h}=1$.}
\label{fig:BHn2qn1}
\end{figure*}

Our numerical procedure for finding black hole solutions is very similar to the soliton case.
We begin our integration close to the event horizon (typically $r-r_{h}\sim 10^{-13}$), using the expansions (\ref{nearhorizonseries}) as initial conditions.
The shooting parameter in this case is $\omega _{h}$.
We fix the event horizon radius $r_{h}=1$, and vary $\mu $ and $\Lambda $.

We showed in Sec.~\ref{sec:nodes} that for black hole solutions the gauge field function $\omega (r)$ must have at least one zero.  Unlike soliton solutions, for black holes the number of zeros of $\omega (r)$ does not have to be even.
Some typical black hole solutions are shown in Figs.~\ref{fig:BHn1qn0} and \ref{fig:BHn2qn1}, where $\omega (r)$ has one or two zeros respectively.
We anticipate that black hole solutions for which $\omega (r)$ has more than two zeros also exist, but they will be increasingly difficult to find numerically.
As for the soliton solutions shown in the previous subsection, in Figs.~\ref{fig:BHn1qn0} and \ref{fig:BHn2qn1} we plot the gauge field function $1+\omega (r)$ and the metric function $m(r)$. In both Figs.~\ref{fig:BHn1qn0} and \ref{fig:BHn2qn1}, the plots on the left-hand-side have fixed Proca field mass $\mu $ and varying cosmological constant $\Lambda $, while those plots on the right-hand-side have fixed $\Lambda $ and varying $\mu $.

With the number of zeros of $\omega (r)$ fixed, we find two branches of black hole solutions analogous to the branches of soliton solutions shown in Fig.~\ref{fig:solitons}.
The shooting parameter $\omega _{h}$ lies in different intervals on the two branches of solutions.
As discussed in the previous subsection, we follow the terminology of \cite{Greene:1992fw,VanderBij:2001ah}, and consider the $n=1$, quasi-$n=0$, $n=2$ and quasi-$n=1$ branches.

When $\omega $ has a single zero, the $n=1$ black hole solutions are depicted in the top row of Fig.~\ref{fig:BHn1qn0}, while the quasi-$n=0$ solutions are shown in the bottom row.
As observed for the soliton solutions, the functions $\omega (r)$ and $m(r)$ for the $n=1$ branch solutions do not vary much as either $\mu $ or $\Lambda $ vary.
Increasing $\mu $ with fixed $\Lambda $ or increasing $\left| \Lambda \right| $ for fixed $\mu $ gives an increased value of $m(r)$ as $r\rightarrow \infty $.
The value of $\omega (r)$ on the event horizon $r=r_{h}$ increases as either $\mu $ increases for fixed $\Lambda $ or $\left| \Lambda \right| $ increases for fixed $\mu $.
In contrast, there is much greater variation in $\omega (r)$ and $m(r)$ for the quasi-$n=0$ solutions.
As either $\mu $ increases for fixed $\Lambda $, or as $\left| \Lambda \right| $ increases for fixed $\mu $, the value of $m(r)$ as $r\rightarrow \infty $ increases.
Furthermore, the location of the zero of $\omega (r)$ moves to larger values of $r$ as either $\mu $ decreases for fixed $\Lambda $ or $\left| \Lambda \right| $ decreases for fixed $\mu $.
Unlike the behaviour seen for the $n=1$ solutions, for the quasi-$n=0$ solutions the value of $\omega _{h}$ decreases as either $\mu $ increases for fixed $\Lambda $ or $\left| \Lambda \right| $ increases for fixed $\mu $.

When $\omega $ has two zeros, in Fig.~\ref{fig:BHn2qn1} we show the $n=2$ black hole solutions in the top row and the quasi-$n=1$ solutions in the bottom row. These two branches of solutions have properties similar to those in Figs.~\ref{fig:solitons} and \ref{fig:BHn1qn0}.
As the parameters $\mu $ and $\Lambda $ vary, the gauge field and metric functions vary more on the quasi-$n=1$ branch than on the $n=2$ branch.
On the $n=2$ branch, as $\left| \Lambda \right| $ increases for fixed $\mu $, or $\mu $ increases for fixed $\Lambda $, the maximum value of $\omega (r)$ increases and the location of this maximum moves to larger $r$.
In contrast, on the quasi-$n=1$ branch, as $\left| \Lambda \right| $ increases for fixed $\mu $ or $\mu $ increases for fixed $\Lambda $, the maximum value of $\omega (r)$ decreases and the location of the maximum moves to smaller $r$.
On the quasi-$n=1$ branch, the outermost zero of $\omega (r)$ also moves to larger $r$ as either $\mu $ decreases for fixed $\Lambda $ or $\left| \Lambda \right| $ decreases for fixed Proca field mass $\mu $.

\begin{figure*}
\includegraphics[width=8.5cm]{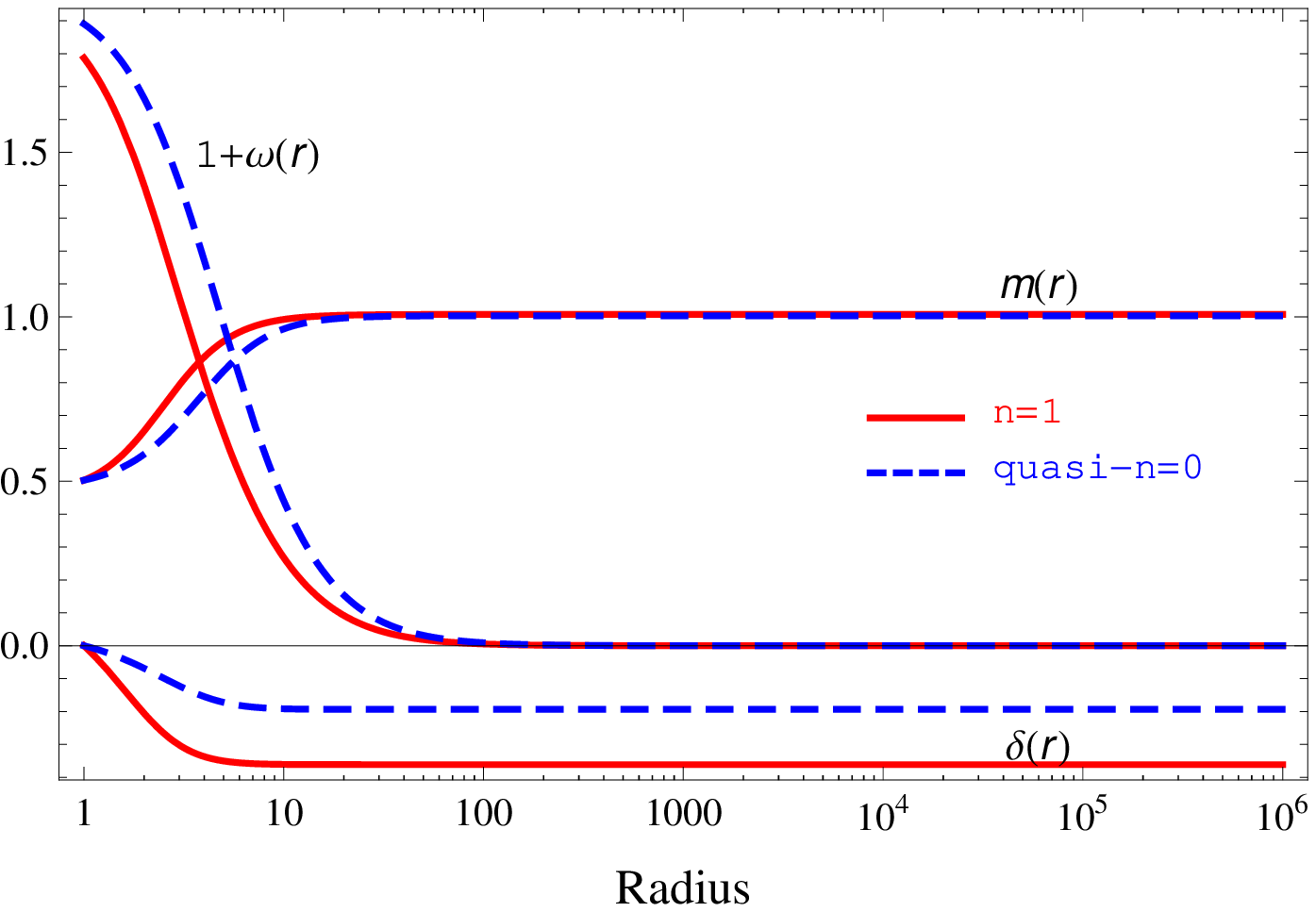}
\includegraphics[width=8.5cm]{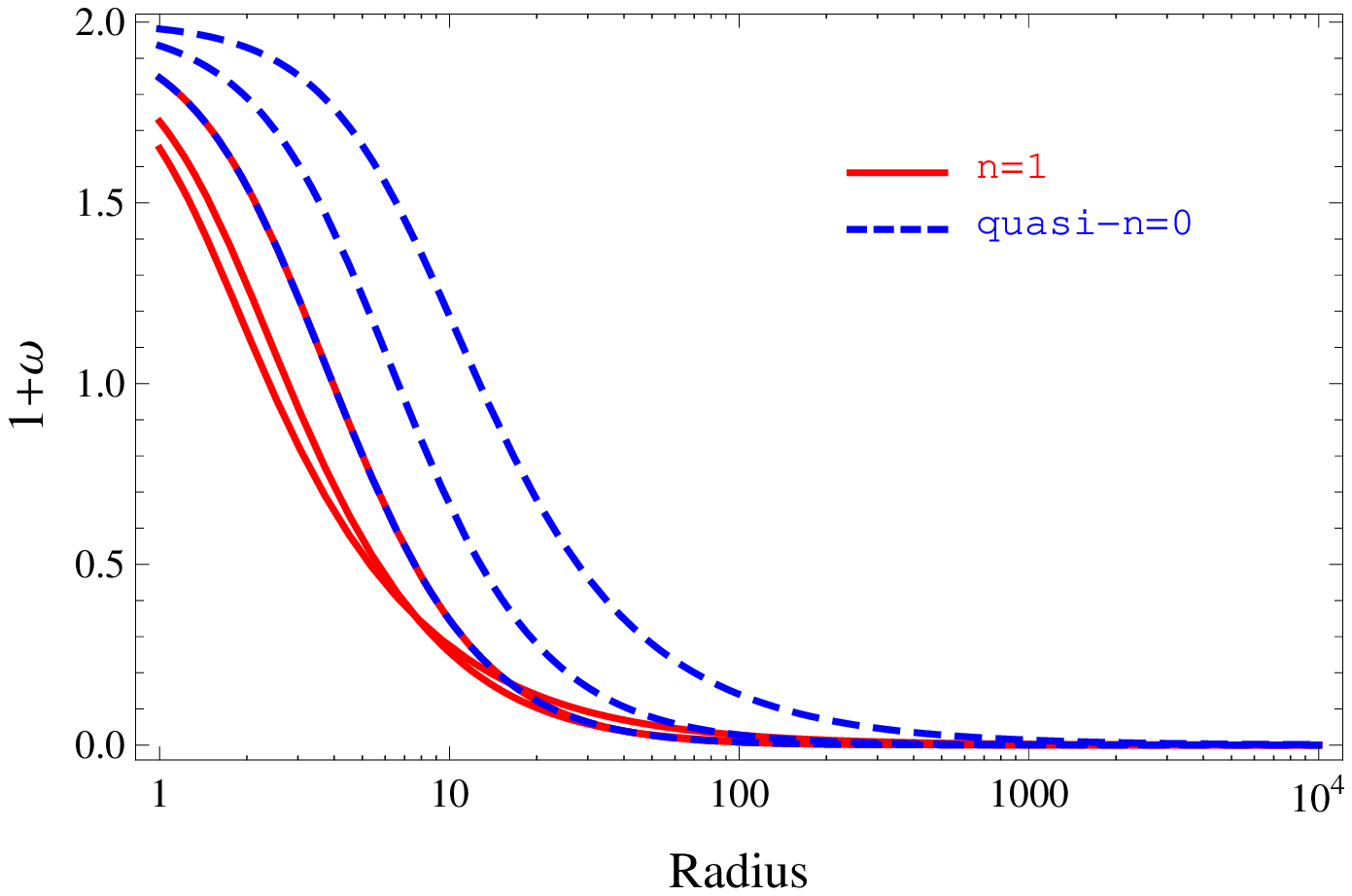}
\caption{Left: Comparison of $n=1$ and quasi-$n=0$ black hole solutions with $\Lambda = -0.02$ and $\mu = 0.1$.  The functions for the $n=1$ black holes are shown in red (solid curves) and those for the quasi-$n=0$ black holes are shown in blue (dashed curves).
Right: Gauge field function $1+\omega (r)$ for a selection of $n=1$ and quasi-$n=0$ black holes with fixed $\Lambda = -0.025$ and varying Proca field mass $\mu = 0.00001$, $0.075$ and $0.1018$.  Quasi-$n=0$ curves are blue (dashed) while those for $n=1$ are red (solid). The two branches of solutions merge when $\mu= \mu _{\text {max}}=0.1018$.}
\label{fig:n1qn0comp}
\end{figure*}

\begin{figure*}
\includegraphics[width=8.5cm]{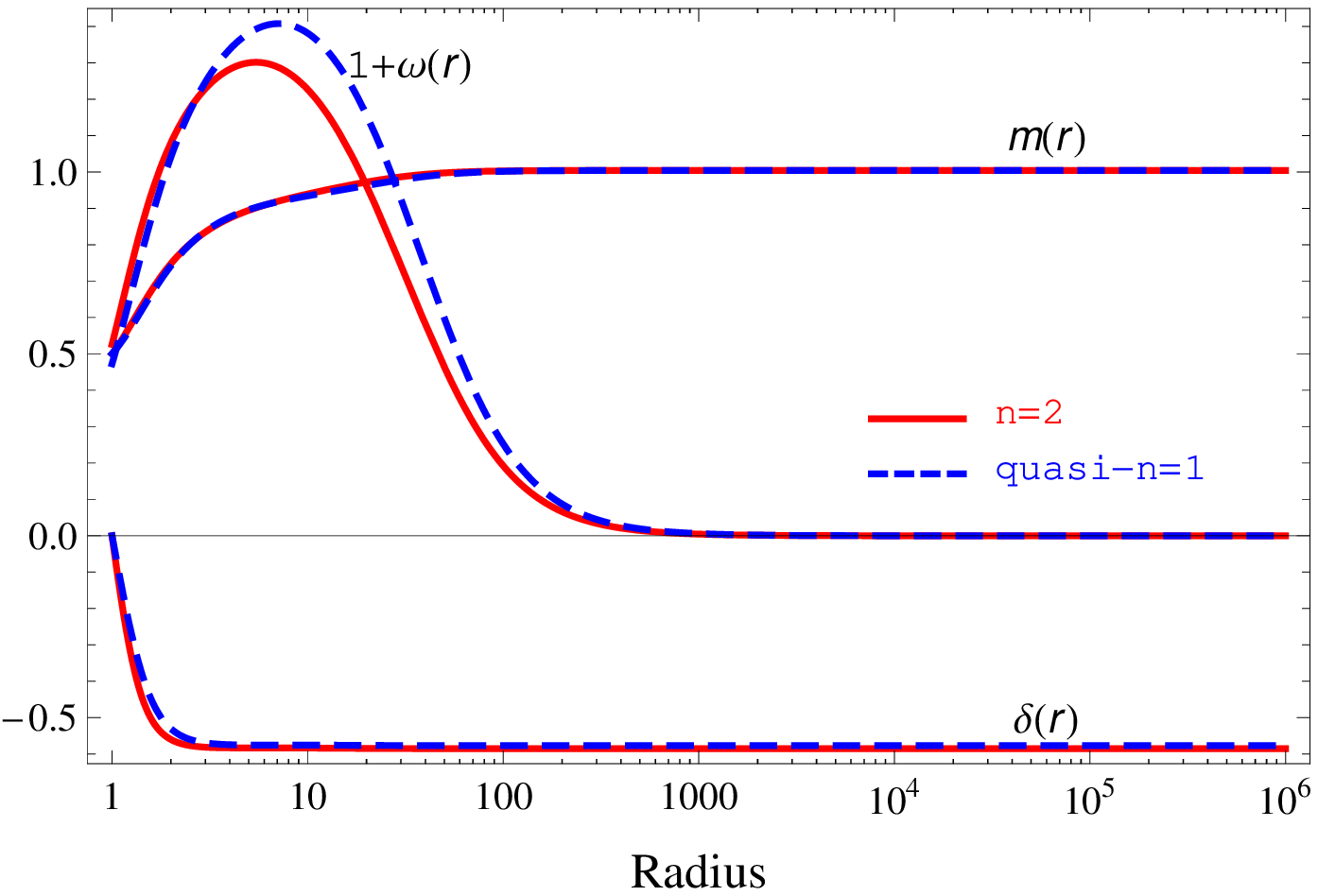}
\includegraphics[width=8.5cm]{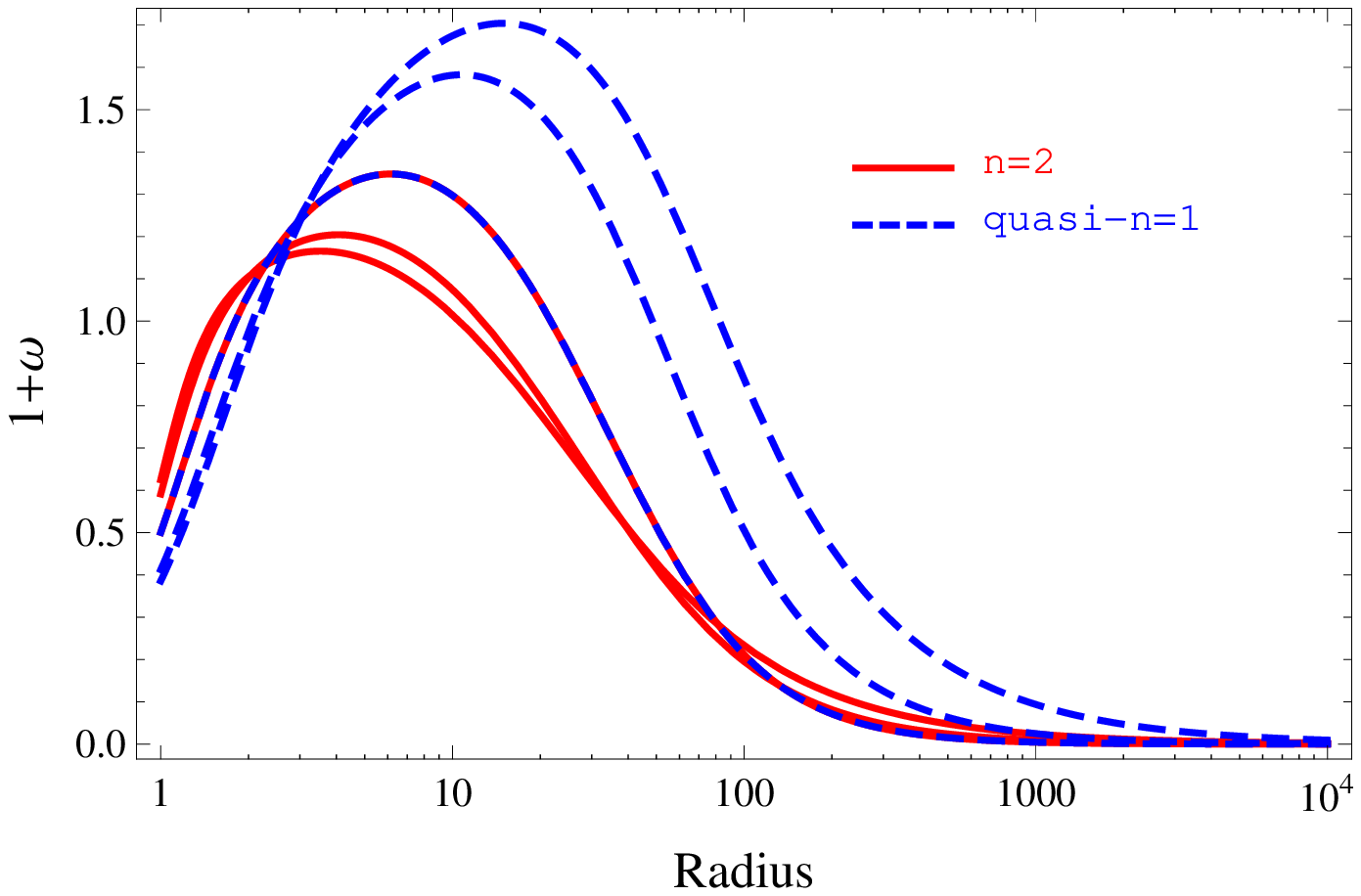}
\caption{Left: Comparison of $n=2$ and quasi-$n=1$ black hole solutions with $\Lambda = -0.0005$ and $\mu = 0.014$.  The functions for the $n=2$ black holes are shown in red (solid curves) and those for the quasi-$n=1$ black holes are shown in blue (dashed curves).
Right: Gauge field function $1+\omega (r)$ for a selection of $n=2$ and quasi-$n=1$ black holes with fixed $\Lambda = -0.0005$ and varying Proca field mass $\mu = 0.00001$, $0.009$ and $0.01442$.  Quasi-$n=1$ curves are blue (dashed) while those for $n=2$ are red (solid). The two branches of solutions merge when $\mu= \mu _{\text {max}}=0.01442$.}
\label{fig:n2qn1comp}
\end{figure*}

We compare the $n=1$ and quasi-$n=0$ branches and the $n=2$ and quasi-$n=1$ branches of black hole solutions in Figs.~\ref{fig:n1qn0comp} and \ref{fig:n2qn1comp} respectively.
As for the soliton solutions in Fig.~\ref{fig:n2qn1}, we plot the gauge potential function $1+\omega (r)$ and metric functions $m(r)$ and $\delta (r)$.
To plot $\delta (r)$, we have set $\delta _{h}=0$ in (\ref{nearhorizonseries}), which means that $\delta (r)$ does not tend to $0$ as $r\rightarrow \infty $.
However, this can be rectified by adding an appropriate constant to $\delta (r)$.

Comparing first the $n=1$ and quasi-$n=0$ solutions with fixed $\mu $ and $\Lambda $, from the left-hand-plot in Fig.~\ref{fig:n1qn0comp} we see that the quasi-$n=0$ soliton has a larger value of $\omega _{h}$ than the corresponding $n=1$ solution.  The difference in values of the metric function $\delta (r)$ at the horizon and infinity is smaller for the quasi-$n=0$ solution than for the $n=1$ black hole. The metric function $m(r)$ is very similar for the two solutions; the quasi-$n=0$ solution has a slightly smaller mass.
As we found for the soliton solutions, for fixed $\Lambda $, the $n=1$ and quasi-$n=0$ branches of solutions merge at $\mu = \mu _{\text {max}}$, this can be seen in the right-hand plot in Fig.~\ref{fig:n1qn0comp}.
We also find that, for fixed $\mu $, there is a maximum value of $\left| \Lambda \right| $ for which there are black hole solutions with $\omega (r)$ having a single zero.

The $n=2$ and quasi-$n=1$ solutions have similar properties, see Fig.~\ref{fig:n2qn1comp}.  The quasi-$n=1$ solutions have smaller values of $\omega _{h}$ and smaller masses than the $n=2$ black holes. For the quasi-$n=1$ solutions, the maximum value of $\omega (r)$ is larger than for the $n=2$ black holes, and the location of this maximum is at larger $r$ for the quasi-$n=1$ black holes than for the $n=2$ solutions.
For fixed $\Lambda $, we find a maximum value of the Proca field mass $\mu =\mu _{\text {max}}$ for which black holes with $\omega (r)$ having two zeros exist; the $n=2$ and quasi-$n=1$ branches merge at this value of $\mu $ (see the right-hand plot in Fig.~\ref{fig:n2qn1comp}).
We also find, for fixed $\mu $, a maximum value of $\left| \Lambda \right| $ for black holes with $\omega (r)$ having two zeros.
The space of values of $\mu $, $\left| \Lambda \right| $ for which there are black holes with $\omega (r)$ having two zeros is considerably smaller than the corresponding space of values for which there are black holes with $\omega (r)$ having a single zero.
This property is shared by the asymptotically flat ENAP and EYMH solutions \cite{Greene:1992fw} and the asymptotically-AdS EYMH solutions \cite{VanderBij:2001ah}.
We anticipate that this trend would continue as the number of zeros of $\omega (r)$ increases, making the solutions increasingly difficult to find numerically.

\section{Stability analysis}
\label{sec:stab}

We now examine the stability of the soliton and black hole solutions of ENAP theory in asymptotically AdS space-time presented in the previous section.
We consider linear, spherically symmetric, perturbations of the metric and gauge field.

\subsection{Perturbation equations}
\label{sec:perts}

The spherically symmetric metric takes the form (\ref{metric}), where the metric functions $R(t,r)$, $S(t,r)$ and $m(t,r)$ (\ref{mdef}) now depend on time $t$ as well as the radial coordinate $r$.
The gauge potential now has the general form (\ref{generalgaugeconnection}), with all quantities $a(t,r)$, $b(t,r)$, $c(t,r)=\omega (t,r)$ and $d(t,r)$ again depending on both $t$ and $r$.
We introduce the notation
\begin{align}
\omega (t,r) &= \omega_0(r) + \epsilon \omega_1(t,r),
\label{1storderpert}
\end{align}
and similarly for  the other quantities in the theory, $m$, $R$, $S$, $a$, $b$ and $d$.
In (\ref{1storderpert}), $\epsilon$ is a small parameter. The zeroth order term, for example, $\omega_{0}(r)$, denotes the equilibrium solution.
The zeroth order quantities $a_{0}$, $b_{0}$ and $d_{0}$ vanish identically.
Variables with the subscript ``$1$'' denote the perturbations.

By substituting (\ref{1storderpert}) and similar expressions for the other field variables into the field equations (\ref{fieldequations}), the zeroth order gives the static field equations (\ref{fieldequations1}) and the linearized NAP field equations (\ref{NAPequation}) up to order $\epsilon$ are given by
\begin{widetext}
\begin{subequations}
\label{NAPpert}
\begin{align}
0 &= -r^{2}R_{0}S_{0}{a}^{\prime\prime}_{1}+rR_{0}\left(rS^{\prime}_{0}-2S_0\right){a}^{\prime}_{1} +S_0\left(r^2\mu^2+2\omega^2_0\right)a_1 + r^2R_{0}S_0\dot{b}^{\prime}_{1}
-rR_{0} \left( r S_{0}^{\prime } -2S_{0} \right) {\dot {b}}_{1}
+2S_0\omega_0\dot{d}_1,
\label{perturbedNAPt}\\
0 &= \frac{1}{R_{0}S^2_0}\dot{a}^{\prime}_{1}-\frac{1}{R_{0}S^2_0}\ddot{b}_{1} - \left(\mu^2+\frac{2\omega^2_0}{r^2}\right)b_1  - \frac{2\omega_0}{r^2}d^{\prime}_{1} + \frac{2\omega^{\prime}_0}{r^2}d_1, \label{perturbedNAPr} \\
0 &= \frac{1}{R_{0}S^2_0}\ddot{d}_{1} - R_{0}d^{\prime\prime}_{1}-
\frac{(R_{0}S_0)^{\prime}}{S_0}d^{\prime}_{1} + \left(\mu^2+\frac{\omega^2_0-1}{r^2}\right)d_1 + \frac{\omega_0}{R_{0}S^2_0}\dot{a}_{1}-R_{0}\omega_0b^{\prime}_{1} -\left(\omega_0\frac{(R_{0}S_0)^{\prime}}{S_0}+2R_{0}\omega^{\prime}_0\right)b_1, \label{perturbedNAPtheta} \\
0 &= -r^3\ddot{\omega}_{1} + r^3R^{2}_{0}S^{2}_0\omega^{\prime\prime}_{1}+r^{3}R_{0}S_0(R_{0}S_0)^{\prime}\omega^{\prime}_1   -rR_{0}S^2_0\left(3\omega^2_0+r^2\mu^2-1\right)\omega_1 -2r^2R_{0}S^{2}_{0}\omega^{\prime}_{0}m^{\prime}_{1} \nonumber \\
&\quad -2rR_{0}S_0\left[\omega^{\prime}_{0}\left(rS^{\prime}_0-S_0\right)+\omega^{\prime\prime}_0S_0r\right]m_{1} + r^3R^2_{0}S_0\omega^{\prime}_0S^{\prime}_{1}-r^3R^2_{0}S^{\prime}_0\omega^{\prime}_0S_1.
\label{perturbedNAPphi}
\end{align}
\end{subequations}
\end{widetext}
The linearized constraint equation (\ref{gaugeCond}) reads
\begin{align}
0 &= \frac{1}{R_{0}S^{2}_0}\dot{a}_{1} - R_{0}b^{\prime}_{1} -\left(\frac{2R_{0}}{r}+\frac{\left(R_{0}S_0\right)^{\prime}}{S_0}\right)b_1 + \frac{2}{r^2}d_1. \label{perturbedgaugecon}
\end{align}
The derivatives of the metric perturbations $m^{\prime}_{1}$ and $S^{\prime}_{1}$ can be found explicitly from the $(tt)$ and $(rr)$ components of the Einstein field equations (\ref{EFE})
\begin{subequations}
\label{EFEpert}
\begin{align}
m^{\prime}_{1} = & \frac{2R_{0}\omega^{\prime}_{0}}{e^2}\omega^{\prime}_{1} + \frac{2\left(1+\omega_0\right)\left(r^2\mu^2+\left(\omega_0-1\right)\omega_0\right)}{e^2r^2}\omega_1
\nonumber \\ &
- \frac{2\omega_0^{\prime 2}}{e^2r}m_1,
\label{perturbedm1P}
\\
S^{\prime}_{1} = & \frac{4S_0\omega_0^{\prime}}{e^2}\omega^{\prime}_1 + \frac{2\omega_0^{\prime 2}}{e^2r}S_1.
\label{perturbedS1P}
\end{align}
\end{subequations}
The final perturbation equation comes from the $tr$ component of the Einstein field equations (\ref{EFE}) and gives the time derivative of the metric perturbation ${\dot {m}}_{1}$.
Performing an integration with respect to time then yields
\begin{align}
m_{1} &= \frac{2R_{0}\omega^{\prime}_0}{e^2}\omega_1 + \mathcal{F}(r),
\label{perturbedm1a}
\end{align}
where $\mathcal{F}(r)$ is an arbitrary function of $r$ only.
Substituting (\ref{perturbedm1a}) into (\ref{perturbedm1P}) gives the following first order equation for ${\mathcal {F}}$:
\begin{align}
\frac{\mathcal{F}^{\prime}}{\mathcal{F}} &= -\frac{2\omega_{0}'^2}{e^2 r},
\end{align}
which has the solution
\begin{align}
\mathcal{F} &= \mathcal{K}\exp\left[-\int_{r_0}^{r} \frac{2\omega_{0}'^2(\tilde{r})}{e^2 \tilde{r}}~d\tilde{r}\right] ,
\label{mathcalFsolution}
\end{align}
where $\mathcal{K}$ is a constant of integration and $r_{0}=0$ for solitons, $r_{0}=r_{h}$ for black holes.
At either the origin or a black hole event horizon  we require that $m_{1}=0$, so that the origin remains regular or the event horizon is not changed by the perturbation.
At the origin, $\omega _{0}'=0$ from (\ref{BCregular}), so we must have ${\mathcal {F}}(0)=0$.
At the horizon, $R_{0}(r_{h})=0$ and again this means that ${\mathcal {F}}(r_{h})=0$.
To have $\mathcal{F}(r_0)\equiv 0$ in (\ref{mathcalFsolution}), we must set $\mathcal{K}=0$. This means that $\mathcal{F}(r)\equiv 0$ for all $r$ and
\begin{equation}
m_{1} = \frac {2R_{0}\omega ^{\prime }_{0}}{e^{2}} \omega _{1}.
\label{m1final}
\end{equation}

The above expression for the metric perturbation $m_{1}$, together with the equations (\ref{EFEpert}), can be used to eliminate the metric perturbations from the final perturbed NAP equation (\ref{perturbedNAPphi}).
This leaves four perturbation equations (\ref{NAPpert}), together with the constraint (\ref{perturbedgaugecon}) which is a consequence of them.
These four equations decouple into two sectors: the {\em {gravitational sector}} comprises the single equation (\ref{perturbedNAPphi}) for the gauge field perturbation $\omega _{1}$, while the {\em {sphaleronic sector}} consists of the remaining three perturbation equations (\ref{perturbedNAPt}, \ref{perturbedNAPr}, \ref{perturbedNAPtheta}) for the perturbations $a_{1}$, $b_{1}$ and $d_{1}$.

\subsection{Gravitational sector}
\label{sec:grav}

We begin our stability analysis by considering the gravitational sector perturbation $\omega _{1}$.
Eliminating the metric perturbations and using the static field equation (\ref{NAP}), the perturbation equation (\ref{perturbedNAPphi}) simplifies to
\begin{align}
\label{gravityperturbationeq}
0 &= -\ddot{\omega}_1 + R_{0}^2S^{2}_0\omega^{\prime\prime}_{1}+R_{0}S_0(R_{0}S_0)^{\prime}\omega^{\prime}_1+ V(r)\omega_1,
\end{align}
where the perturbation potential $V(r)$ is given by
\begin{align}
V(r) = & R_{0}S^{2}_0\left[\frac{1}{r^2}-\mu^2-\frac{3\omega^{2}_0}{r^2}-\frac{8\mu^2\omega^{\prime}_0}{e^2r}+\frac{8\omega_0\omega^{\prime}_0}{e^2r^3}
\right. \nonumber \\ & \left.
-\frac{8\mu^2\omega_0\omega^{\prime}_0}{e^2r} -\frac{8\omega^{3}_0\omega^{\prime}_0}{e^2r^3}+\frac{4R_{0}\omega_0^{\prime 2}}{e^2r^2}+\frac{4R_{0}^{\prime}\omega_0^{\prime 2}}{e^2r}
\right. \nonumber \\ & \left.
+\frac{8R_{0}\omega_0^{\prime 4}}{e^4r^2}\right].
\label{Vdef}
\end{align}
Setting the Proca field mass $\mu =0$, the perturbation potential (\ref{Vdef}) reduces to that for the gravitational sector of pure ${\mathfrak {su}}(2)$ EYM theory \cite{Straumann:1989tf,Winstanley:1998sn,Bjoraker:1999yd}.

We consider time periodic perturbations
\begin{align}
\label{timedependenceperturbation}
\omega_{1}(t,r) &= e^{-i\sigma t}\omega_{1}(r),
\end{align}
and introduce the usual tortoise coordinate $r^\ast$ such that
\begin{align}
\label{tortoisedefinition}
\frac{d r_{\ast}}{dr} &= \frac{1}{R_{0}S_0}.
\end{align}
By choosing an appropriate constant of integration, the tortoise coordinate $r_{\ast}$ ranges from $0<r_{\ast}<r_c$ (where $r_c>0$ is a positive constant) for solitons and $-\infty<r_{\ast}<0$ for black holes.
Then the gravitational perturbation equation (\ref{gravityperturbationeq}) takes the standard Schr\"odinger form
\begin{align}
\label{gravityperturbationeq1}
\sigma^2\omega_1 &= -\frac{d^2\omega_1}{dr^{2}_{\ast}}-V(r)\omega_1.
\end{align}
Since (\ref{gravityperturbationeq1}) is in self-adjoint form, the eigenvalue $\sigma ^{2}$ is real and standard Sturm-Liouville theory applies, so that for each eigenvalue $\sigma _{i}^{2}$, where $\sigma _{1}^{2}<\sigma _{2}^{2}< \sigma _{3}^{2}<\ldots $, the corresponding eigenfunction has $i-1$ zeros.
In particular, the lowest eigenvalue $\sigma _{1}^{2}$ will correspond to an eigenfunction which is nodeless.
To prevent confusion with $n$ (the number of zeros of the equilibrium gauge field function $\omega _{0}(r)$), we denote the number of zeros of the perturbation $\omega _{1}$ by $N$.

Before integrating (\ref{gravityperturbationeq1}) numerically, we need to impose suitable boundary conditions  on the perturbation $\omega _{1}$.
We require that $\omega _{1}$ vanishes at the origin (for soliton solutions), event horizon (for black hole solutions) and as $r\rightarrow \infty $, so that
\begin{subequations}
\label{pertBCs}
\begin{equation}
\label{powerofboundarycondition0}
\omega_1 \sim
\begin{cases}
r^\alpha & \mbox{for $r\rightarrow0$}, \\
(r-r_h)^\beta & \mbox{for $r\rightarrow r_h$},  \\
r^\rho & \mbox{for $r\rightarrow \infty$,} \\
\end{cases}
\end{equation}
where we require that $\alpha $ and  $\beta $ have positive real part and $\rho $ has negative real part.
By substituting (\ref{powerofboundarycondition0}) into the perturbation equation (\ref{gravityperturbationeq1}), we find
\begin{align}
\label{powerofboundarycondition}
\alpha &= 2, \nonumber\\
\beta &= \pm \frac{i e^2 r^3_h\sigma}{\left(1-\omega^2_{0h}\right)^2-e^2r^2_h+e^2r^4_h\Lambda+2(1+\omega_{0h})^2r^2_h\mu^2}, \nonumber \\
\rho &= -\frac{1}{2}\left(1+\sqrt{1-\frac{12\mu^2}{\Lambda}}\right),
\end{align}
\end{subequations}
where $\omega_{0h}=\omega_0(r_h)$.
It is clear that $\alpha >0$ and $\rho <0$, as required.
For $\beta $, the sign is chosen so that $\text{Re} (\beta )>0$.

To integrate (\ref{gravityperturbationeq1}) numerically, we use a standard shooting method.
For a given equilibrium solution, we search for values of $\sigma ^{2}$ such that the boundary conditions (\ref{pertBCs}) are satisfied by the perturbation $\omega _{1}$.
If $\sigma ^{2}>0$, then $\sigma $ is real and the perturbation (\ref{timedependenceperturbation}) is periodic in time, but if $\sigma ^{2}<0$, then $\sigma $ is purely imaginary and there is a perturbation (\ref{timedependenceperturbation}) which grows exponentially in time. In the latter situation we deduce that the corresponding equilibrium configuration is unstable.
We now study the perturbations $\omega _{1}$ for a selection of equilibrium ENAP solitons and black holes.
For the rest of this section we set the gauge coupling constant $e=1$.

\subsubsection{Solitons}
\label{sec:solstab}

\begin{figure*}[p]
\includegraphics[width=8.5cm]{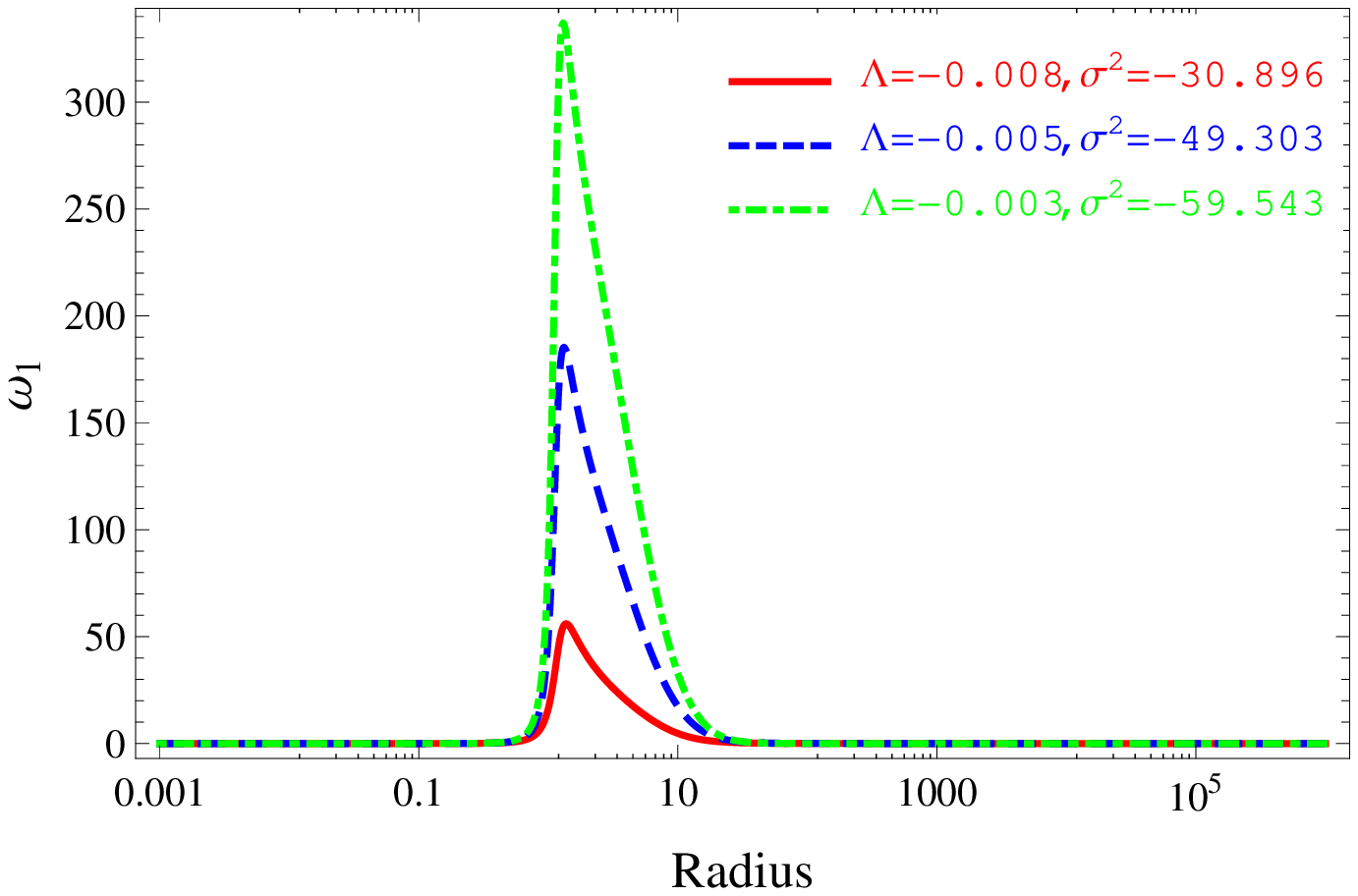}
\includegraphics[width=8.5cm]{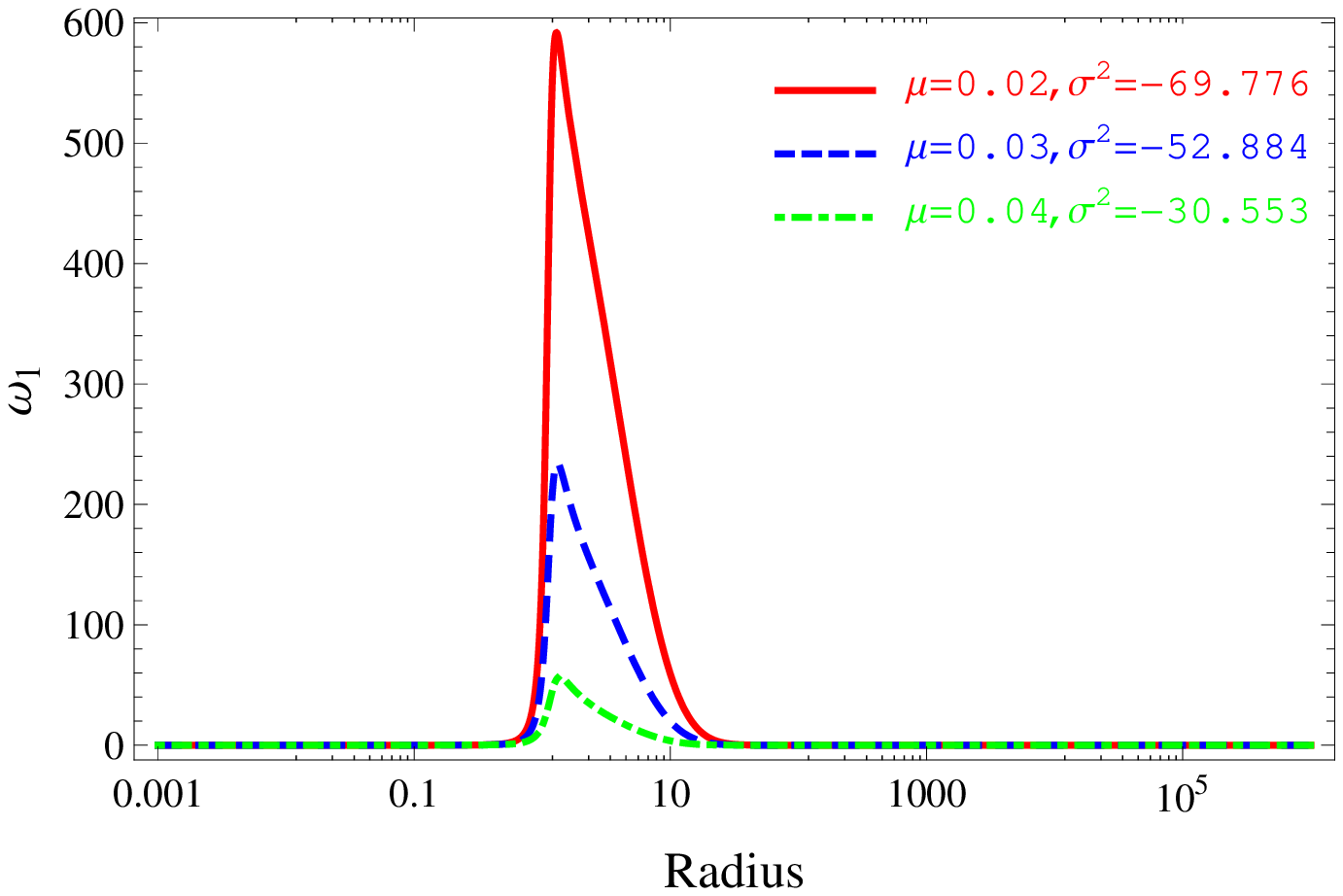}
\includegraphics[width=8.5cm]{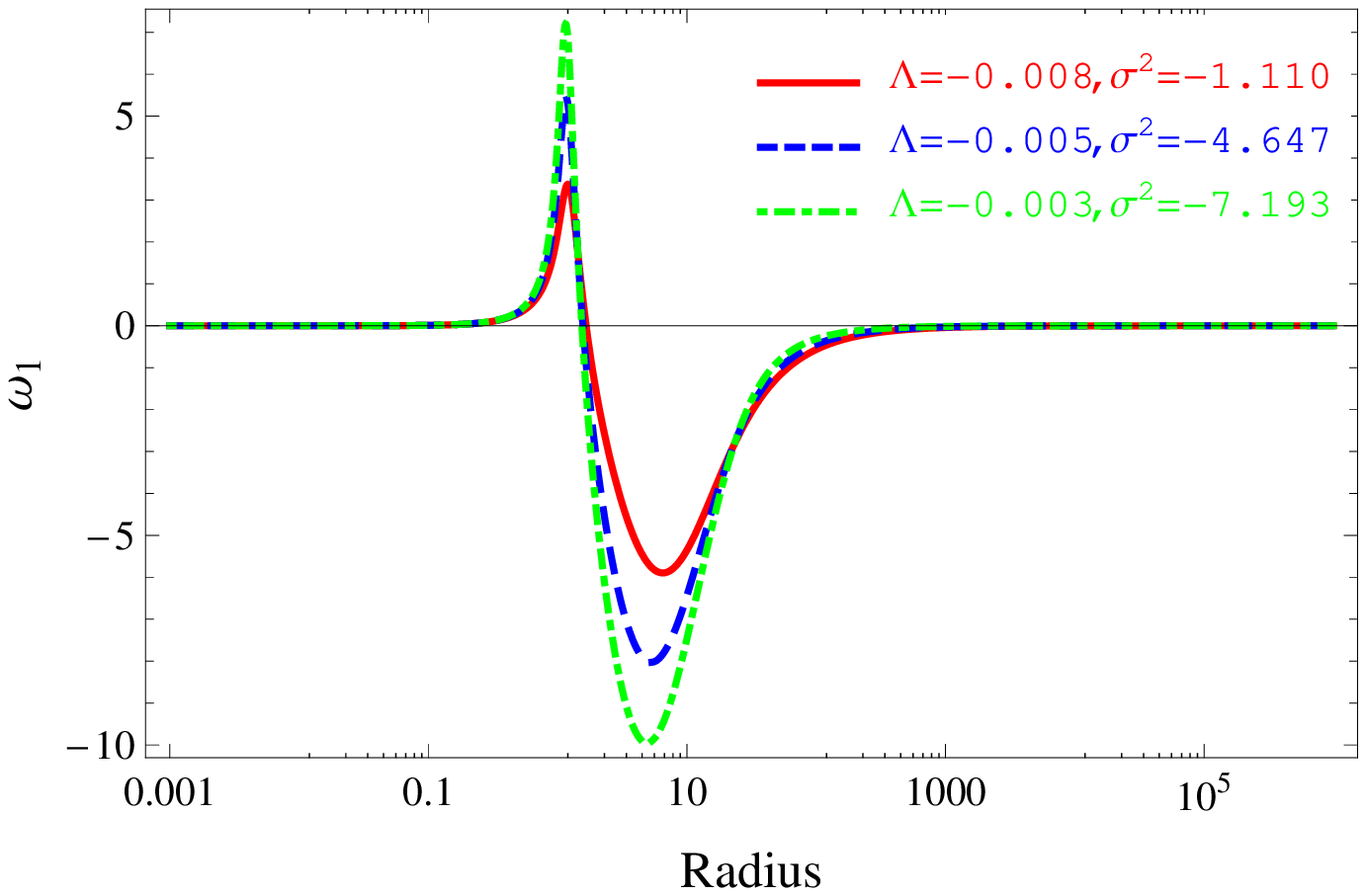}
\includegraphics[width=8.5cm]{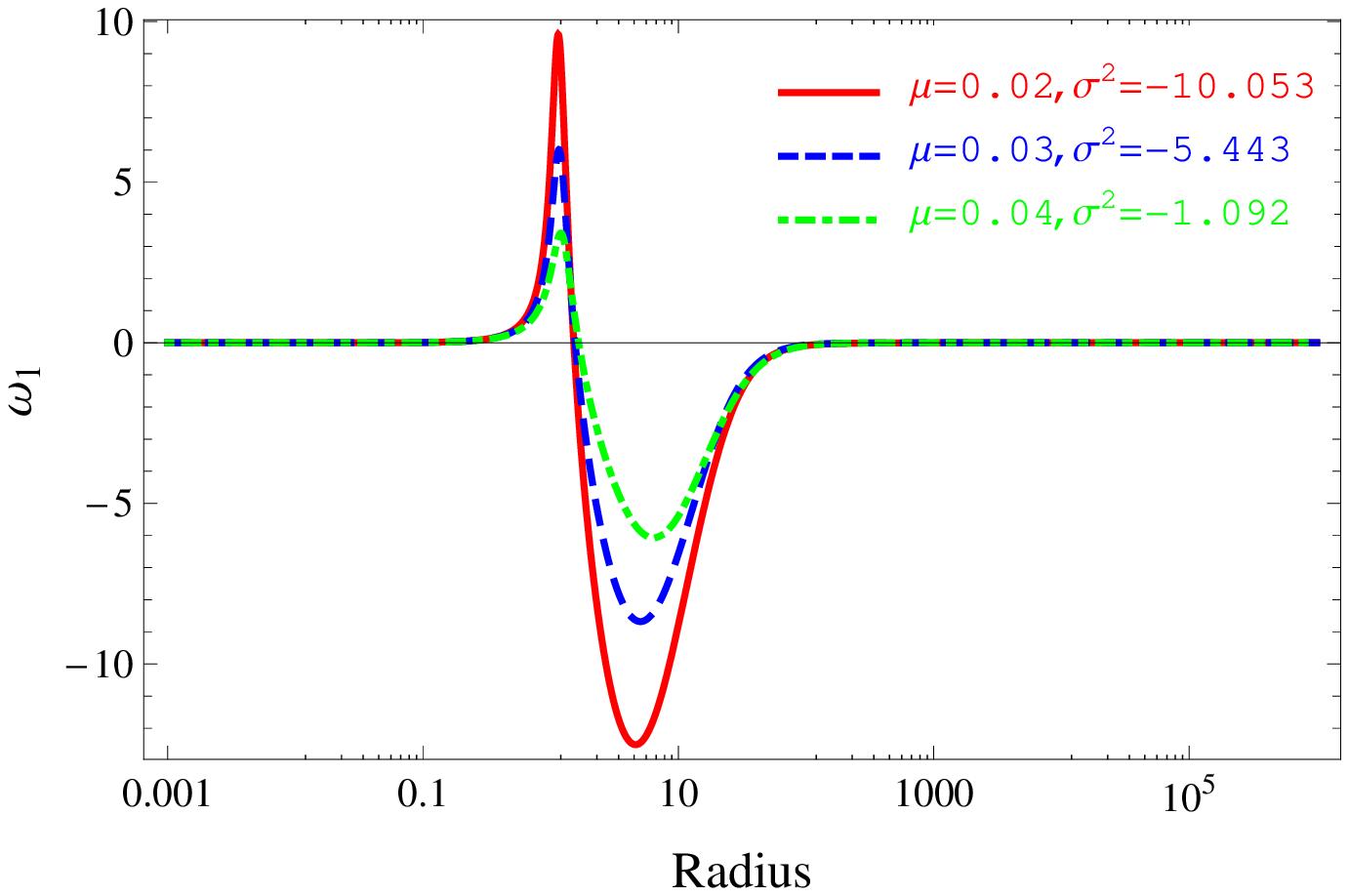}
\caption{Unstable perturbations $\omega _{1}$ for $n=2$ equilibrium solitons. Top row: nodeless perturbations with $N=0$.  Bottom row: perturbations having a single node, $N=1$.
Left-hand plots: fixed Proca field mass $\mu =0.02$ and varying cosmological constant $\Lambda $.
Right-hand plots: fixed $\Lambda = -0.001$ and varying $\mu $.}
\label{fig:soln2grav}
\end{figure*}

\begin{figure*}[p]
\includegraphics[width=8.5cm]{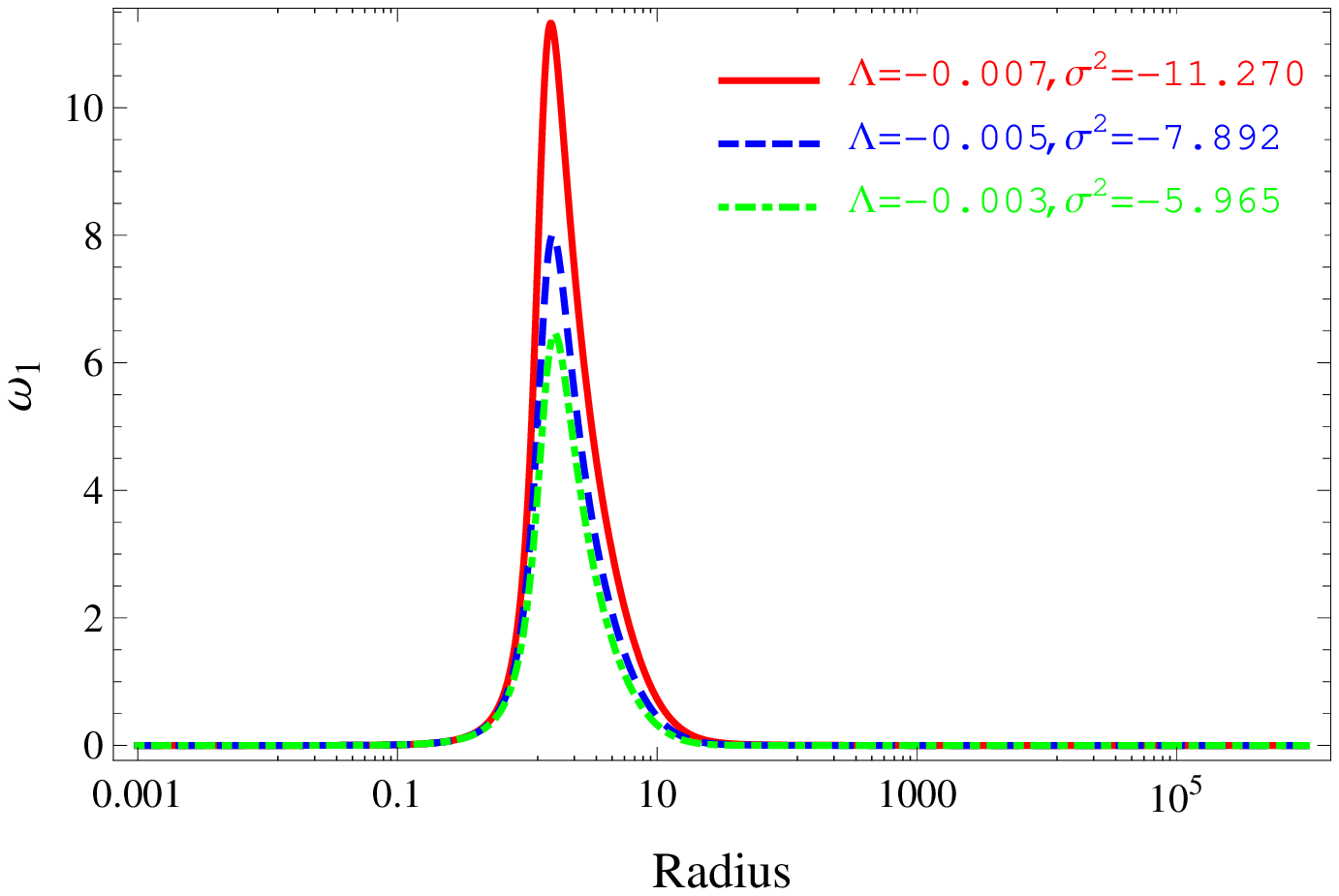}
\includegraphics[width=8.5cm]{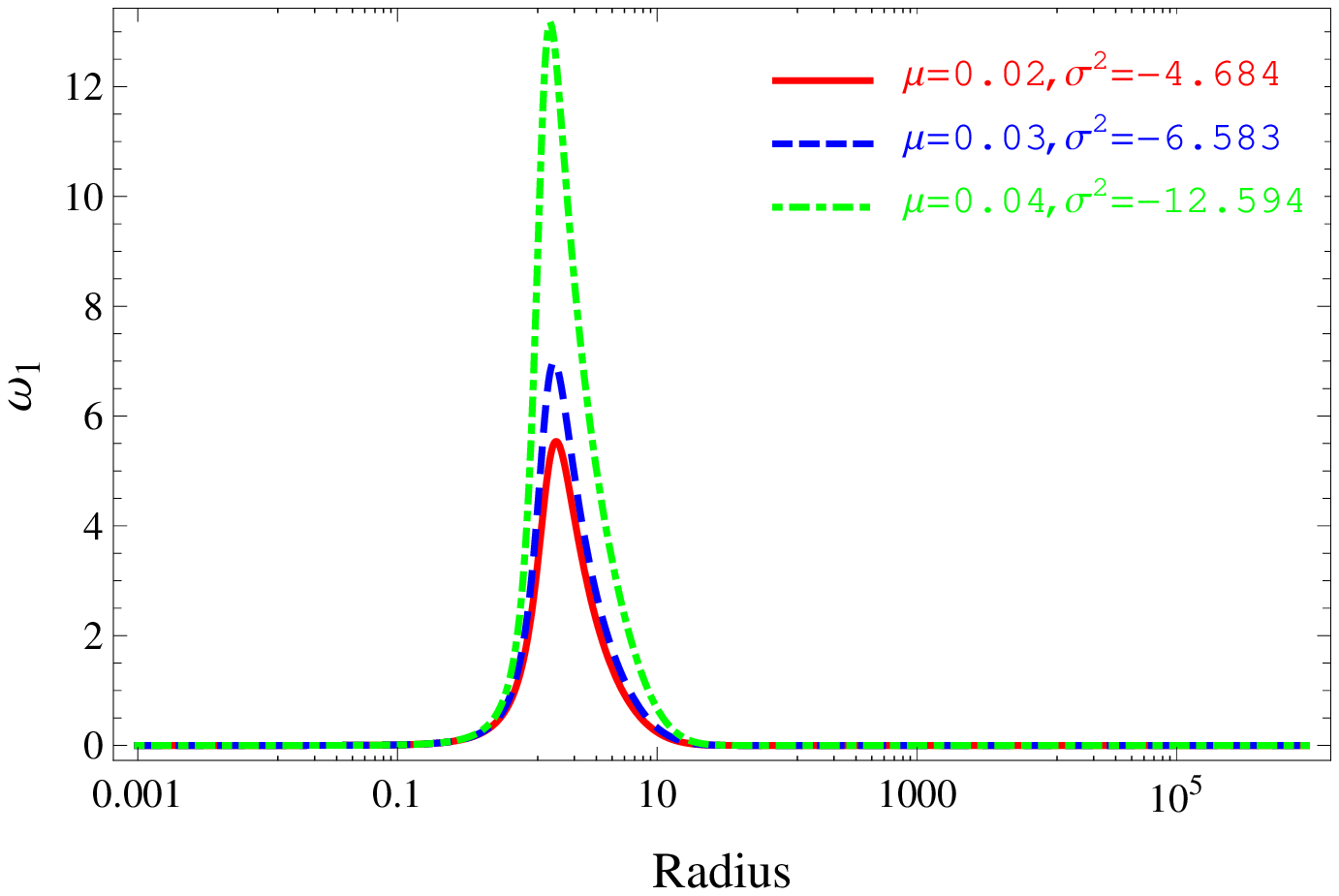}
\caption{Unstable perturbations $\omega _{1}$ for quasi-$n=1$ equilibrium solitons. All perturbations shown are nodeless, with $N=0$.
Left-hand plot: fixed $\mu = 0.02$ and varying $\Lambda $. Right-hand plot: fixed $\Lambda = -0.001$ and varying $\mu $.}
\label{fig:solqn1grav}
\end{figure*}

We begin with the perturbations of the $n=2$ branch of soliton solutions. For each of the $n=2$ solitons we investigated, we found two unstable
modes with $\sigma ^{2}<0$, one with no zeros ($N=0$) and one with a single zero ($N=1$).
Some examples of these unstable perturbations are shown in Fig.~\ref{fig:soln2grav}.
The nodeless perturbations with $N=0$ are shown in the top row of plots, and the $N=1$ perturbations with a single zero are shown in the bottom row of plots.
In the left-hand plots we have fixed the Proca field mass $\mu $ and varied the cosmological constant $\Lambda $, while in the right-hand plots we have fixed $\Lambda $ and varied $\mu $.
In accordance with standard Sturm-Liouville theory, we find that the nodeless perturbations correspond to values of the eigenvalue $\sigma ^{2}$ which are more negative than those for the $N=1$ perturbations.
Considering the nodeless $N=0$ perturbations, we find that with the Proca field mass $\mu $ fixed, the eigenvalue $\sigma ^{2}$ decreases (becomes more negative) as $\left| \Lambda \right| $ decreases.
With $\Lambda $ fixed, we find that $\sigma ^{2}$ increases as $\mu $ increases.
Similar trends are observed for the $N=1$ perturbations.

Since the perturbation equation (\ref{gravityperturbationeq1}) is linear, the overall scale of the perturbations shown in Fig.~\ref{fig:soln2grav}
is not important.
All the $N=0$ perturbations have a very similar shape, with a peak which rises sharply at roughly the same location for all $\mu $ and $\Lambda $, and which decays away more slowly for larger $r$.
For the $N=1$ perturbations, in Fig.~\ref{fig:soln2grav} we see that the location of the zero of $\omega _{1}$ does not vary much as either $\mu $ or $\Lambda $ varies.
The overall shape of the $N=1$ perturbations, like the $N=0$ perturbations, is roughly the same for all $\mu $ and $\Lambda $.  There is a peak at smaller values of $r$ followed by a deeper trough at larger values of $r$.

Next we consider the perturbations of the quasi-$n=1$ branch of solitons.  For this branch, we find just one unstable perturbation for each equilibrium soliton, and that perturbation has no zeros, see Fig.~\ref{fig:solqn1grav}. In the left-hand plot in Fig.~\ref{fig:solqn1grav} we have fixed $\mu $ and varied the cosmological constant $\Lambda $, while in the right-hand plot we have fixed $\Lambda $ and varied the Proca field mass $\mu $.
The values of $\sigma ^{2}$ that we find for these $N=0$ unstable modes are less negative than those for the $N=0$ perturbations of the $n=2$ branch of solitons, and are  of a similar magnitude to those that we find for the $N=1$ unstable modes of the $n=2$ equilibrium solitons.
With $\mu $ fixed, we find that the eigenvalue $\sigma ^{2}$ decreases as $\left| \Lambda \right| $ increases, and for fixed $\Lambda $ we find that $\sigma ^{2}$ decreases as $\mu $ increases.
Both these trends are the opposite of that observed for perturbations of the $n=2$ branch of solitons.
Again, the overall scale of the perturbations is not important.  The shape of the perturbations is very similar for all $\mu $ and $\Lambda $. The slopes on the left- and right-hand sides of the peaks of the perturbations are more even than for the $N=0$ perturbations of the $n=2$ branch of equilibrium solutions (see the top row of Fig.~\ref{fig:soln2grav}).

\begin{figure*}
\includegraphics[width=8.5cm]{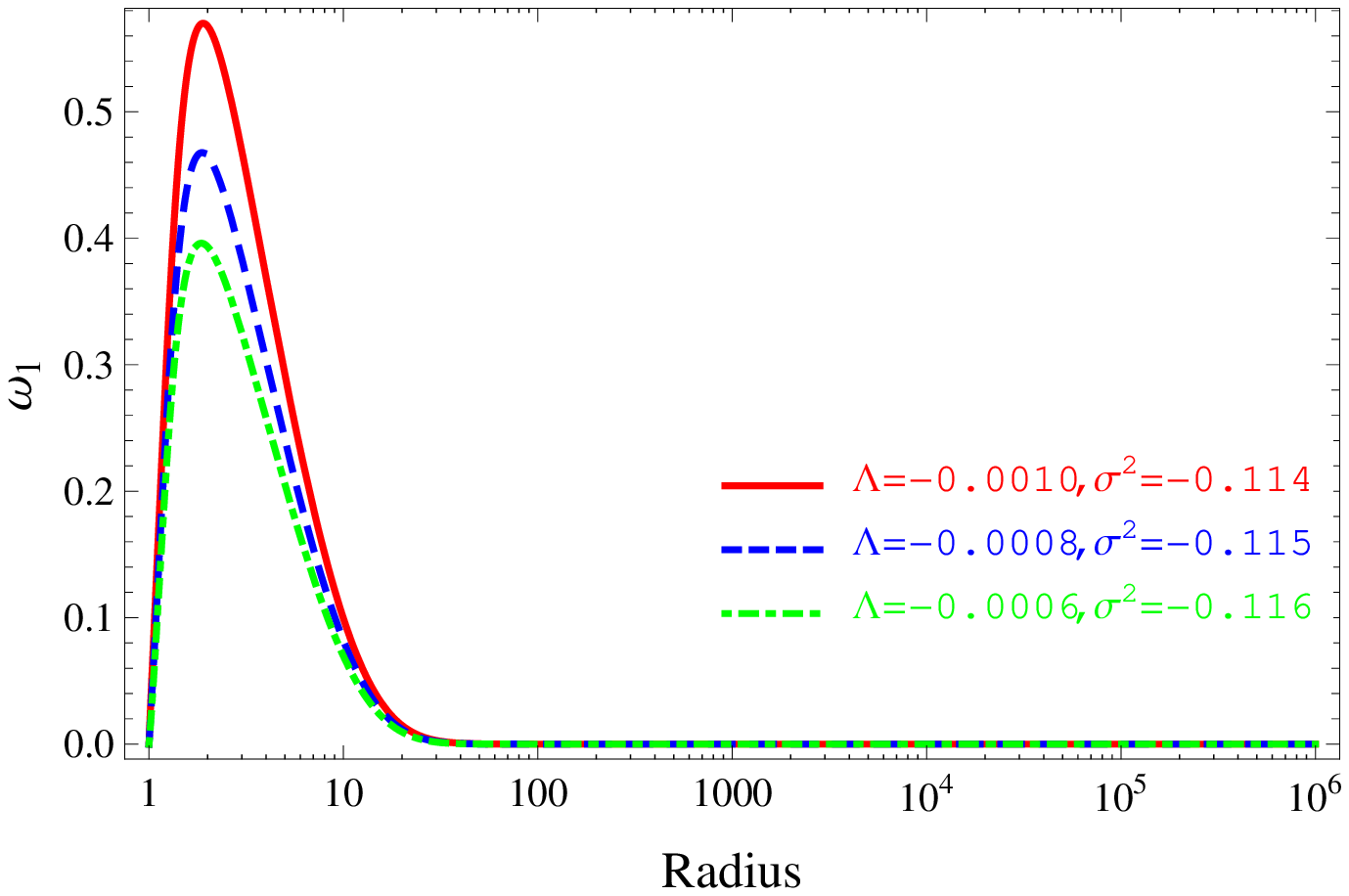}
\includegraphics[width=8.5cm]{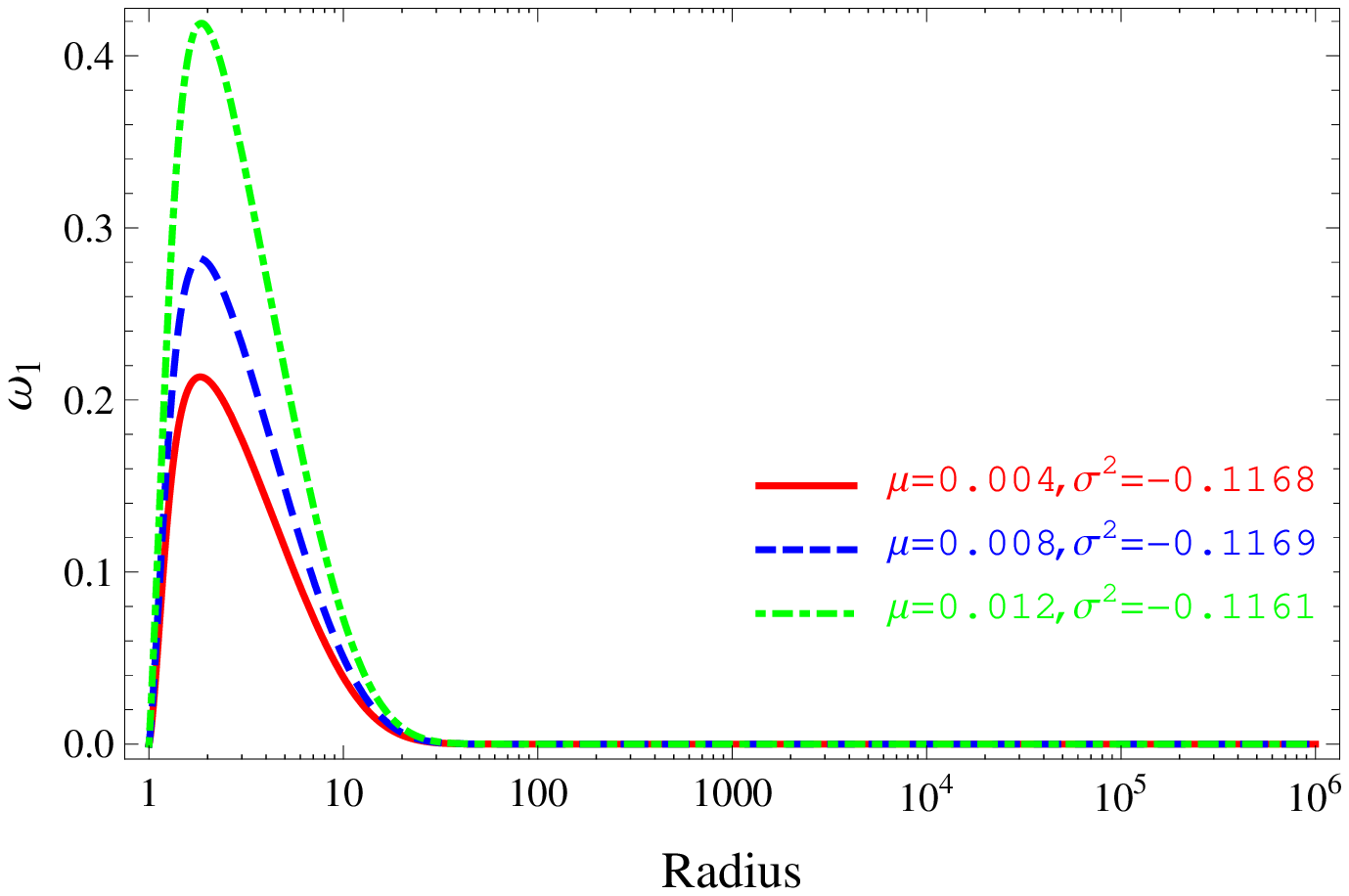}
\includegraphics[width=8.5cm]{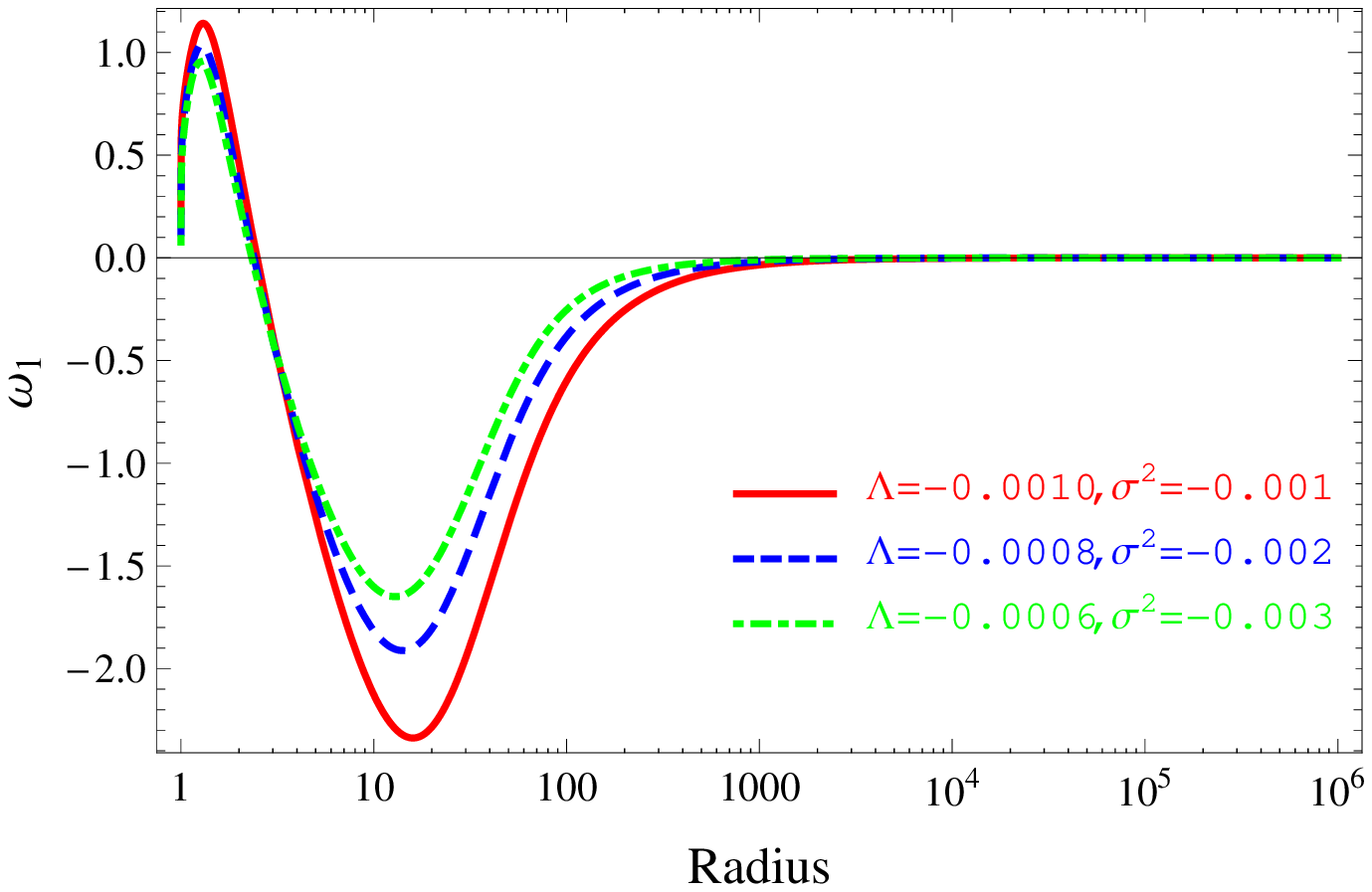}
\includegraphics[width=8.5cm]{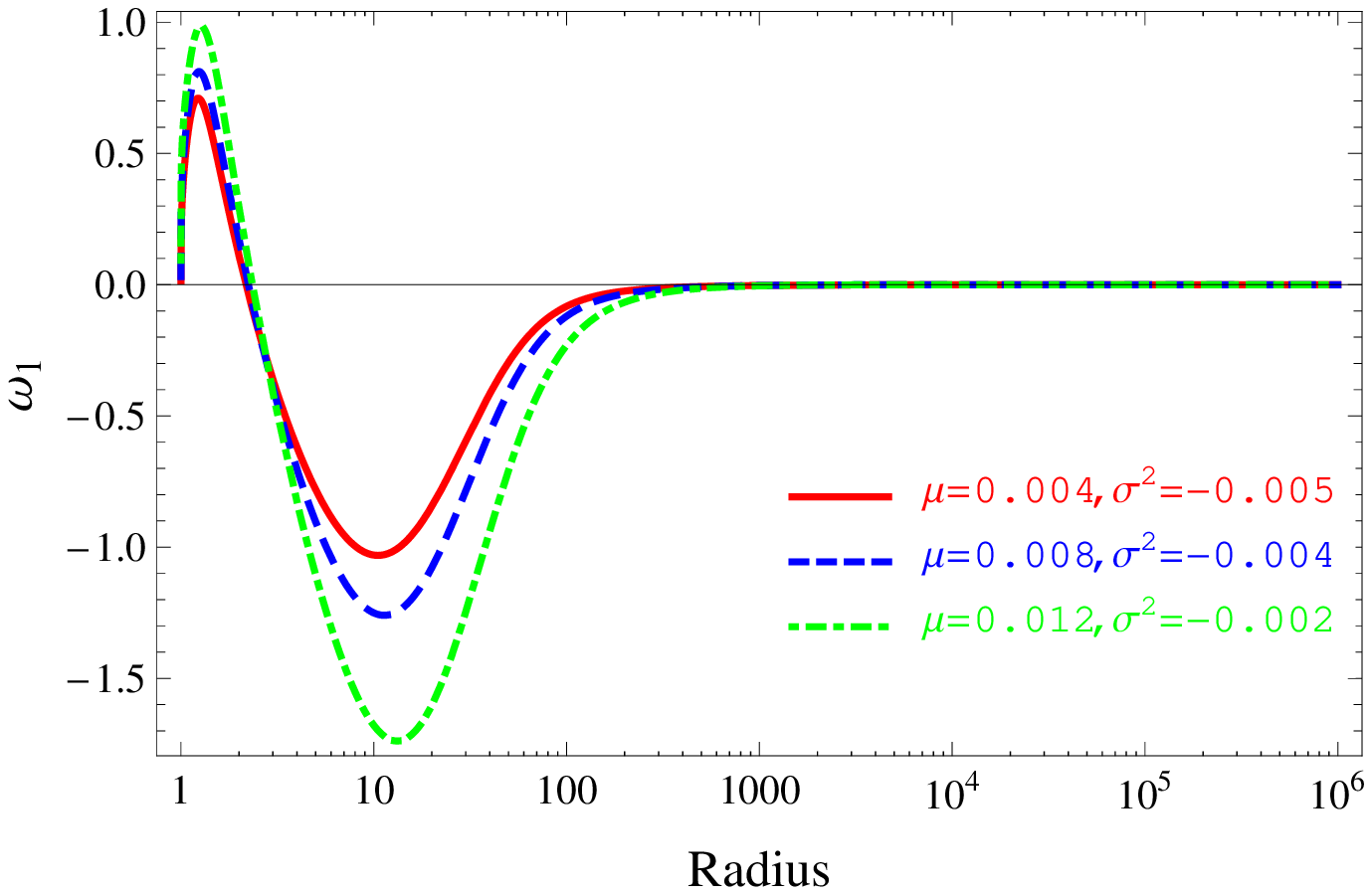}
\caption{Unstable perturbations $\omega _{1}$ for $n=2$ equilibrium black holes. Top row: nodeless perturbations with $N=0$.  Bottom row: perturbations having a single node, $N=1$.
Left-hand plots: fixed Proca field mass $\mu =0.01$ and varying cosmological constant $\Lambda $.
Right-hand plots: fixed $\Lambda = -0.0004$ and varying $\mu $.
The event horizon radius is fixed to be $r_{h}=1$.}
\label{fig:bhn2grav}
\end{figure*}

\begin{figure*}
\includegraphics[width=8.5cm]{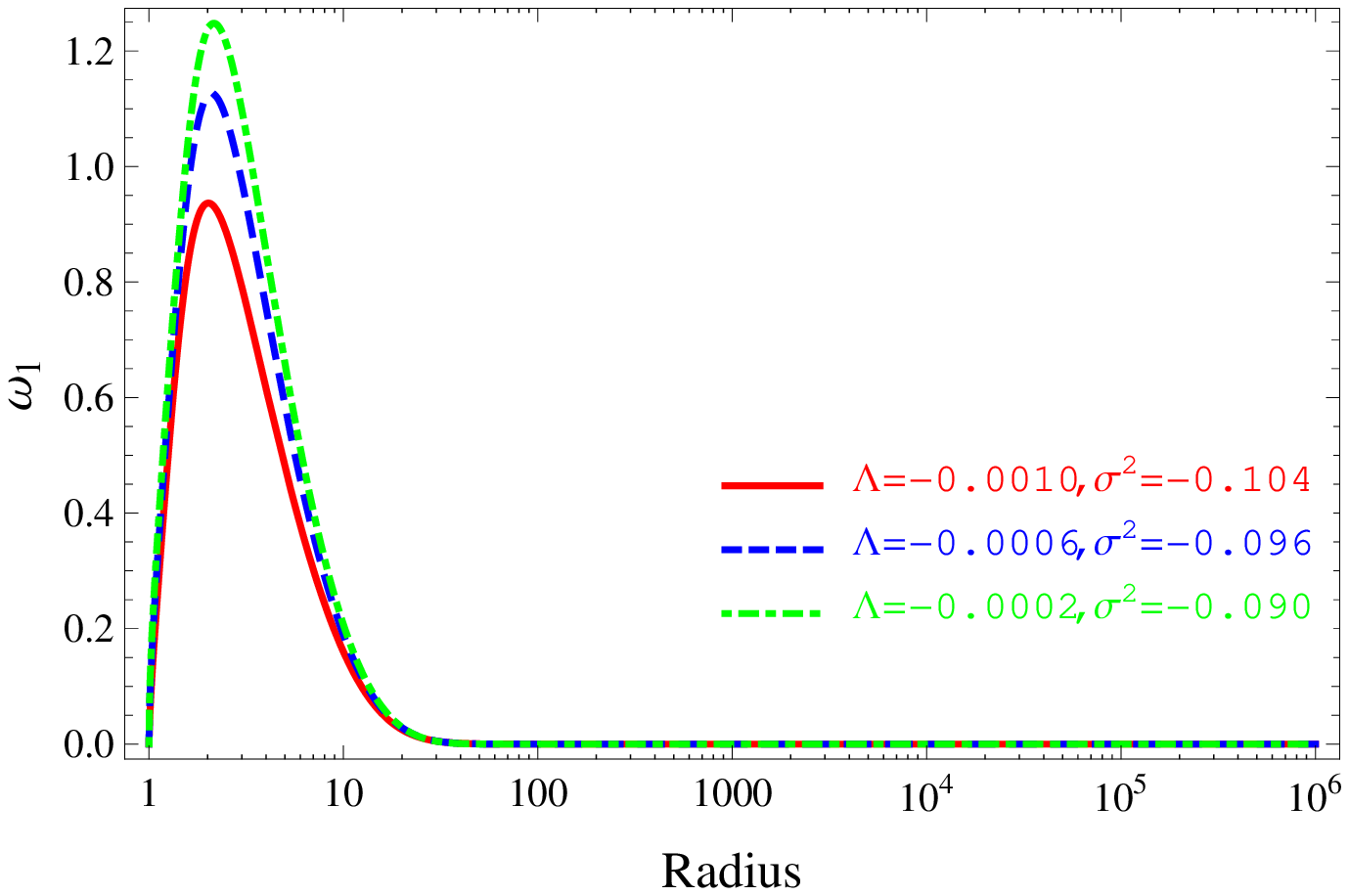}
\includegraphics[width=8.5cm]{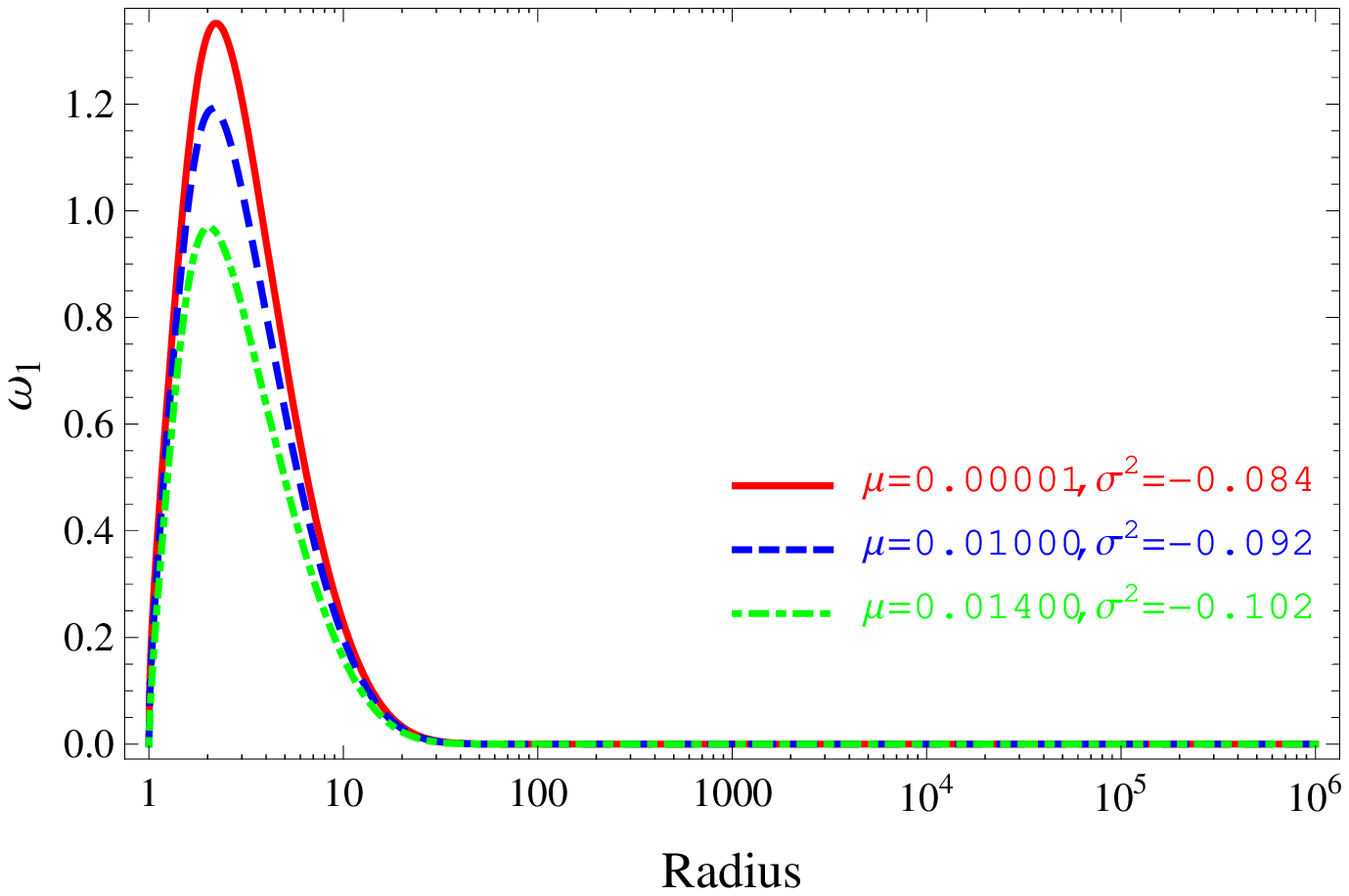}
\caption{Unstable perturbations $\omega _{1}$ for quasi-$n=1$ equilibrium black holes. All perturbations have $N=0$.
Left-hand plot: fixed Proca field mass $\mu =0.01$ and varying cosmological constant $\Lambda $.
Right-hand plot: fixed $\Lambda = -0.0004$ and varying $\mu $.
The event horizon radius is fixed to be $r_{h}=1$.}
\label{fig:bhqn1grav}
\end{figure*}

\begin{figure*}
\includegraphics[width=8.5cm]{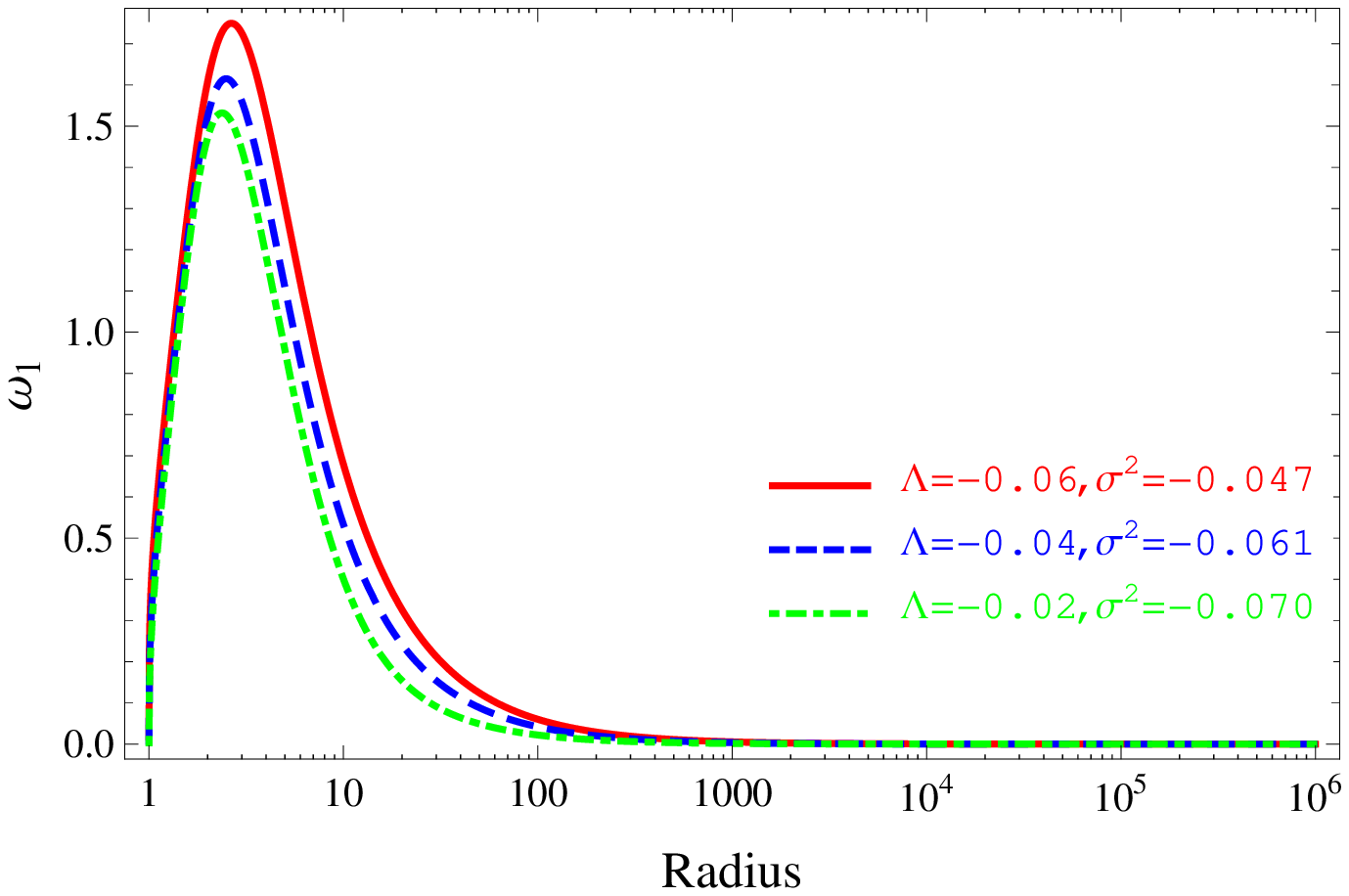}
\includegraphics[width=8.5cm]{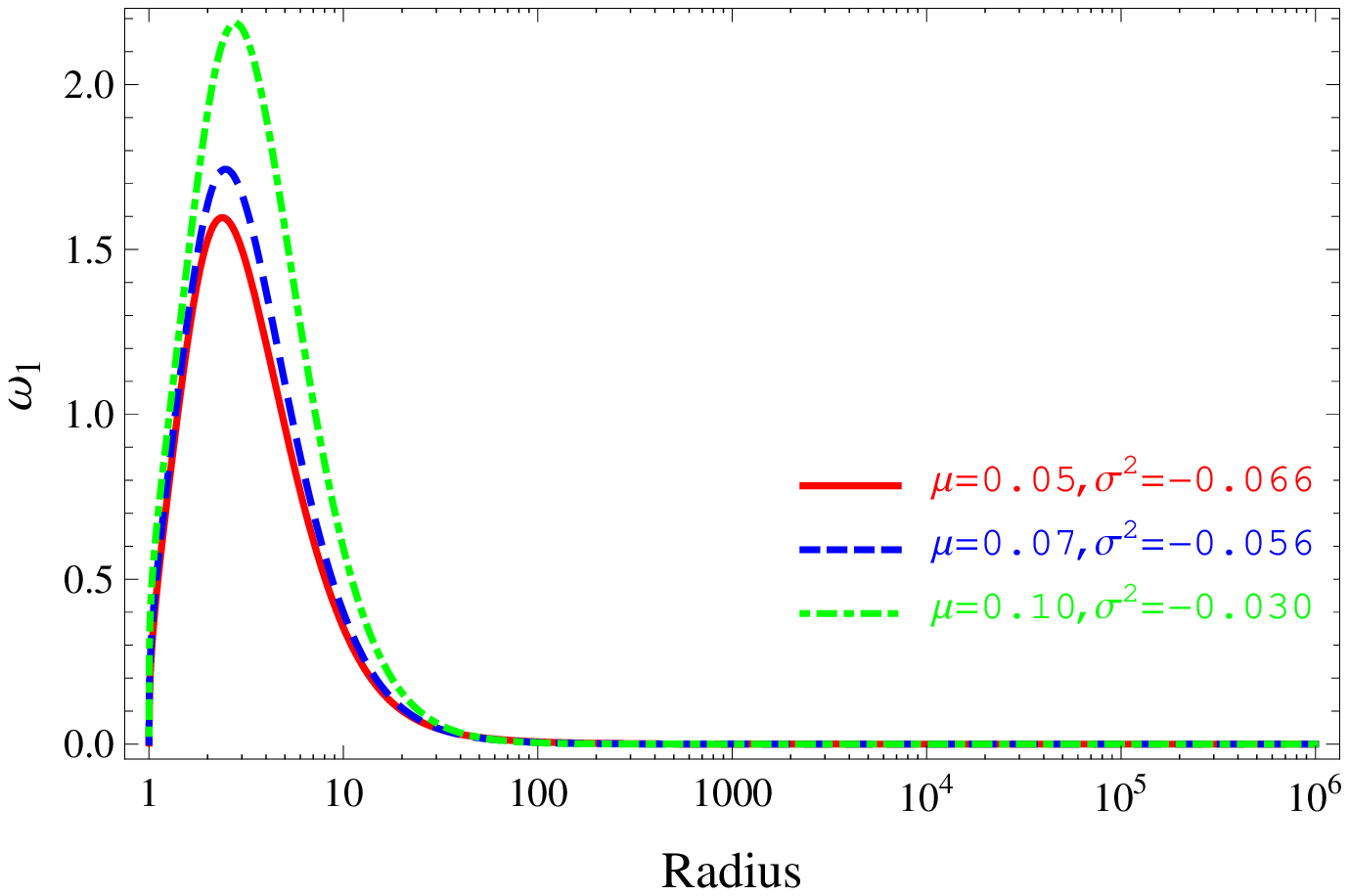}
\caption{Unstable perturbations $\omega _{1}$ for $n=1$ equilibrium black holes. All perturbations have $N=0$.
Left-hand plot: fixed Proca field mass $\mu =0.03$ and varying cosmological constant $\Lambda $.
Right-hand plot: fixed $\Lambda = -0.01$ and varying $\mu $.
The event horizon radius is fixed to be $r_{h}=1$.}
\label{fig:bhn1grav}
\end{figure*}

\subsubsection{Black holes}
\label{sec:BHstab}

We now turn to the stability of the equilibrium black hole solutions, beginning with the $n=2$ branch of solutions.
We fix the event horizon radius $r_{h}=1$ throughout this section.
All the black holes studied on the $n=2$ branch are unstable, and we find two unstable modes for each equilibrium black hole, one with no zeros and one with a single zero.
In Fig.~\ref{fig:bhn2grav} we show some unstable perturbations $\omega _{1}$ for $n=2$ black holes.
The top row shows perturbations for which the number of zeros of $\omega _{1}$ is zero, $N=0$, while the bottom row shows perturbations with $N=1$.
In the left-hand plots we have fixed the Proca field mass $\mu $ and varied $\Lambda $; in the right-hand plots the cosmological constant $\Lambda $ is fixed and $\mu $ varies.
As expected, the values of the eigenvalue $\sigma ^{2}$ are lower for the $N=0$ perturbations than they are for the $N=1$ perturbations.
We find a general trend for both the $N=0$ and $N=1$ perturbations, that the absolute value of $\sigma^{2}$ decreases as either $\left| \Lambda \right| $ increases for fixed $\mu $ or $\mu $ increases for fixed $\Lambda $.
For both the $N=1$ and $N=0$ perturbations, the overall shape of the perturbations does not change much as either $\mu $ or $\Lambda $ varies.
The $N=0$ perturbations have a peak close to the event horizon, while the $N=1$ perturbations have a peak close to the horizon and then a deeper trough at larger values of $r$. As observed for the $n=2$ branch of soliton solutions, the location of the zero of the $N=1$ perturbations does not change much as either $\mu $ or $\Lambda $ varies.

Next we consider the quasi-$n=1$ branch of black hole solutions.
As with the quasi-$n=1$ solitons, for all the quasi-$n=1$ black holes studied, we found a single unstable perturbation mode with $N=0$.
Some example perturbations are shown in Fig.~\ref{fig:bhqn1grav}, where in the left-hand plot we have fixed the Proca field mass $\mu $ and varied the cosmological constant $\Lambda $, while in the right-hand plot $\Lambda $ is fixed and $\mu $ varies.
The values of $\sigma^{2}$ that we find are slightly smaller in magnitude than those of the $N=0$ perturbations of the black holes on the $n=2$ branch of solutions.
The general shape of the perturbations is also similar to those of the $N=0$ perturbations of the $n=2$ branch of black holes.
We find that the absolute value of the eigenvalue $\sigma ^{2}$ decreases as either $\left| \Lambda \right| $ decreases for $\mu $ fixed, or $\mu $ decreases for $\Lambda $ fixed.
We found similar behaviour for the quasi-$n=1$ branch of soliton solutions.

As discussed in Sec.~\ref{sec:BH}, there are black hole solutions for which the equilibrium gauge potential function $\omega _{0}(r)$ has an odd number of zeros as well as black hole analogues of the soliton solutions with $\omega _{0}(r)$ having an even number of zeros.
We therefore next consider the stability of the $n=1$ branch of black holes.
On this branch we find that each equilibrium black hole has a single unstable gravitational sector perturbation, see Fig.~\ref{fig:bhn1grav} for some examples.
On this branch of solutions, the eigenvalues $\sigma ^{2}$ that we find have magnitudes which are smaller than those for the $N=0$ perturbations of both the $n=2$ and quasi-$n=1$ branches of black hole solutions.
In a similar way to the other branches of black hole solutions, $\left| \sigma ^{2} \right| $ increases as $\left| \Lambda \right| $ decreases for fixed $\mu $, and decreases as $\mu $ increases for fixed $\Lambda $.
Once again the shape of the perturbations does not change much as either $\mu $ or $\Lambda $ varies, and is very similar to the other $N=0$ perturbations of black holes, shown in Figs.~\ref{fig:bhn2grav} and \ref{fig:bhqn1grav}.

It remains to consider the stability of the quasi-$n=0$ branch of black hole solutions.
However, for all equilibrium solutions considered on this branch, we were unable to find any perturbations satisfying the boundary conditions (\ref{pertBCs})
with $\sigma ^{2}<0$.
Thus it appears that black hole solutions on the quasi-$n=0$ branch have no unstable modes in the gravitational sector.
We will examine the stability of this branch of solutions under sphaleronic sector perturbations in Sec.~\ref{sec:sphal}.

\subsubsection{General properties of the gravitational sector perturbations}
\label{sec:gengrav}

Before discussing the sphaleronic sector perturbations in the next section, we now summarize our results on the gravitational sector perturbations.
All the equilibrium solitons studied possess unstable gravitational sector perturbations.
For equilibrium black holes, we found unstable gravitational sector perturbations for all solutions on the $n=2$, quasi-$n=1$ and $n=1$ branches, but were unable to find any unstable modes for solutions on the quasi-$n=0$ branch.

Asymptotically flat solutions of pure ${\mathfrak {su}}(2)$ EYM theory for which the gauge potential function $\omega (r)$ has $n$ zeros possess $n$ unstable modes in the gravitational sector \cite{Lavrelashvili:1994rp}.
In contrast, for asymptotically flat solutions of ${\mathfrak {su}}(2)$ EYMH the number of unstable modes in the gravitational sector depends on the branch of solutions under consideration as well as the number of zeros of $\omega (r)$ \cite{Mavromatos:1995kc}.
Solitons and black holes on the $n=i$ branch of solutions (with $i=1,2,\ldots $) have $i$ unstable modes in the gravitational sector; while those on the quasi-$n=i-1$ branch have $i-1$ unstable gravitational sector modes.
We have found that the various branches of ENAP-AdS solitons and black holes have the same number of unstable modes in the gravitational sector as the asymptotically flat EYMH solutions.
This includes the quasi-$n=0$ branch of black hole solutions, where we have not found any unstable modes in the gravitational sector.

We find that the quasi-$n=i-1$ branches of solutions, as well as having fewer unstable gravitational sector modes than the corresponding $n=i$ branches of solutions, also have lowest eigenvalues $\sigma ^{2}$ which have a smaller magnitude than the lowest $\sigma ^{2}$ for the $n=i$ branch of solutions.
Therefore the timescales for the instability of the quasi-$n=i-1$ branches of solutions are longer than for the corresponding $n=i$ branches of solutions.
For black hole solutions, we also find that the $n=1$ branch has lowest eigenvalues $\sigma ^{2}$ with smaller absolute values than either the $n=2$ or quasi-$n=1$ branches.

Interestingly, the lowest eigenvalues $\sigma ^{2}$ that we find for all the unstable black holes considered (the $n=2$, quasi-$n=1$ and $n=1$ branches) have a much smaller absolute value than those for the corresponding solitons.
Thus it appears that the black hole solutions decay on rather longer timescales than the solitons.

\subsection{Sphaleronic sector}
\label{sec:sphal}

We now turn to the sphaleronic sector of perturbations $(a_{1}, b_{1}, d_{1})$, governed by the equations (\ref{perturbedNAPt}, \ref{perturbedNAPr}, \ref{perturbedNAPtheta}).

In pure EYM theory \cite{Straumann:1989tf,Winstanley:1998sn,Bjoraker:1999yd} and in EYMH theory \cite{Maeda:1993ap,Mavromatos:1995kc,VanderBij:2001ah,Winstanley:2016taz}
it is possible to make a gauge transformation of the form (\ref{gaugetransform1}) to set $a_{1}\equiv 0$, which simplifies the perturbation equations in the sphaleronic sector.  In ENAP theory, the additional constraint (\ref{perturbedgaugecon}) restricts our choice of gauge for the perturbations. We therefore
take an alternative approach, following \cite{Nolan:2015vca}.

First we introduce new variables $(\psi, \xi, \eta, \gamma )$, defined by
\begin{align}
\psi & =  a^{\prime}_1-\dot{b}_1 ,
\nonumber \\
\xi & = a_1+\dot{\gamma} ,
\nonumber \\
\eta & = b_1+\gamma^{\prime},
\nonumber \\
\omega_0 \gamma & = d_{1}.
\label{newvars}
\end{align}
Under an infinitesimal gauge transformation of the form (\ref{gaugetransform1}), we have
\begin{equation}
\gamma \rightarrow \gamma + \beta ,
\end{equation}
while $\psi $, $\xi $ and $\eta $ are unchanged.

The sphaleronic sector perturbation equations (\ref{perturbedNAPt}, \ref{perturbedNAPr}, \ref{perturbedNAPtheta}) can be rewritten compactly in terms of these new variables as
\begin{subequations}
\label{sphalset1}
\begin{align}
0 = & -r^2R_0 S_0\psi^{\prime} + rR_0\left(rS^{\prime}_0-2S_0\right)\psi -S_0r^2\mu^2\dot{\gamma}
\nonumber \\
& +S_0\left(r^2\mu^2+2\omega^{2}_0\right)\xi,
\label{NAPtGauge}\\
0 = & \frac{1}{R_0S^{2}_0}\dot{\psi}-\left(\mu^2+\frac{2\omega^{2}_0}{r^2}\right)\eta +\mu^2\gamma^{\prime},
\label{NAPrGauge} \\
0 = & -\mu^2\gamma -R_0\omega_0\eta^{\prime}-\left(2R_0\omega^{\prime}_0+\frac{\omega_0(R_0S_0)^{\prime}}{S_0}\right)\eta
\nonumber \\
 & +\frac{\omega_0}{R_0S^{2}_0}\dot{\xi}.
\label{NAPthetaGauge}
\end{align}
\end{subequations}
Setting the Proca field mass $\mu =0$, the equations (\ref{sphalset1}) reduce to those for sphaleronic sector perturbations in pure EYM theory, written in terms of $\psi $, $\xi $ and $\eta $.
The key difference here is the presence of the non-gauge-invariant quantity $\gamma $.
Our strategy  is  therefore to eliminate the gauge-dependent variable $\gamma $ from the sphaleronic sector perturbations, leaving a set of perturbation equations for gauge-independent quantities only.
Using the new variables (\ref{newvars}), the constraint (\ref{perturbedgaugecon}) takes the form
\begin{align}
0 = &
\frac {1}{R_{0}S_{0}^{2}} \left( {\dot {\xi }} - {\ddot {\gamma }} \right)
- R_{0} \left( \eta ^{\prime } - \gamma ^{\prime \prime } \right)
\nonumber \\ &
- \left( \frac {2R_{0}}{r} + \frac {\left( R_{0}S_{0} \right) ^{\prime }}{S_{0}} \right) \left( \eta - \gamma ' \right)
+ \frac {2\omega _{0}}{r^{2}} \gamma .
\end{align}
However, we do not need to consider this equation further since it is a consequence of the equations (\ref{sphalset1}).

\begin{figure*}
\includegraphics[width=8.5cm]{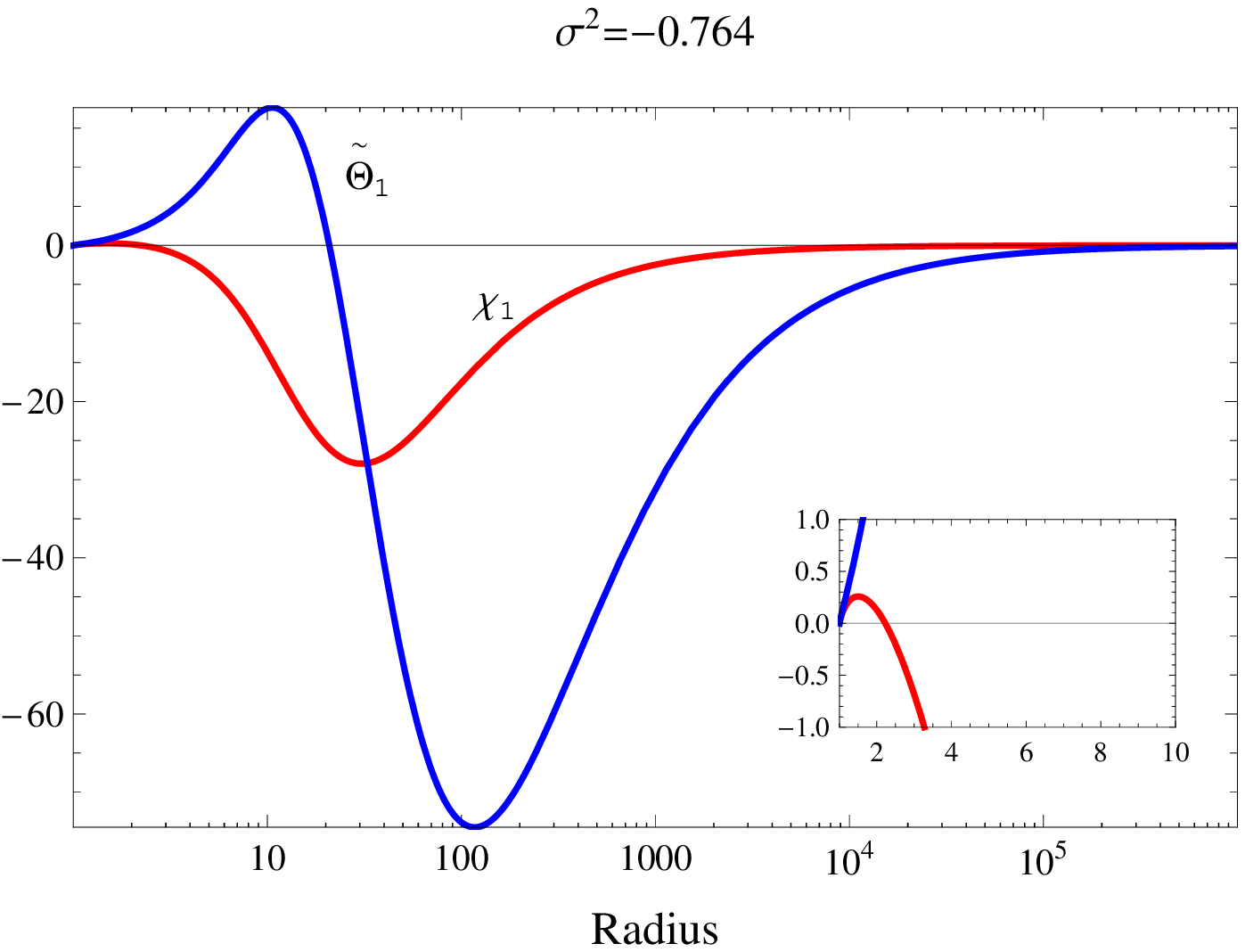}
\includegraphics[width=8.5cm]{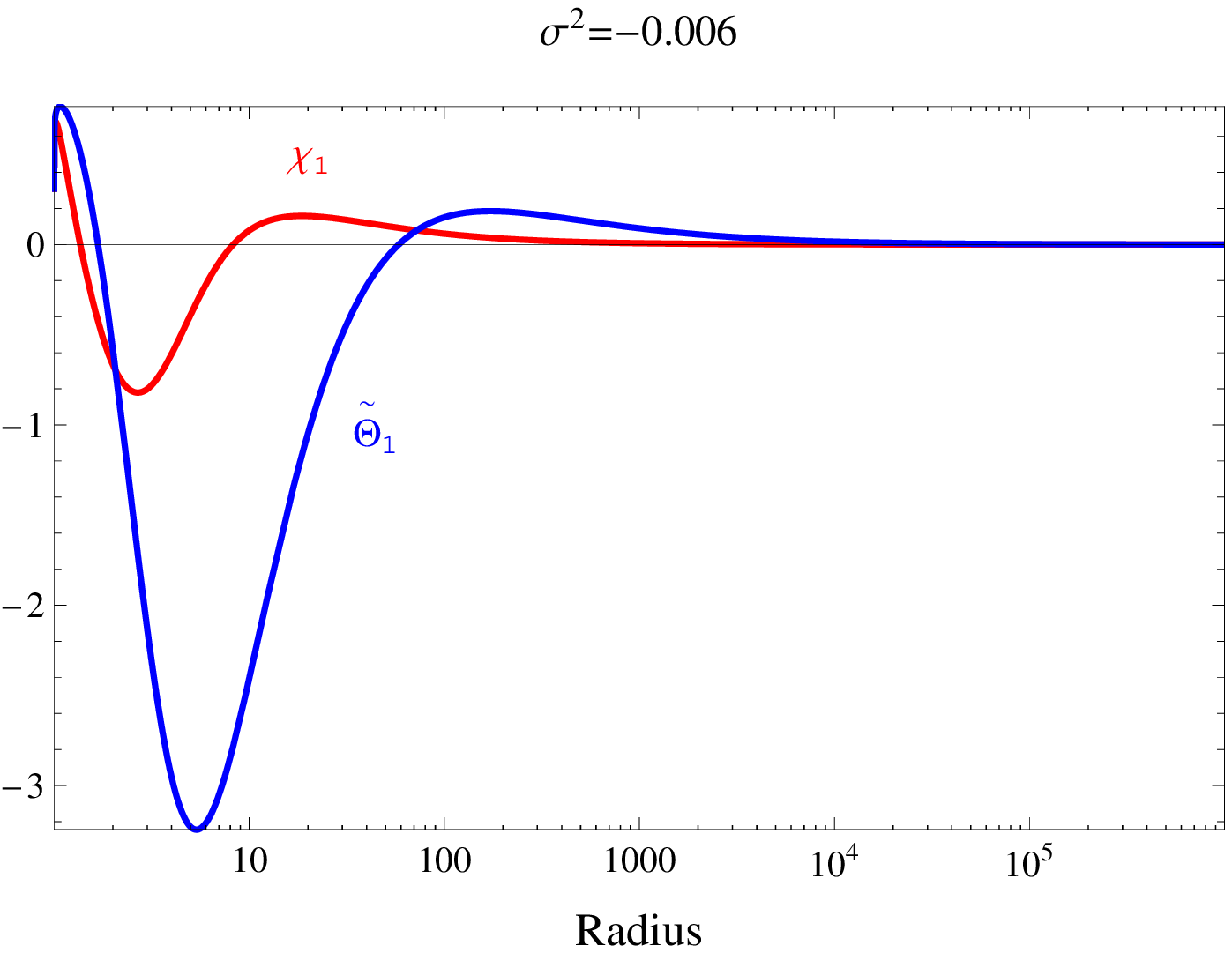}
\caption{$N=1$ (left) and $N=2$ (right) unstable perturbations of a quasi-$n=0$ black hole solution with $\mu = 0.03$ and $\Lambda = -0.075$. A subplot in the left-hand panel shows the behaviour of the perturbations near the event horizon.  The quantities $\chi _{1}$ and ${\tilde {\Theta }}_{1}$ are denoted by red and blue colours respectively.  The quantity ${\tilde {\Theta }}_{1}$ has a deeper trough than $\chi _{1}$.
The event horizon radius is fixed to be $r_{h}=1$.}
\label{fig:bhsex}
\end{figure*}

Next we define further new variables $\chi $ and $\Theta $ by
\begin{equation}
\psi = {\dot {\chi }}, \qquad
\xi = {\dot {\Theta }},
\label{chiTheta}
\end{equation}
where we are free to add an arbitrary function of the radial coordinate $r$ only to $\chi $ and $\Theta $.
Substituting in for $\psi $ and $\xi $ from (\ref{chiTheta}) into (\ref{NAPtGauge}), and performing an integration with respect to time gives
\begin{align}
\label{NAPtGauge1}
\mu^2\gamma &= \left(\mu^2 + \frac{2\omega^2_0}{r^2}\right)\Theta + R_0\left(\frac{S'_0}{S_0} - \frac{2}{r}\right)\chi - R_0 \chi',
\end{align}
where we have used the freedom in the definition of $\chi $ and $\Theta $ to set an arbitrary function of $r$ to zero.

Next we note that (\ref{newvars}, \ref{chiTheta}) imply that
\begin{equation}
{\dot {\eta }} = -{\dot {\chi }} + {\dot {\Theta }}^{\prime },
\end{equation}
and hence
\begin{equation}
\eta = - \chi + \Theta ^{\prime },
\label{eta}
\end{equation}
where we have used the remaining freedom in the definition of $\chi $ and $\Theta $ to set an arbitrary function of $r$ to zero.

We now use (\ref{NAPtGauge1}, \ref{eta}) to eliminate $\gamma $ and $\eta $ from the perturbation equations (\ref{NAPrGauge}, \ref{NAPthetaGauge}), obtaining the following pair of coupled perturbation equations:
\begin{widetext}
\begin{subequations}
\label{sphaltemp}
\begin{align}
0 = & -\frac{\ddot{\chi}}{R_0S^2_0} + R_0\chi'' + \left(R'_0 + \frac{2R_0}{r} - \frac{R_0S'_0}{S_0}\right)\chi' - \left(\mu^2 + \frac{2\omega^2_0}{r^2} + \frac{2R_0}{r^2} -\frac{2R'_0}{r}+ \left(\frac{R_0S'_0}{S_0}\right)^{\prime}\right)\chi
- \left(\frac{2\omega^2_0}{r^2}\right)^{\prime}\Theta,
\label{chiequation} \\
0 = & \frac{\ddot{\Theta}}{R_0S^2_0} - R_0\Theta'' - \left(\frac{\left(R_0S_0\right)'}{S_0} + \frac{2R_0\omega'_0}{\omega_0}\right)\Theta' - \left(\mu^2 + \frac{2\omega^2_0}{r^2}\right)\frac{\Theta}{\omega_0} +\frac{R_0(1+\omega_0)}{\omega_0}\chi'
\nonumber \\ &
 + \left(\frac{\left(R_0S_0\right)'}{S_0} + \frac{2R_0\omega'_0}{\omega_0} + \frac{2R_0}{r\omega_0} - \frac{R_0S'_0}{\omega_0S_0}\right)\chi .
\label{Thetaequation}
\end{align}
\end{subequations}
\end{widetext}
The equations (\ref{sphaltemp}) have a singularity when $\omega _{0}(r)$ has a zero. To eliminate this, we define a further new variable ${\tilde {\Theta }}$ by
\begin{equation}
{\tilde {\Theta }}(t,r) = \omega _{0}(r) \Theta (t,r).
\end{equation}
Finally, we assume that the perturbations are periodic in time:
\begin{align}
\chi(t,r) &= e^{-i\sigma t}\chi_1(r), \qquad {\tilde {\Theta }}(t,r) = e^{-i\sigma t}{\tilde {\Theta }}_1(r).
\end{align}
The sphaleronic sector perturbation equations (\ref{sphaltemp}) then become
\begin{widetext}
\begin{subequations}
\label{sphalfinal}
\begin{align}
\sigma^2\chi_1 = & - R^2_0S^2_0\chi''_1 - R_0S^2_0\left(\frac{2R_0}{r}+R'_0 -\frac{R_0S'_0}{S_0}\right)\chi_1' + R_0S^2_0\left(\mu^2 + \frac{2\omega^2_0}{r^2} + \frac{2R_0}{r^2}
-\frac{2R'_0}{r}
+ \left(\frac{R_0S'_0}{S_0}\right)^{\prime}\right)\chi_1
\nonumber \\ &
 + \omega_0R_0S^2_0\left(\frac{2\omega^2_0}{r^2}\right)'\tilde{\Theta}_1,
 \label{matterperturbationeq1a} \\
\sigma^2\tilde{\Theta}_1 = &  -R^2_0S^2_0\tilde{\Theta}_1^{\prime\prime} - R_0S_0(R_0S_0)'\tilde{\Theta}_1' + R_0S^2_0\left(\mu^2 - \frac{1}{r^2} - \frac{2\omega_0}{r^2}+\frac{\omega^2_0}{r^2}\right)\tilde{\Theta}_1 + R^2_0S^2_0(1+\omega_0)\chi_1'
\nonumber \\
& + R_0S^2_0\left(\frac{\left(R_0S_0\right)'\omega_0}{S_0} + 2R_0\omega'_0 + \frac{2R_0}{r} - \frac{R_0S'_0}{S_0}\right)\chi_1 .
\label{matterperturbationeq3}
\end{align}
\end{subequations}
\end{widetext}

To integrate the equations (\ref{sphalfinal}) numerically, boundary conditions must be imposed on the quantities $\chi _{1}$ and ${\tilde {\Theta }}_{1}$.
We set:
\begin{align}
\label{sphalBCs}
\chi_1, {\tilde {\Theta }}_{1} \sim &
\begin{cases}
r^{3} & {\mbox {for $r\rightarrow 0$,}} \\
(r-r_h)^\beta &  \mbox{for $r \rightarrow r_h$,} \\
r^\rho & \mbox{for $r \rightarrow \infty$.}
\end{cases}
\end{align}
The constants $\beta$ and $\rho$ are defined in the same way as (\ref{powerofboundarycondition}).
The behaviour of the quantities $\chi _{1}$ and ${\tilde {\Theta }}_{1}$ as $r\rightarrow r_{h}$ and $r\rightarrow \infty $ is the same as that of the gravitational sector perturbation $\omega _{1}$ (\ref{powerofboundarycondition0}).
However, the behaviour of the sphaleronic sector quantities is different from that of $\omega _{1}$ as $r\rightarrow 0$.

\begin{figure*}
\includegraphics[width=8.5cm]{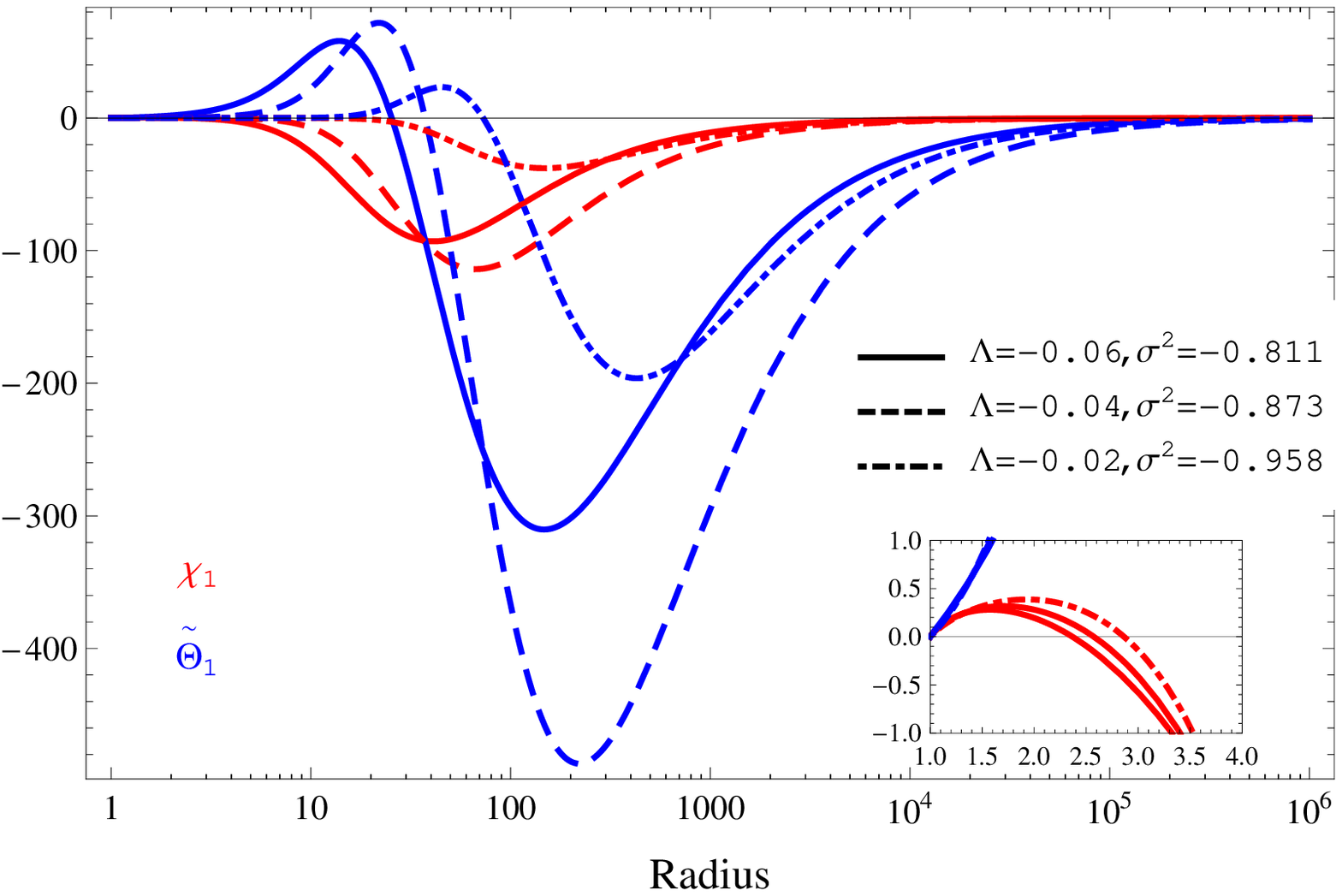}
\includegraphics[width=8.5cm]{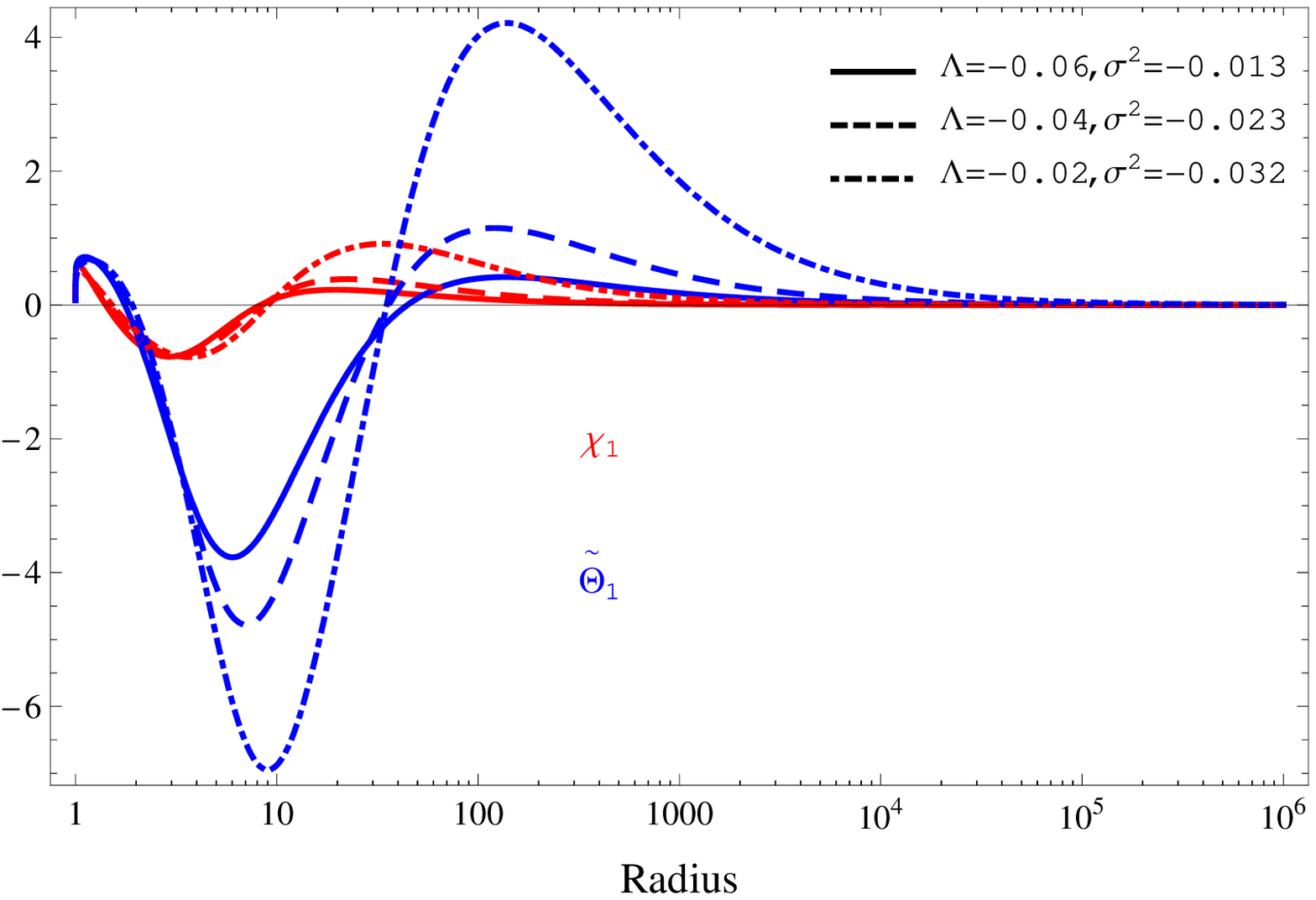}
\caption{$N=1$ (left) and $N=2$ (right) unstable perturbations of quasi-$n=0$ black hole solutions with fixed $\mu = 0.03$ and varying $\Lambda $. A subplot in the left-hand panel shows the behaviour of the perturbations near the event horizon.  The quantities $\chi _{1}$ and ${\tilde {\Theta }}_{1}$ are denoted by red and blue colours respectively.  The quantity ${\tilde {\Theta }}_{1}$ has a deeper trough than $\chi _{1}$.
The event horizon radius is fixed to be $r_{h}=1$.}
\label{fig:bhsvarylambda}
\end{figure*}

\begin{figure}
\includegraphics[width=8.5cm]{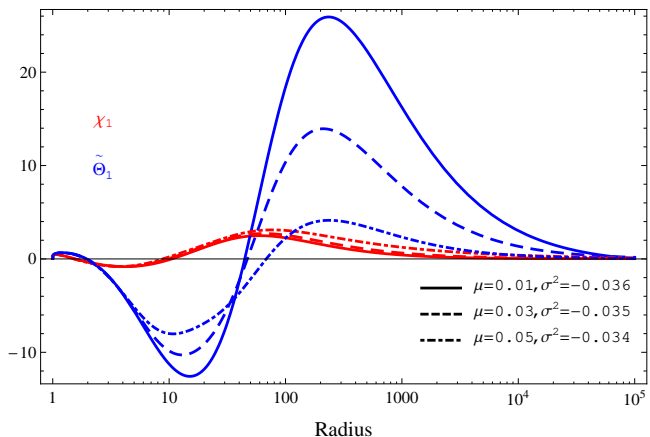}
\caption{$N=2$ unstable perturbations of quasi-$n=0$ black hole solutions with fixed $\Lambda  = -0.01$ and varying $\mu $. The quantities $\chi _{1}$ and ${\tilde {\Theta }}_{1}$ are denoted by red and blue colours respectively.  The quantity ${\tilde {\Theta }}_{1}$ has a deeper trough than $\chi _{1}$.
For these solutions we do not find any $N=1$ perturbation modes.
The event horizon radius is fixed to be $r_{h}=1$.}
\label{fig:bhsvarymu}
\end{figure}

By studying gravitational sector perturbations, in Sec.~\ref{sec:grav} we showed that solitons on both the $n=2$ and quasi-$n=1$ branches were unstable. We also showed that black hole solutions on the $n=2$, quasi-$n=1$ and $n=1$ branches are also unstable.
However, we did not find any unstable modes in the gravitational sector of perturbations for black holes on the quasi-$n=0$ branch of solutions.
Therefore in this section we consider just the quasi-$n=0$ branch of black hole solutions, since we already know that the other equilibrium solutions are unstable.
We set the event horizon radius $r_{h}=1$ for the rest of this section.
We follow the standard shooting method, seeking eigenvalues $\sigma ^{2}$ such that the perturbations $\chi _{1}$, ${\tilde {\Theta }}_{1}$ satisfy the boundary conditions (\ref{sphalBCs}).
The pair of coupled perturbation equations (\ref{sphalfinal}) are not in self-adjoint form, and we have been unable to find a transformation yielding a set of self-adjoint equations.
Therefore it is not {\it {a priori}} necessarily the case that the eigenvalue $\sigma ^{2}$ is real.
However, for all solutions investigated, we find that $\sigma ^{2}$ is real.
Since the perturbation equations are not self-adjoint, it is also not necessarily the case that the eigenfunctions corresponding to the lowest eigenvalue $\sigma ^{2}$ have no zeros.

All the black holes studied on the quasi-$n=0$ branch are unstable, in each case we find an eigenvalue $\sigma ^{2}<0$, corresponding to a perturbation mode which grows exponentially in time.
In Fig.~\ref{fig:bhsex} we consider one particular quasi-$n=0$ equilibrium black hole solution with fixed $\mu = 0.03$ and $\Lambda =-0.075$.  For this
particular solution, we find two negative eigenvalues $\sigma ^{2}$.
The perturbations $\chi _{1}$ and ${\tilde {\Theta }}_{1}$ corresponding to the lowest value of $\sigma ^{2}$ each have one zero and are shown in the left-hand plot in Fig.~\ref{fig:bhsex}; those corresponding to the higher value of $\sigma ^{2}$ each have two zeros and are shown in the right-hand plot.
In the left-hand plot, the zero of $\chi _{1}$ is very close to the horizon and can be seen in the subplot.
As with the gravitational sector perturbations, we denote the number of zeros of either $\chi _{1}$ or ${\tilde {\Theta }}_{1}$ by $N$.
The $N=1$ perturbations shown in Fig.~\ref{fig:bhsex} have a peak close to the event horizon, and then a minimum at larger values of $r$.
The $N=2$ perturbations also have a peak close to the event horizon, then a trough and finally another maximum further away from the horizon.
In both cases the perturbation ${\tilde {\Theta }}$ has a deeper trough than $\chi _{1}$.

In Figs.~\ref{fig:bhsvarylambda} and \ref{fig:bhsvarymu} we investigate how the perturbations and $\sigma ^{2}$ change as either $\Lambda $ or $\mu $ varies.
Fig.~\ref{fig:bhsvarylambda} shows the perturbations $\chi _{1}$ and ${\tilde {\Theta }}_{1}$ for a selection of quasi-$n=0$ black holes with fixed Proca field mass $\mu = 0.03$ and varying cosmological constant $\Lambda $.
For the black holes shown in Fig.~\ref{fig:bhsvarylambda}, we find two negative values of $\sigma ^{2}$ for each equilibrium solution, corresponding to perturbations with $N=1$ (left-hand plot) and $N=2$ (right-hand plot).
The perturbations have the same general shape as those shown in Fig.~\ref{fig:bhsex}, but show more variation as $\Lambda $ varies than the gravitational sector perturbations discussed in Sec.~\ref{sec:grav}.
The $N=1$ perturbations correspond to eigenvalues $\sigma ^{2}$ with larger magnitudes than those for the $N=2$ perturbations.
We also find that as $\left| \Lambda \right| $ increases, the absolute value of $\sigma ^{2}$ decreases.

Finally in this section, Fig.~\ref{fig:bhsvarymu} shows the perturbations for a selection of quasi-$n=0$ black holes with fixed $\Lambda = -0.01$ and varying $\mu $.
For this selection of black holes, we were only able to find a single negative value of $\sigma ^{2}$, with corresponding perturbations having two zeros.
It is not clear whether this is a numerical issue or whether the $N=1$ perturbations shown in Figs.~\ref{fig:bhsex} and \ref{fig:bhsvarylambda} do not exist for these black holes.
Our main conclusion that all the quasi-$n=0$ black holes are unstable is however unchanged.
In Fig.~\ref{fig:bhsvarymu} the $N=2$ perturbations have the same general shape as those in Figs.~\ref{fig:bhsex} and \ref{fig:bhsvarylambda}.
We find that the magnitude of the eigenvalue $\sigma ^{2}$ decreases as the Proca field mass $\mu $ increases.
For all the quasi-$n=0$ black holes studied in this section, the values of $\sigma ^{2}$ that we find have similar magnitudes to those found in the gravitational sector in Sec.~\ref{sec:BHstab}.

In this section, all the quasi-$n=0$ equilibrium black holes we have studied are unstable.
Combining this with the results of Sec.~\ref{sec:grav}, we deduce that the solitons on the $n=2$ and quasi-$n=1$ branches are unstable, as are black holes on the $n=2$, quasi-$n=1$, $n=1$ and quasi-$n=0$ branches.
We expect that these results would extend to branches of solutions in which the equilibrium gauge potential function $\omega _{0}$ has more than two zeros, so that all spherically symmetric soliton and black hole solutions of ENAP theory in asymptotically AdS space-time are unstable.

\section{Conclusions}
\label{sec:conc}

In this paper we have presented new soliton and black hole solutions of ENAP theory in asymptotically AdS space-time.
With gauge group ${\mathfrak {su}}(2)$, we have shown that nontrivial static, spherically symmetric gauge field configurations must be purely magnetic and can be described by a single function $\omega (r)$.
Furthermore, we have shown that $\omega (r)$ must have at least one zero. For soliton solutions, the number of zeros of $\omega (r)$ is even, but there is no such restriction for black hole solutions.
For the configurations to have finite total energy, the asymptotically AdS boundary conditions imply that the gauge field must be in its vacuum state at infinity, so that $\omega (r)\rightarrow -1 $ as $r\rightarrow \infty $.

Numerical solutions of the field equations representing regular solitons and black holes are found using a standard shooting method.
Solutions are found at discrete points in the parameter space for fixed Proca field mass $\mu $ and cosmological constant $\Lambda <0$.
For fixed $\Lambda $, there is a maximum value of $\mu $ for which we find nontrivial solutions; similarly for fixed $\mu $ there is a maximum value of $\left| \Lambda \right| $ for which nontrivial solutions exist.

Fixing $n$, the number of zeros of the gauge field function $\omega (r)$, we find two branches of solutions, which, by analogy with the asymptotically flat ENAP and EYMH solutions \cite{Greene:1992fw} and the asymptotically-AdS EYMH solutions \cite{VanderBij:2001ah} we dub the $n=i$ and quasi-$n=i-1$ branches.
For fixed $\Lambda $, the two branches merge at the maximum value of $\mu $.
We have explored in detail the soliton solutions for which $\omega (r)$ has two zeros; and the black hole solutions where $\omega (r)$ has either one or two zeros.
We anticipate that solutions for which $\omega (r)$ has more than two zeros also exist.

As with the pure EYM and EYMH systems, the linearized ENAP perturbation equations decouple into two sectors, the gravitational and sphaleronic sectors.
All the soliton solutions studied, and the black holes on the $n=2$, quasi-$n=1$, and $n=1$ branches have instabilities in the gravitational sector of perturbations.
However, we were unable to find any unstable gravitational sector perturbations for black hole solutions lying on the quasi-$n=0$ branch.
We therefore studied the sphaleronic sector of perturbations for quasi-$n=0$ black holes, and all solutions studied had unstable modes in this sector.
The perturbation equations are sufficiently complicated that numerical analysis is necessary, and therefore our stability analysis only applies to equilibrium solutions for which the gauge potential function $\omega (r)$ has either one or two zeros.
We expect that equilibrium solutions with $\omega (r)$ having more than two zeros will also be unstable.

Since the gauge field is in the vacuum configuration at infinity, far from the event horizon the black hole solutions we find are indistinguishable from Schwarzschild-AdS black holes.
These black holes are therefore counter-examples to the ``no-hair'' conjecture, in a similar way to the asymptotically flat pure EYM coloured black holes \cite{Bizon:1990sr}.
However, the pure EYM coloured black holes are unstable \cite{Straumann:1989tf,Lavrelashvili:1994rp} and, as a result, Bizon formulated a generalized no-hair conjecture, which states that \cite{Bizon:1994dh}
\begin{quotation}
Within a given matter model, a {\em {stable}} stationary black hole is uniquely determined by global charges.
\end{quotation}
The pure EYM coloured black holes satisfy this generalized no-hair conjecture since they are unstable.
The asymptotically flat ENAP and EYMH solitons and black holes are also unstable \cite{Greene:1992fw,Maeda:1993ap,Mavromatos:1995kc},
as are the asymptotically AdS EYMH solitons \cite{VanderBij:2001ah} and black holes \cite{Winstanley:2016taz}.
As conjectured in \cite{VanderBij:2001ah,Winstanley:2016taz}, it is therefore not surprising that all the asymptotically AdS ENAP solitons and black holes studied in this paper are also unstable.

In contrast with the ENAP and EYMH systems, there exist stable soliton and black hole solutions of pure EYM theory in asymptotically AdS space-time \cite{Winstanley:1998sn,Bjoraker:1999yd}.
A natural question is what is special about EYM theory which permits the existence of stable solutions, while its generalizations ENAP and EYMH do not?
We argue that the boundary conditions satisfied by the gauge field at infinity are crucial.
In the EYM case, for asymptotically AdS space-times the boundary conditions on the gauge field as $r\rightarrow \infty $ are not very restrictive: the gauge function $\omega (r)$ must tend to a constant, but that constant is arbitrary.
This means that the gauge field near the AdS boundary is not necessarily in its vacuum configuration.
On the other hand, for EYM in asymptotically flat space-time, it must be the case that $\omega (r) \rightarrow \pm 1$ as $r\rightarrow \infty $, which is much more restrictive and in particular means that the gauge field is in its vacuum configuration at infinity.
In the ENAP and EYMH models, in both asymptotically flat and asymptotically AdS space-times, as $r\rightarrow \infty $ it must be the case that $\omega (r)$ approaches its vacuum value $-1$ (there is also a boundary condition on the Higgs field in the EYMH model, but that is less important for our discussion here).

From the point of view of Bizon's generalized no-hair conjecture, if the non-Abelian gauge field is in the vacuum configuration at infinity, there can be no nonzero non-Abelian charges to distinguish the black holes from the embedded Schwarzschild or Schwarzschild-AdS solutions.
On the other hand, if the gauge field has a nonvacuum configuration at infinity, then one would expect the existence of nonzero charges defined far from the black hole.
Indeed, in the pure EYM case, it has been argued \cite{Shepherd:2012sz} that such non-Abelian charges uniquely characterize at least a subset of stable asymptotically AdS hairy black holes.
Thus we have a consistent picture of black holes in the EYM/ENAP/EYMH models, in accordance with the generalized no-hair conjecture:  black holes which are indistinguishable from Schwarzschild(-AdS) at infinity are unstable; and there appears to be a set of global charges uniquely characterizing stable hairy black holes.

From a physical point of view, the massless nature of the non-Abelian gauge field in the EYM model seems to be crucial in asymptotically AdS space-times.
Although the AdS boundary is at an infinite proper distance from the origin, it can be reached in a finite affine parameter by a null geodesic.
Therefore massless fields can extend all the way to the AdS boundary, as is the case in the EYM model.
This can be seen in the boundary condition on the gauge field as $r\rightarrow \infty $ \cite{Winstanley:1998sn,Bjoraker:1999yd}, which in the pure EYM case means that the gauge potential function $\omega (r)$ approaches its asymptotic value slowly, so that $\omega (r) -\omega (\infty ) \sim r^{-1}$ as $r\rightarrow \infty $.
Massive fields are however confined to the interior of AdS.
In both the ENAP and EYMH models, the gauge field has a mass (an effective term in the Lagrangian in the ENAP case, and dynamically generated in the EYMH case), localising the field either near the origin or in the vicinity of the event horizon.
Again this is reflected in the boundary conditions on the gauge field, which now decays more quickly to its asymptotic value, $\omega (r)-\omega (\infty )\sim r^{-\Delta }$ where $\Delta >1$  (\ref{BCinf3}) \cite{VanderBij:2001ah}.

In asymptotically flat space-time, the instability of the pure EYM solitons and black holes can be understood as resulting from an unstable balance between the attractive gravitational force and a repulsive force due to the non-Abelian gauge field.
In this picture the gauge field will tend to either collapse under gravity or radiate away to infinity if it is perturbed (as borne out by nonlinear simulations of the evolution of the unstable asymptotically flat EYM solitons and black holes \cite{Zhou:1991nu}).
In asymptotically AdS space-time, the gauge field is unable to radiate away to infinity, either because it will be reflected at the time-like AdS boundary (in the massless case), or because the diverging effective gravitational potential on the boundary means that the field is unable to escape to infinity (in the massive case).
For massive gauge fields in the ENAP and EYMH models, since the gauge field is localized either near the origin or event horizon, the balance between the gravitational attraction and gauge field repulsion is unstable, and we conjecture that the gauge field will collapse under gravity and the end-point of the instability of both solitons and black holes will be a Schwarzschild-AdS black hole.
However, in the massless pure EYM case, it is possible to have a stable balance between the gauge field repulsion and gravitational attraction, as the gauge field can extend all the way out to infinity.
In this case the gauge field configuration could be thought of as being analogous to a fundamental standing wave for a bounded system.
A fully nonlinear study of the evolution of both the stable EYM solutions and the unstable ENAP/EYMH solutions would be required to investigate this picture further, and we leave this to future work.

\begin{acknowledgments}
The work of EW is supported by the Lancaster-Manchester-Sheffield Consortium for Fundamental Physics under STFC grant ST/L000520/1,
and by an Erskine Visiting Fellowship at the University of Canterbury, Christchurch, New Zealand.
\end{acknowledgments}

\end{document}